\documentclass[12pt]{article}
\pdfoutput=1
\usepackage{float}
\usepackage[final]{graphicx}
\usepackage{amssymb,amsmath}
\usepackage{epsfig}
\usepackage{epstopdf}
\usepackage{latexsym}
\usepackage{graphicx}
\usepackage{subfigure}
\usepackage{booktabs}
\usepackage{bbm}
\usepackage{bbold}
\usepackage[margin=20pt,small]{caption}

\usepackage[toc]{appendix}

\usepackage{color}
\usepackage{datetime}
  
\ifpdf
\fi

\def\cF{{\cal F}}

\def\cM{{\cal M}}
\def\cN{{\cal N}}
\def\cO{{\cal O}}

\def\cL{{\cal L}}

\def\cS{{\cal S}}

\def\cX{{\cal X}}
\def\cL{{\cal L}}
\def\cA{{\cal A}}

\def\cF{{\cal F}}
\def\cG{{\cal G}}

\def\cM{{\cal M}}
\def\cX{{\cal X}}
\def\cZ{{\cal Z}}
\def\S{{\cal S}}
\def\F{{\cal F}}
\def\G{{\cal G}}
\def\T{{\cal T}}

\definecolor{cardinal}{rgb}{0.6,0,0}
\definecolor{darkgreen}{rgb}{0,0.5,0}
\definecolor{golden}{rgb}{0.92, 0.7, 0}
\definecolor{midnight}{rgb}{0, 0, 0.5}
\definecolor{darkblue}{rgb}{0.2, 0, 0.8}

\newcommand{\be}{\begin{equation}}
\newcommand{\ee}{\end{equation}}
\newcommand{\bea}{\begin{eqnarray}}
\newcommand{\eea}{\end{eqnarray}}

\begin{document}

\begin{titlepage}

\bigskip
\bigskip
\bigskip
\centerline{\Large \bf Effective Thermal Physics in Holography:}
\bigskip
\centerline{\Large \bf A Brief Review}
\bigskip
\bigskip
\centerline{{\bf Arnab Kundu}}
\bigskip

\bigskip
\centerline{Theory Division}
\centerline{Saha Institute of Nuclear Physics, HBNI,}
\centerline{1/AF Bidhannagar, Kolkata 700064, India.} 
\bigskip
\bigskip
\bigskip
\centerline{arnab.kundu[at]saha.ac.in}
\bigskip
\bigskip

\begin{abstract}

\noindent It is well-known that a Rindler observer measures a non-trivial energy flux, resulting in a thermal description in an otherwise Minkowski vacuum. For systems consisting of large number of degrees of freedom, it is natural to isolate a small subset of them, and engineer a steady state configuration in which these degrees of freedom act as Rindler observers. In Holography, this idea has been explored in various contexts, specifically in exploring the strongly coupled dynamics of a fundamental matter sector, in the background of adjoint matters. In this article, we briefly review some features of this physics, ranging from the basic description of such configurations in terms of strings and branes, to observable effects of this effective thermal description.

\end{abstract}

\newpage

\tableofcontents

\end{titlepage}

\newpage

\setcounter{equation}{0}
\section{Introduction}

Thermodynamics is ubiquitous. Typically, for a collection of large number of degrees of freedom, be it strongly interacting or weakly interacting, a thermodynamic description generally holds across various energy scales and irrespective of whether it is classical or quantum mechanical. The underlying assumption here is a notion of at least a {\it local thermal equilibrium}, for which such a formulation is possible. Intuitively, this is simple to define: a thermal equilibrium occurs when there is no net flow of energy. Typically this can be characterized by an intensive variable, temperature, with zero or very small time variation. The {\it smallness} needs to be established in terms of the smallest time-scale that is present in the corresponding system.

While thermodynamics has a remarkable reach of validity, equilibrium is still an approximate description of Nature, at best. Most natural events are dynamical in character. Of these, a particular class of phenomena can be easily factored out, that of systems at steady state. While steady state systems are not strictly in thermodynamic equilibrium, they can be described in terms of stationary macroscopic variables. For such systems, there is a non-vanishing expectation value of a flow, such as an energy flow or a current flow, which does not evolve with time. Typically, such states can be reached {\it asymptotically} starting from a generic initial state, or they appear as {\it transient} states before time evolution begins.

In this article, we will consider a similar situation. The prototype will consists of a {\it bath} degrees of freedom, which is assumed to be infinitely large and will serve the purpose of a reservoir. In this {\it bath} background, we will consider the dynamics of a {\it probe} sector. In this sector, a stationary configuration can be easily constructed, by dumping all the excess energy into the reservoir. For example, consider a non-vanishing current flow in the probe sector. There will be work done to maintain the constant current flow, therefore it is expected that the actual description is dynamical. However, if we engineer the bath sector as a source of providing this energy, or a sink in which this energy is deposited, the resulting configuration remains stationary.

In the framework of quantum field theory (QFT), a similar construction was considered by Feynman-Vernon in \cite{Feynman:1963fq}, based on the Schwinger-Keldysh formalism of \cite{Schwinger:1960qe, Keldysh:1964ud}. For some review on this formalism, see {\it e.g.}~\cite{Chou:1984es, Landsman:1986uw, Kamenev:2009jj}. In recent times, much work has gone into the reformulation of the Schwinger-Keldysh formalism, see {\it e.g.}~\cite{Haehl:2016pec, Haehl:2016uah, Haehl:2017qfl}. While many of our subsequent discussion potentially has a leg deep inside this formalism, we will not make explicit use of the formalism, to keep a terse discourse. The basic idea is based on the so-called thermofield double construct, which has appeared long time back in \cite{1975tu}. Subsequently, in \cite{Israel:1976ur}, a connection of classical black holes with the thermofield double construct was also established. For us, these two ideas are sufficient.

Consider the basic idea behind the thermofield double. Consider a quantum mechanical system, with a Hamiltonian $H$ and a complete set of eigenstates $\left | n \right  \rangle$, such that 
\begin{eqnarray}
H \left | n \right  \rangle = E_n \left | n \right  \rangle \ , 
\end{eqnarray}
where $E_n$ is the energy of the corresponding eigenstate. Evidently, $\{ \left | n \right  \rangle \}$ constitute a basis of the Hilbert space. Let us now double the total degrees of freedom, by considering two copies of the same system. At the level of the dynamics, these two copies of degrees of freedom are non-interacting between them.\footnote{The total Hamiltonian of the doubled system can be defined as $\left( H_1 + H_2 \right) $ or $ \left( H_1 - H_2 \right)$, With the latter choice, the thermofield double state does not evolve with time.} Therefore, the Hilbert space of the doubled quantum system is spanned by $\{ \left | m \right  \rangle_1 \left | n \right  \rangle_2 \}$. Given this, the thermofield double state is defined as:
\begin{eqnarray}
\left | {\rm TFD} \right  \rangle = \frac{1}{\sqrt{Z\left( \beta \right) }} \sum_n {\rm exp} \left( - \frac{\beta E_n}{2} \right)  \left | n \right  \rangle_1 \left | n \right  \rangle_2 \ ,  \label{tfd}
\end{eqnarray}
This is certainly a special state in the doubled quantum system. We can assign a density matrix corresponding to this state: $\rho_{\rm TFD} = \left | {\rm TFD} \right  \rangle  \left \langle {\rm TFD} \right | $. This is a pure density matrix, as can be explicitly checked by establishing: $ \rho_{\rm TFD}^2 =   \rho_{\rm TFD}$.

Given such a pure density matrix, let us compute the reduced density matrix while integrating out one copy of the system. Thus we obtain:
\begin{eqnarray}
\rho_{\rm th} = {\rm Tr}_2 \rho_{\rm TFD} = \frac{1}{Z\left( \beta \right) } \sum_n {\rm exp} \left( - \beta E_n \right) \left | n \right  \rangle_1  \left \langle n \right |_1 \ .
\end{eqnarray}
The reduced density matrix, denoted above by $\rho_{\rm th}$ appears thermal in nature, with a temperature $\beta^{-1}$. Thus, given the thermofield double state, one can construct an equivalent thermal description. The process of integrating out one copy of the system may be conducted in various ways: this can be thought of as integrating out a subsystem to compute entanglement between the two. In the context of a black hole, similar to \cite{Israel:1976ur}, or in the presence of a causal horizon, one can construct a Kruskal extension of the geometry. This maximal extension of the geometry can be thought of as the thermofield double and by integrating out one side, an effective thermal density matrix is obtained. It is also clear from the above discussion that, given any gauge invariant observable or a collection of such operators acting on the untraced system, denoted by $\cO$, the expectation value is simply given by
\begin{eqnarray}
\left \langle {\rm TFD} \right  | \cO  \left | {\rm TFD} \right  \rangle = \frac{1}{Z\left( \beta \right) } \sum_n {\rm exp} \left( - \beta E_n \right) \left \langle  n \right  | \cO  \left | n \right  \rangle \ ,
\end{eqnarray}
which is the thermal expectation value.

\begin{figure}[h!]
\centering
\includegraphics[scale=0.4]{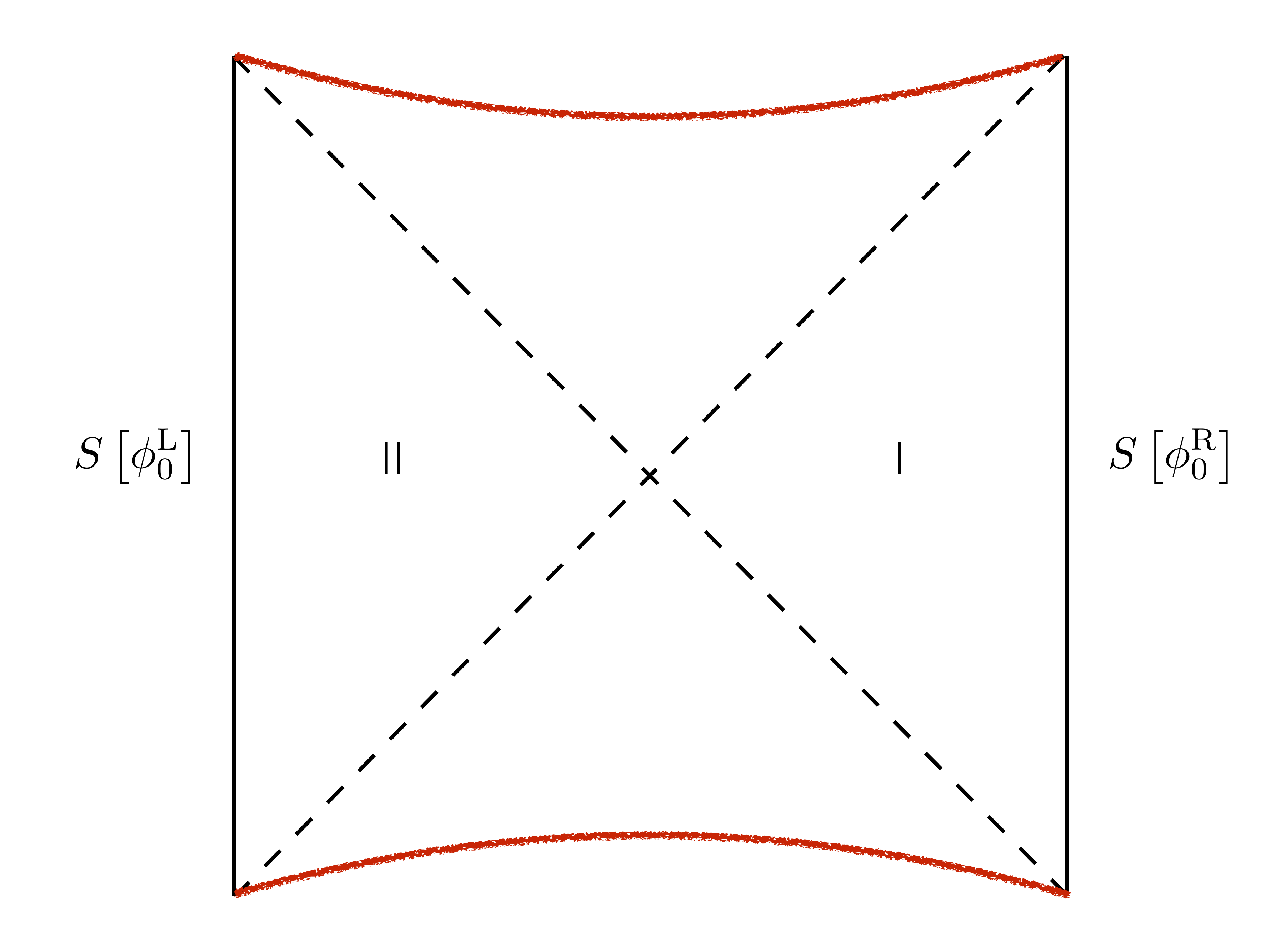}
\caption{\small The Penrose diagram corresponding to an eternal black hole in AdS. The left and the right boundaries are where the dual CFT is defined, which are schematically denoted by $S\left[ \phi_{\rm L}^0\right]$ and $\left[ \phi_{\rm R}^0\right]$, respectively. }
\label{figkts}
\end{figure}

A similar picture holds true in the Holographic framework, which will be the primary premise in our subsequent discussions. In \cite{Maldacena:2001kr}, eternal black holes in an asymptotically AdS geometry were proposed to be dual to two copies of the conformal field theory (CFT). Each CFT corresponds to the dual CFT that is defined on the conformal boundary of AdS. The basic picture is represented in figure \ref{figkts}. 
In the Euclidean signature, the corresponding thermofield double state is created by the Euclidean path integral over an interval of $\beta/2$. The thermofield double state, defined in (\ref{tfd}), is maximally entangled from the point of view of the doubled degrees of freedom. Tracing over one copy produces a thermal effective description and this seems to lie at the core of the construction. Motivated by this, one can surmise that a qualitative emergent description of a thermofield double state ensures an effective thermodynamics. For this to happen, the essential ingredients is a black hole like causal structure. An intriguing idea relating quantum entanglement and the existence of an Einstein-Rosen bridge has recently been proposed in \cite{Maldacena:2013xja}.

Let us now crystalize our discussion towards our specific goal. All of the above discussions are assumed to go through the framework of QFT. In general, of course, for weakly coupled QFT systems, explicit perturbative calculations are sometimes feasible, although those are certainly not simple for processes involving real-time dynamics. Furthermore, the existence of such a small coupling is far from guaranteed in Nature, {\it e.g.}~the quark-gluon plasma (QGP) at the Relativistic Heavy Ion Collider (RHIC) and the Large Hadron Collider (LHC) at TeV-scale (see {\it e.g.}~\cite{Aamodt:2010pa}), or the cold atoms at unitarity at eV-scale (see {\it e.g.}~\cite{O'Hara:2002}). In general, at strong coupling, powerful symmetry constraints in terms of Ward-Takahashi identities or non-perturbative Schwinger-Dyson equations can sometimes help. However, for many explicit cases, these approaches have limitations. One can certainly try to formulate such complicated issues on a computer, using lattice-methods and its modern generalizations, at the cost of analytical control over the physics.

An interesting avenue is to explore the Gauge-Gravity duality, or the Holographic principle, or the AdS/CFT correspondence\cite{Maldacena:1997re, Gubser:1998bc, Witten:1998qj, Aharony:1999ti}. While all these words, in a precise sense, carry different meaning, we will not distinguish between them. The basic statement here is: A large class of strongly coupled QFTs, such as the $\cN=4$ super Yang-Mills (SYM) theory with an SU$(N_c)$ gauge group in the limit $N_c \to \infty$, is dual to classical (super) gravity in an AdS$_5 \times S^5$ geometry. By this duality, one translates questions of strongly coupled large $N_c$ gauge theories into questions of classical gravity. The latter is a more familiar and tractable framework for explicit calculations. Although this class of QFTs are not what one would like to understand for experimental processes in the RHIC or the LHC, they certainly can serve the purpose of instructive toy examples. Mathematically, the duality statement schematically takes the form:
\begin{eqnarray}
\left \langle {\rm exp} \left( \int_{\partial {\rm AdS}} d^dx \phi_0 \cO \right) \right \rangle \equiv \cZ_{\rm gravity} \left[ \phi_{\rm AdS} | \phi_{\partial{\rm AdS}} = \phi_0 \right] \ , 
\end{eqnarray}
where the left hand side is an expression in the CFT and the right hand side is defined as the gravity partition function in AdS, subject to a specific boundary condition. Correspondingly, correlations in the CFT can be calculated by taking functional derivatives on the LHS, with respect to $\phi_0$. The duality relates this correlation function to a similar calculation in the bulk gravity scenario on the RHS. For a detailed discussion on correlation functions in this framework, see \cite{Son:2002sd} for Minkowski-space correlators, \cite{Herzog:2002pc} for a Schwinger-Keldysh framework, \cite{Iqbal:2009fd} for a real-time analysis, \cite{Skenderis:2008dh, Skenderis:2008dg} for a detailed account of the real-time correlation functions.

During the past couple of decades, a wide range of research has been carried out in this framework, in the context of quantum chromodynamics (QCD) as well as several condensed matter-type systems. Popularly, these efforts are sometimes dubbed as AdS/QCD or AdS/CMT literature. By no stretch of imagination, we will attempt to be extensive: for some recent reviews, see {\it e.g.}~\cite{CasalderreySolana:2011us, DeWolfe:2013cua, Gubser:2009md} for the former, and \cite{Hartnoll:2009sz, Hartnoll:2016apf} for the latter.

The Holographic framework, which can also be viewed as the most rigorous definition of a theory of quantum gravity, is a promising avenue to explore complicated gauge theory dynamics, qualitatively. There already exists a large literature analyzing various dynamical features of thermalization, quench dynamics for strongly coupled systems. We will not attempt to enlist the references here, however, present one example of new results, such as scaling laws in quantum quench processes\cite{Das:2016eao}. For a more general discourse on non-equilibrium aspects in Holography, see {\it e.g.}~\cite{Hubeny:2010ry}.

Even though remarkable progress has been made in understanding dynamical issues, they remain rather involved and, in general, far more complex than equilibrium physics. Thermal equilibrium is particularly simple since it can me macroscopically described in terms of a small number of intensive and extensive variables. Intriguingly, for steady state configurations, which is neither in precise thermal equilibrium nor fullly dynamical, an {\it effective} thermal description may hold. See for example, \cite{PhysRevLett.95.267001, PhysRevLett.97.227003, Karch:2010kt}, in systems at quantum criticality\cite{PhysRevLett.103.206401} or aging glass systems\cite{Cugliandolo.97}. In this review, we will briefly summarize a similar construct for a wide class of strongly coupled gauge theories, within the Holographic framework.

As mentioned earlier, we have one {\it bath} sector and one {\it probe} sector. These are made of the adjoint matter of an SU$(N_c)$ gauge theory and a fundamental matter sector, respectively. A canonical example of this is to consider an $\cN =2$ hypermultiplet matter as a probe introduced in the $\cN=4$ SYM system. We will work in the limit $N_c \ll N_f$, where $N_f$ is the number of the fundamental matter. This limit ensures the suppression of backreaction by the matter sector and we can safely treat them as probes. In a more familiar language, this limit is similar to the {\it quenched} limit in the lattice literature, wherein one ignores loop effects of quark degrees of freedom (which are the fundamental sector here), but includes loop effects in the gluonic matter (which is the adjoint matter).

The adjoint matter sector in $(p+1)$-dimension comes from the low energy limit of a stack of D$p$-branes, from the open string description of the brane. Equivalently, the closed string low energy description of the same stack of branes is given by a classical supergravity background in ten-dimensions. Gauge-Gravity duality is an equivalence between the two descriptions. Now, in this picture, one can introduce an additional set of $N_f$ D$q$ branes, in the limit $N_f \ll N_c$, which introduces an additional set of open string degrees of freedom. In the probe limit, the gravitational backreaction of the D$q$-branes are ignored and therefore these only have an open string sector description. This open string sector, geometrically, can be studied by analyzing the embedding problem of D$q$ branes in the given supergravity background.

Within this framework, many interesting physics have been uncovered within the probe fundamental sector, specially the thermal physics, see {\it e.g.}~\cite{Babington:2003vm, Mateos:2006nu, Albash:2006ew, Karch:2006bv}. There is a vast literature on this, and we will not attempt to provide a substantial reference here. For us, the steady state configuration will be engineered in this probe sector, by exciting a U$(1)$-flux on the probe D-brane worldvolume. This steady state will be maintained by working in the $N_f \ll N_c$ limit. 
\begin{figure}[h!]
\centering
\includegraphics[scale=0.45]{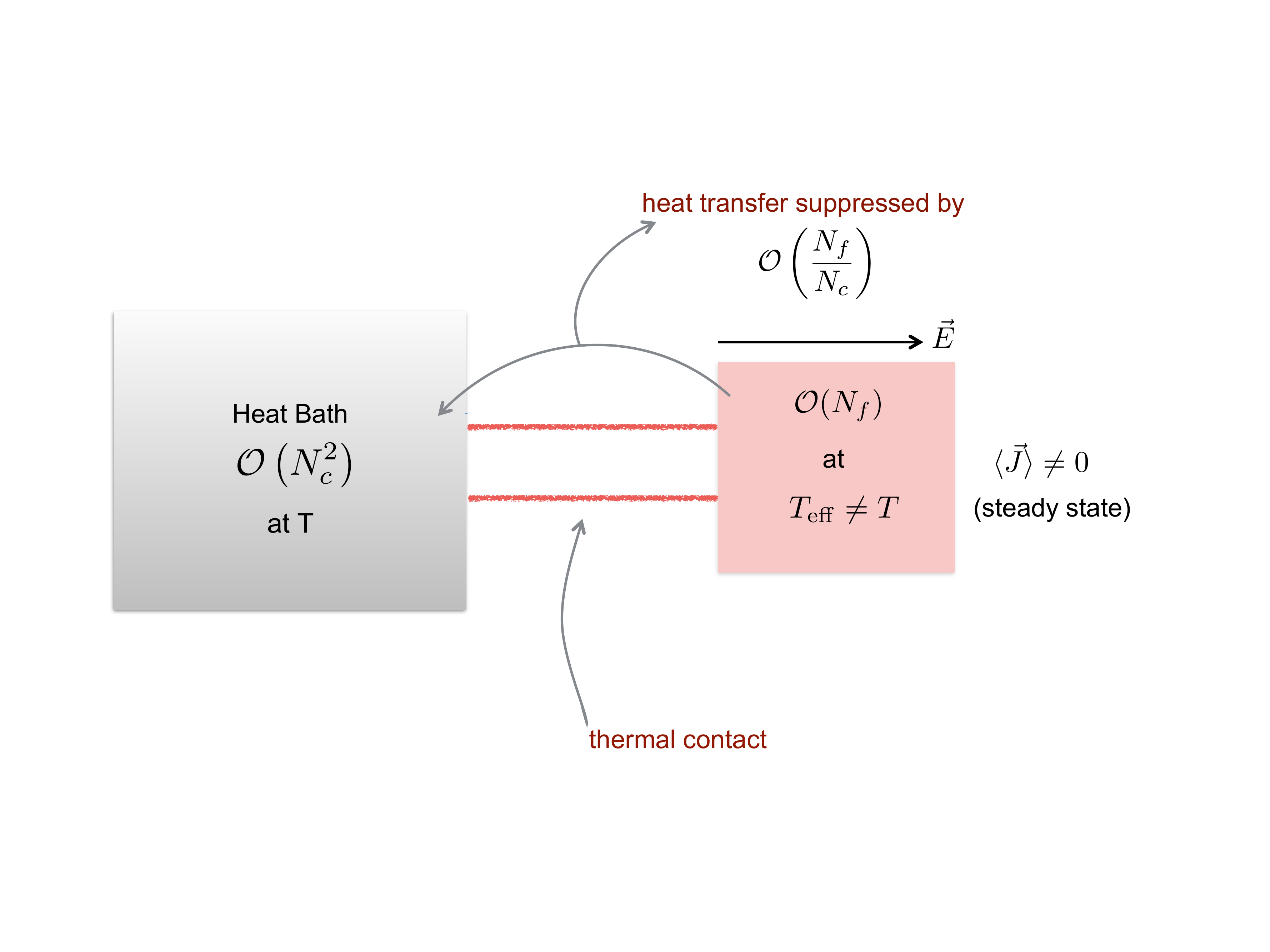}
\caption{\small A schematic arrangement of the steady state. The $\cO(N_f)$ fundamental degrees of freedom are in a steady-state, situated in a heat bath of $\cO(N_c^2)$ degrees of freedom,}
\label{ssprobe}
\end{figure}
Pictorially, this is demonstrated in figure \ref{ssprobe}.

The non-linear dynamics of the brane, along with the U$(1)$-flux will induce an event horizon to which only the brane degrees of freedom are coupled, with an open string analogue of equivalence principle.  Qualitatively, therefore, one inherits a black hole like causal structure and a corresponding thermofield double description. As we have discussed above, we are thus led to a thermal density matrix starting from a maximally entangled state. This construction lies at the core of our subsequent discussions. Interestingly, this description can be explicitly realized both on a string worldsheet, in which one studies a probe long open string in a supergravity background, by studying the dynamics of Nambu-Goto (NG) action; as well as on a probe D-brane by studying the Dirac-Born-Infeld (DBI) action. In the latter, although open strings are present, they may not appear explicit.

Based on these ideas, a lot of interesting physics has been explored over the years. Here we will merely present a few broad categories and some representative references, which we will not explicitly discuss in this review. In \cite{Gubser:2006bz, Herzog:2006gh}, the drag force on an external quark passing through the $\cN=4$ SYM was calculated, meson dissociation by acceleration was explored in \cite{Peeters:2007ti}, the causal structure on the string worldsheet was discussed in details in \cite{Hubeny:2014kma}. Stochasticity and Brownian motion associated with the worldsheet event horizon was explored in \cite{deBoer:2008gu, Son:2009vu, Giecold:2009cg, CasalderreySolana:2009rm}. For a review on the hard probe dynamics, see {\it e.g.}~\cite{DeWolfe:2013cua, Gubser:2009md}, for a quark dynamics review, see {\it e.g.}~\cite{Chernicoff:2011xv}, and for a general review, see {\it e.g.}~\cite{CasalderreySolana:2011us}. The idea of ER=EPR has been explored on the string worldsheet in {\it e.g.}~\cite{Jensen:2013ora, Sonner:2013mba, Jensen:2014lua, Chernicoff:2013iga}. On the probe brane, a non-equilibrium description of phase transition and effective temperature is explored in {\it e.g.}~\cite{Nakamura:2012ae, Nakamura:2013yqa, Hoshino:2014nfa, Matsumoto:2018ukk, Hoshino:2018vne}. Relatedly, thermalization on this probe sector is discussed in {\it e.g.}~\cite{Das:2010yw, AliAkbari:2012hb, AliAkbari:2012vt}.

This article is divided into the following parts: In the next section we begin with a brief description of how non-linearity can result in a black hole like causal structure, and how this is currently being investigated to understand features of QCD. In section $3$, we introduce the Holographic framework. In the next three sections, we discuss some explicit and instructive features of the effective temperature on a string worldsheet as well as a D-brane worldvolume, based on analytically controllable examples. Section $7$ is devoted to a generic discussion of the physics, without reference to any explicit example. Finally, we conclude in section $8$.

\setcounter{equation}{0}
\section{Temperature, Outside the Folklore}

The standard folklore concept of temperature is certainly in describing equilibrium thermal systems, at least in a local sense. In physics, there are various ways to define the temperature of a system: In thermodynamics, temperature is defined as an intensive variable that encodes the change of entropy with respect to the internal energy of the system. In kinetic theory, the definition of temperature can be given in terms of the equipartition theorem for every microscopic degree of freedom. In linear response theory, temperature can also be defined in terms of the fluctuation-dissipation relation.

Outside the realm of equilibrium thermal description, the notion of a temperature can sometimes be generalized. This is a vast and evolving topic in itself and we will refer the interested to reader to {\it e.g.}~\cite{tempreview}. One particular method, which is also relevant for our discussions, is to use the fluctuation-dissipation relation to define temperature in a non-equilibrium process with a slow dynamics, see {\it e.g.}~\cite{2011JPhA...44V3001C}. In particular, these apply reasonably well within classical and quantum mechanical systems at steady state. However, it is unclear, even within such systems, whether these ideas hold at strong coupling.

Intriguingly though, strong coupling seems to suggest a much simpler scenario. Most of our review will concern with the standard strongly coupled systems (and toy models) within the framework of gauge-gravity duality, where a fluctuation-dissipation based temperature is already explored in \cite{CaronHuot:2011dr}. However, in this section, let us briefly review some recent interests and activities in strongly coupled quantum chromodynamics (QCD) itself. The Schwarzschild radius of a typical hadron of mass $~ 1$ GeV turns out to be $\sim 10^{-39}$ fm. This estimate assumes gravitational interaction in determining the Schwarzschild. Instead, one replaces the gravitational interaction by the strong interaction, which amounts to multiplying the above estimate by a factor of $\alpha_s / G_N$, where $\alpha_s \sim {\cal O}(1)$ is the strong coupling constant and $G_N$ is the Newton's constant. This yields a Schwarzschild radius $\sim 1$ fm\cite{Castorina:2007eb}. Let us discuss further motivation, following \cite{Castorina:2007eb}.

As described in \cite{Novello:1999pg}, the non-linear effects of a medium for simple electrodynamics can lead to the trapping of photons. This is described in terms of an effective Lagrangian, denoted by $\cL(F)$, and an emergent metric of the form:
\begin{eqnarray}
g_{\mu\nu} = \eta_{\mu\nu} \cL '  - 4 F_{\alpha\mu} F_\nu^\alpha \cL'' \ .
\end{eqnarray}
The photon trapping surface is simply obtained by solving the equation $g_{00} =0$. Thus, even with a simple U$(1)$ theory, non-linear effects lead to an black hole like configuration.

Now, QCD or any such non-Abelian gauge theory, is inherently non-linear and produces a non-trivial medium for itself. In this case, the effective action can be fixed to be (see the discussion in \cite{Castorina:2007eb}):
\begin{eqnarray}
\cL_{\rm QCD} = \frac{1}{4} F_{\mu\nu} F^{\mu\nu} \epsilon \left(g_{\rm QCD} F \right) \ , \label{effqcd}
\end{eqnarray}
where $\epsilon \left(g_{\rm QCD} F \right)$ represents a dielectric variable of the medium and $g_{\rm QCD}$ is the bare QCD coupling. Perturbatively, this function can be evaluated and at one-loop we get:
\begin{eqnarray}
\epsilon \left(g_{\rm QCD} F \right) = 1 - \frac{11 N_c - 2 N_f}{48 \pi^2} \left( \frac{g_{\rm QCD}^2}{4\pi }\right) \log \left( \frac{\Lambda^2}{g_{\rm QCD} F}\right) \ ,  
\end{eqnarray}
where $\Lambda$ is the cut-off scale. Setting $\epsilon = 0$ yields an algebraic solution for $g_{\rm QCD} F$, which is given by
\begin{eqnarray}
g_{\rm QCD} F = \Lambda^2 {\rm exp} \left[ - \frac{4\pi}{g_{\rm QCD}^2} \frac{48 \pi^2} {11 N_c - 2 N_f} \right]  \ .
\end{eqnarray}
This already indicates a possible event horizon structure, since this zero is the locus where the effective kinetic term in (\ref{effqcd}) changes sign. Note that, this event horizon structure is already visible at the perturbative level with a non-linearly interacting theory. However, this is not strictly rigorous, since the perturbative result for $\epsilon$ assumes a small $g_{\rm QCD}$. Thus near the $\epsilon =0$ locus, perturbative calculations are not valid.

Based on the hints above, the basic conjecture was made in \cite{Castorina:2007eb}: due to confinement, the physical vacuum is equivalent to an event horizon for quarks and gluons. This event horizon can be penetrated only through quantum processes such as tunnelling. This constitutes a QCD analogue of the Hawking radiation, which is hence thermal. While the basic idea is essentially based on the thermofield double type construction, the above also makes certain predictions in terms of an universality in thermal hadron production in high energy collisions, in terms of an effective temperature that is defined in terms of the QCD confinement-scale. This is based primarily on two inputs: (i) A thermal description of high energy collision processes with an effective temperature determined in terms of the QCD string tension, or the gluon saturation momentum\cite{Kharzeev:2005iz, Kharzeev:2006zm}; (ii) The experimental data of hadron production across a large energy scale from GeV to TeV exhibits a universal thermal behaviour, with an effective universal temperature $T_{\rm eff} \sim 150-200$ MeV\cite{BraunMunzinger:2003zd, Cleymans:1992zc, Becattini:1997rv}.

The resulting essential consequences are as follows: Color confinement and vacuum pair production leads to the event horizon. The only information that can escape this horizon is a color-neutral thermal description with an effective temperature. The resulting hadronization is essentially a result of such successive tunnelling processes. Also, there is no clear notion of ``thermalization" through standard kinetic theoretic collision processes. A thermal description emerges as a consequences of the strongly coupled dynamics and the pair production. We will end our brief review here, since this is still an active field of research and is an evolving story.

\setcounter{equation}{0}
\section{The Holographic Framework}

In this article, we will primarily review the progress made and important results obtained in the framework of holography, the AdS/CFT correspondence\cite{Maldacena:1997re} to be precise. AdS/CFT correspondence can be viewed as a successful marriage between 't Hooft's early ideas on large $N$ gauge theories as string theories\cite{tHooft:1973alw} and, later, the idea of holography pioneered by 't Hooft and Susskind in \cite{tHooft:1993dmi, Susskind:1994vu}. This correspondence arises from a more fundamental duality between open and closed strings, in string theory, and has become the cornerstone of quantum gravitational physics. So much so that the modern understanding of the correspondence identifies it with the most rigorous definition of a quantum theory of gravity, in an asymptotically AdS-spacetime. For our purpose, we will only require the technical aspects of this correspondence.

To do so, we briefly review the standard understanding of how this correspondence (or, duality) emerges. Closed string theories describe a consistent theory, and in the low energy limit, this consistent description can be truncated to various supergravity theories, in general. On the other hand, open string theories naturally arise as boundary conditions which contain the information of the string end point.   The canonical construction begins with a stack of $N$ D$p$ branes and analyzing the corresponding low energy description of the system. Let us recap the best understood example of this, {\it i.e.}~when $p=3$.

For a stack of $N$ D$3$-branes, the massless spectrum arising from open strings can be readily obtained. Assuming that the open strings are oriented, the degrees of freedom at the open string end points can be shown to transform under the adjoint representation of U$(N)$ group. Furthermore, global symmetry of this stack of D$3$-branes uniquely determine the interacting description in the massless sector of the open string spectrum: the ${\cal N}=4$ super Yang-Mills (SYM) in $(3+1)$-dimension. Also, the U$(N)$ gauge group splits naturally into a global U$(1)$ and an SU$(N)$. For the gauge theoretic description of the system, we can safely ignore the overall U$(1)$ mode, since this corresponds to an overall excitation, see {\it e.g.}~\cite{Aharony:1999ti} for more details. 
\begin{figure}[h!]
\centering
\includegraphics[scale=0.45]{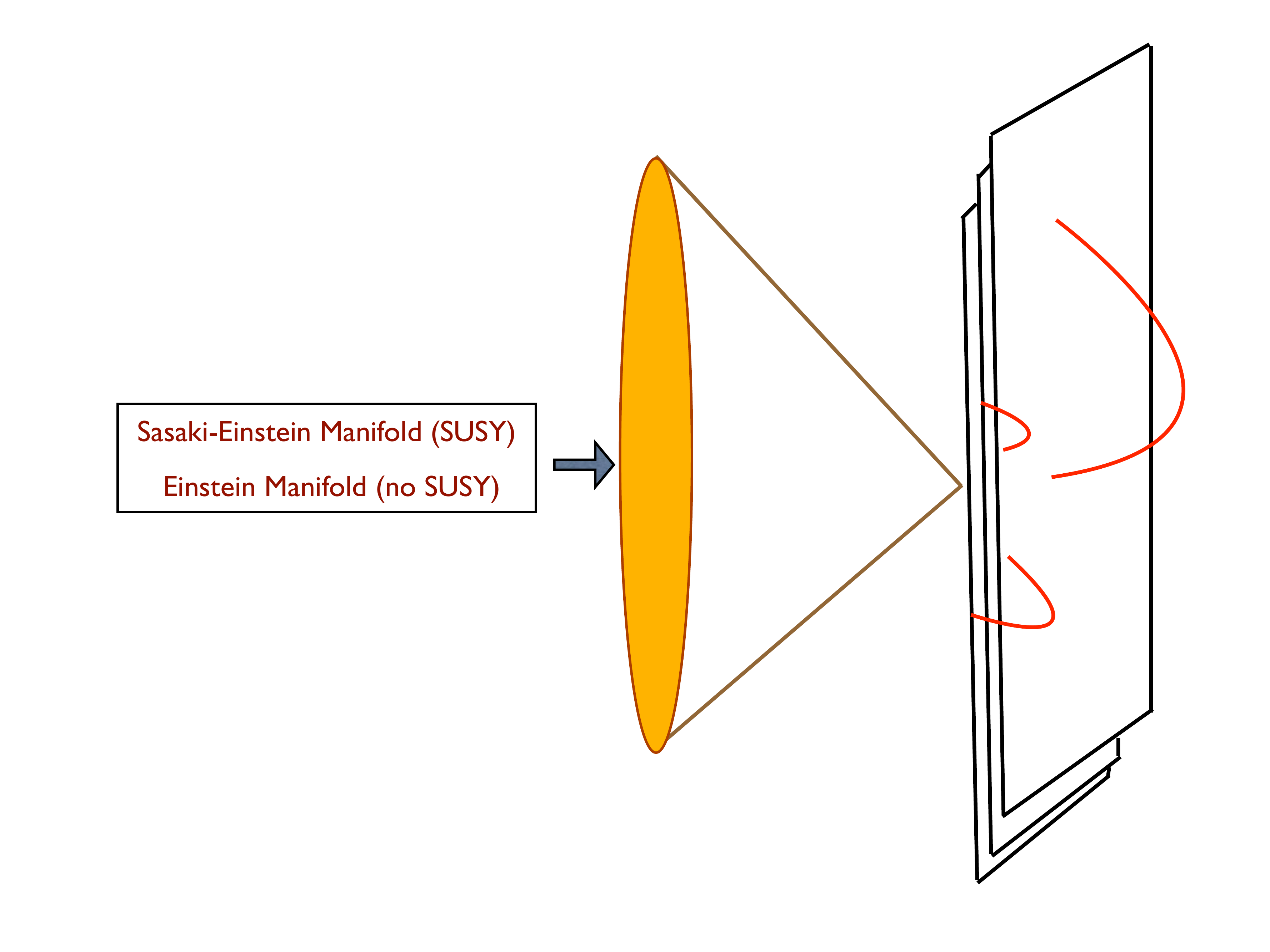}
\caption{\small A schematic presentation of a generic D-brane construction, A certain $N$ number of D$3$-branes are placed on the tip of a cone, whose base could be a Sasaki-Einstein manifold (susy preserving), or simply an EInstein manifold (susy breaking). This picture assumes the cone is a six-dimensional manifold. This condition is not required and can be relaxed easily for other dimensional cases, at least, in principle. The red curves are strings beginning and ending on the stack of the D-branes, whose low energy limit can be captured by a standard gauge theory, similar to the ${\cal N}=4$ SYM.}
\label{dstack}
\end{figure}

Alternatively, a stack of D$3$-branes will source gravity, which is captured by the closed string excitation of this system. In this case, such a stack of $N$ D$3$-branes can be obtained explicitly by solving the low energy description of the corresponding closed strings, which is given by type IIB supergravity in ten dimensions. This solution is given by
\begin{eqnarray}
&& ds^2 = H^{-1/2} \eta_{\mu\nu} dx^\mu dx^\nu + H^{1/2} \delta_{ij} dx^i dx^j  \ , \label{d31} \\
&& e^{\phi} = g_s \ , \label{d32} \\
&& C_{(4)} = H^{-1} g_s^{-1} \, dx^0 \wedge \ldots \wedge dx^3 \ ,  \quad H = 1 + \frac{4\pi g_3 N \alpha'^2}{r^4} \ , \label{d33} 
\end{eqnarray}
where $\mu, \nu = 0, \ldots , 3$; $i, j = 4, \ldots , 9$; $\alpha'$ yields the inverse string tension, $g_s$ is the string coupling. The dilaton field, denoted by $\phi$, sets the string coupling constant and the D$3$-brane sources the so-called Ramond-Ramond $4$-form, denoted by $C_{(4)}$. Defining a new radial coordinate, $u = r/\alpha'$ and taking the $\alpha' \to 0$ limit then decouples the near horizon physics of the geometry in (\ref{d31}) from the asymptotic flat infinity. The geometry becomes an AdS$_5 \times S^5$ and this corresponds to the low energy description of the D$3$ branes' physics. Thus, the correspondence reads: type IIB superstring theory in AdS$_5\times S^5$ background is equivalent to ${\cal N}=4$ SU$(N)$ SYM theory in $(3+1)$-dimension. This $(3+1)$-dimensional geometry is simply the conformal boundary of the AdS$_5$-background. A similar construction can be obtained for a generic D$3$-brane, placed on the tip of a cone, as shown in figure \ref{dstack} and correspondingly obtain a similar duality. In fact, this can be further generalized for a stack of D$p$-branes\cite{Itzhaki:1998dd}.

\begin{figure}[h!]
\centering
\includegraphics[scale=0.45]{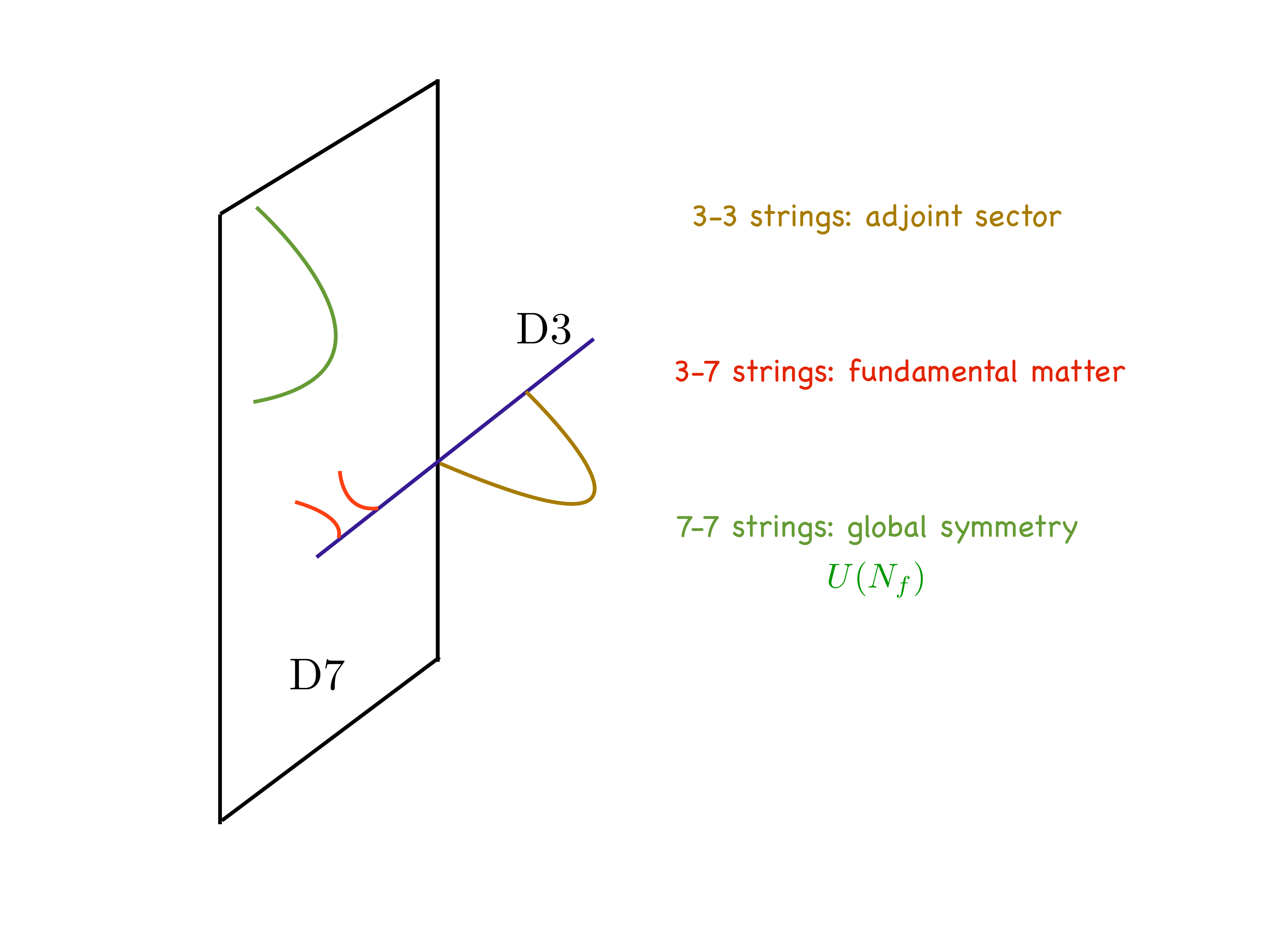}
\caption{\small A schematic presentation of a generic D-brane construction: A certain $N$ number of D$3$-branes, $N_f$ number of D$7$-branes, before taking the decoupling limit. The low energy degrees of freedom can be simply obtained by considering string spectrum. The $3-3$ strings yield the matter content of ${\cal N}=4$ SYM, the $3-7$ and $7-3$ strings yield the ${\cal N}=2$ hypermultiplet, and the $7-7$ string sector is non-dynamical, {\it i.e.}~it yields a global symmetry.}
\label{dstackfund}
\end{figure}

The construction above can now be generalized by introducing additional degrees of freedom. In the dual gauge theory, this corresponds to introducing additional matter sector transforming in different representations of the SU$(N)$ gauge group; in the gravitational description this corresponds to the inclusion of new gravitational degrees of freedom. A specially interesting case is to consider a matter sector that transforms in the fundamental representation of the gauge group. This can be done by adding a ${\cal N}=2$ hypermultiplet to the ${\cal N}=4$ SYM. Gravitationally, this can be realized by introducing a stack of $N_f$ D$7$ branes in the near-horizon ({\it i.e.}~decoupled) AdS$_5 \times S^5$ geometry of (\ref{d31}). A particularly instructive limit to explore is to take $N_f \ll N$ such that gravitational backreaction of the additional D$7$ branes can be safely ignored. Pictorially, this is represented in figure \ref{dstackfund}.
Evidently, these constructions can be generalized for a wide class of examples, following the approach in \cite{Karch:2002sh}. We will elaborate more on these constructions in later sections.

The prototype model in our subsequent discussion is based on the dynamics of this additional set of branes, which capture the dynamics of open long strings. There is a ``bath" with a large number of degrees of freedom which is provided by the stack of $N$ D$3$-branes, in which $N_f$ D$7$-branes are embedded, in the limit $N_f \ll N$. Generically, we intend to study the embeddings of a probe D$q$-brane in the background of a large number of D$p$-branes. In this probe sector, a steady-state can be engineered by simply pumping energy into the system, which can be dissipated into the bath without causing any energy change of the bath. This is facilitated by the $N_f \ll N$ limit, and away from this limit the above approximation breaks down. We will concentrate only on this probe regime.

Before moving further, let us note the following: The fundamental matter sector, essentially, is described by long, open strings. It is not unreasonable to assume that there is a sensible limit in which the probe D-brane becomes irrelevant and the essential physics can be captured by the dynamics of explicit strings. Indeed, for specific models, this limit can be made precise: For examples, if we introduce a mass of the ${\cal N}=2$ hypermultiplet sector, in the large mass limit, the probe D$7$ brane is pushed far away at the UV, leaving a family of strings that connect between the UV D$7$-brane and the IR D$3$-brane. In this limit, one can simply consider the open string sector only. Motivated by this, we will review the physics of open strings in an AdS-background, in the next section.

\setcounter{equation}{0}
\section{The String Worldsheet Description}

We begin with reviewing a general description of a string worldsheet which is embedded in an AdS-background, following closely the treatment of \cite{Caceres:2010rm, Chernicoff:2010yv}. The simplest case is to take an AdS$_5 \times S^5$ geometry, in which the AdS/CFT dictionary is very well-known. The discussion below, however, is far more generic and applies to a more general supergravity background which can be bisected by an AdS$_2$. Let us begin with the Poincar\'{e} patch:
\begin{eqnarray}
ds^2 = \frac{R^2}{z^2} \left( - dt^2 + d\vec{x}^2 + dz^2 \right)  + ds_{S^5}^2 \ , \label{back}
\end{eqnarray}
where $R$ is the curvature scale of the AdS$_5$, $\{t, \vec{x}\}$ represent the ${\mathbb R}^{1,3}$ in which the dual ${\cal N}=4$ SYM theory is defined, and $z$ is the AdS-radial coordinate. The conformal boundary is located at $z\to 0$ limit and the infra-red (IR) is located at $z\to \infty$. In the presence of a bulk event horizon, the IR is cut off at some finite location, which we denote by $z_{\rm h}$. Note that, it is possible to select general radial foliations of the bulk AdS geometry, such that the dual CFT (in this case, the $\cN =4$ SYM) is defined on the corresponding Lorentzian manifold $\cM^{1,3}$, which is realized as the conformal boundary of the bulk spacetime. This is best expressed in the so-called Fefferman-Graham patch\cite{FG}, which is suitable to describe any asymptotically AdS geometry:
\begin{eqnarray}
ds^2 = \frac{R^2}{z^2}  \left( dz^2 + g_{AB} dx^A dx^B \right)  \ , \label{adsfg}
\end{eqnarray}
where $g_{AB}$ is a function of the radial co-ordinate $z$, as well as the $x^A$ co-ordinates. The corresponding CFT is defined on the background whose metric is given by $ds_{\rm CFT}^2 = \left. g_{AB} dx^A dx^B \right|_{z \to 0}$, which defines the line-element on $\cM^{1,3}$. The background in (\ref{adsfg}) is uniquely determined in terms of the boundary data $\left. g_{AB} \right|_{z\to 0}$ and a sub-leading mode of the function $g_{AB}(z)$, in a $z$-expansion around the conformal boundary. This sub-leading mode, essentially, contains the data of the CFT stress-tensor. The detailed procedure of extracting the resulting CFT stress-tensor, which is based on holographic renormalization, is discussed in {\it e.g.}~\cite{deHaro:2000vlm, Skenderis:2000in, Skenderis:2002wp}. The corresponding 't Hooft coupling is given by $\lambda \equiv  g_{\rm YM}^2 N_c = R^4 / \alpha'^2$, where $g_{\rm YM}$ is the gauge theory coupling and $N_c$ determines the rank of the gauge group, while $\alpha'$ is the inverse tension of the string.

The dynamics of the string is governed by the Nambu-Goto action:
\begin{eqnarray}
S_{\rm NG} = - \frac{1}{2\pi\alpha'} \int d\tau d\sigma \sqrt{- {\rm det} \gamma} + S_{\rm boundary}\ , \quad \gamma_{ab} = G_{\mu\nu}(X) \partial_a X^\mu \partial_b X^\nu \ , \label{ngact}
\end{eqnarray}
where $\{\tau, \sigma\}$ represent the worldsheet coordinates, $\alpha' = l_s^2$ sets the string tension (while $l_s$ is the string length), and $S_{\rm boundary}$ is a generic boundary term. The background manifold is described by the metric $G_{\mu\nu}(X)$, while $X$ represents the coordinate patch chosen to describe the manifold. The various indices are chosen to represent the following: $a, b, \ldots$ denote the worldsheet coordinates, $A, B, \ldots$ denote the coordinates on the manifold where the CFT is defined (in general, denoted by $\cM^{1,3}$) and $\mu, \nu, \ldots$ denote the full ten-dimensional supergravity background. We will consider cases in which the string is extended along the radial direction, and therefore the boundary term can capture the coupling of the end-point with an applied external field:
\begin{eqnarray}
S_{\rm boundary} = \int d\tau A_B \partial_\tau X^B \ .
\end{eqnarray}
Before proceeding further, let us offer some comments regarding the boundary term. The physical picture is as follows: Given the ten-dimensional geometry, one considers such {\it long} strings which extend from the IR of the geometry all the way to the UV. Note that, these open strings must have an end point on a D-brane and therefore it makes sense to think of the end point as a point on a D-brane at the UV. There are various ways to think about such configurations, for our purpose we can introduce a UV cut-off $z_{\rm UV}$ where the string ends. Such a string certainly carries an energy, as a function of $\{z_{\rm IR}, z_{\rm UV}\}$, where $z_{\rm IR}$ is the IR end of the geometry. Irrespective of the details, one can naturally assign a physical mass-scale, $M$, associated to the heavy string with a static and constant profile:
\begin{eqnarray}
M = \frac{\lambda^{1/2}}{2\pi z_{\rm UV}} \ .
\end{eqnarray}
The $\sqrt{\lambda}$ behaviour simply comes from the string tension, which is proportional to $\alpha'^{-1}$. Thus, we simply introduce a fundamental degree of freedom ({\it i.e.}~like a quark) with a non-vanishing mass. In the context of $\cN=4$ SYM, this fundamental matter can sit inside an $\cN=2$ hypermultiplet, for example.

\begin{figure}[h!]
\centering
\includegraphics[scale=0.45]{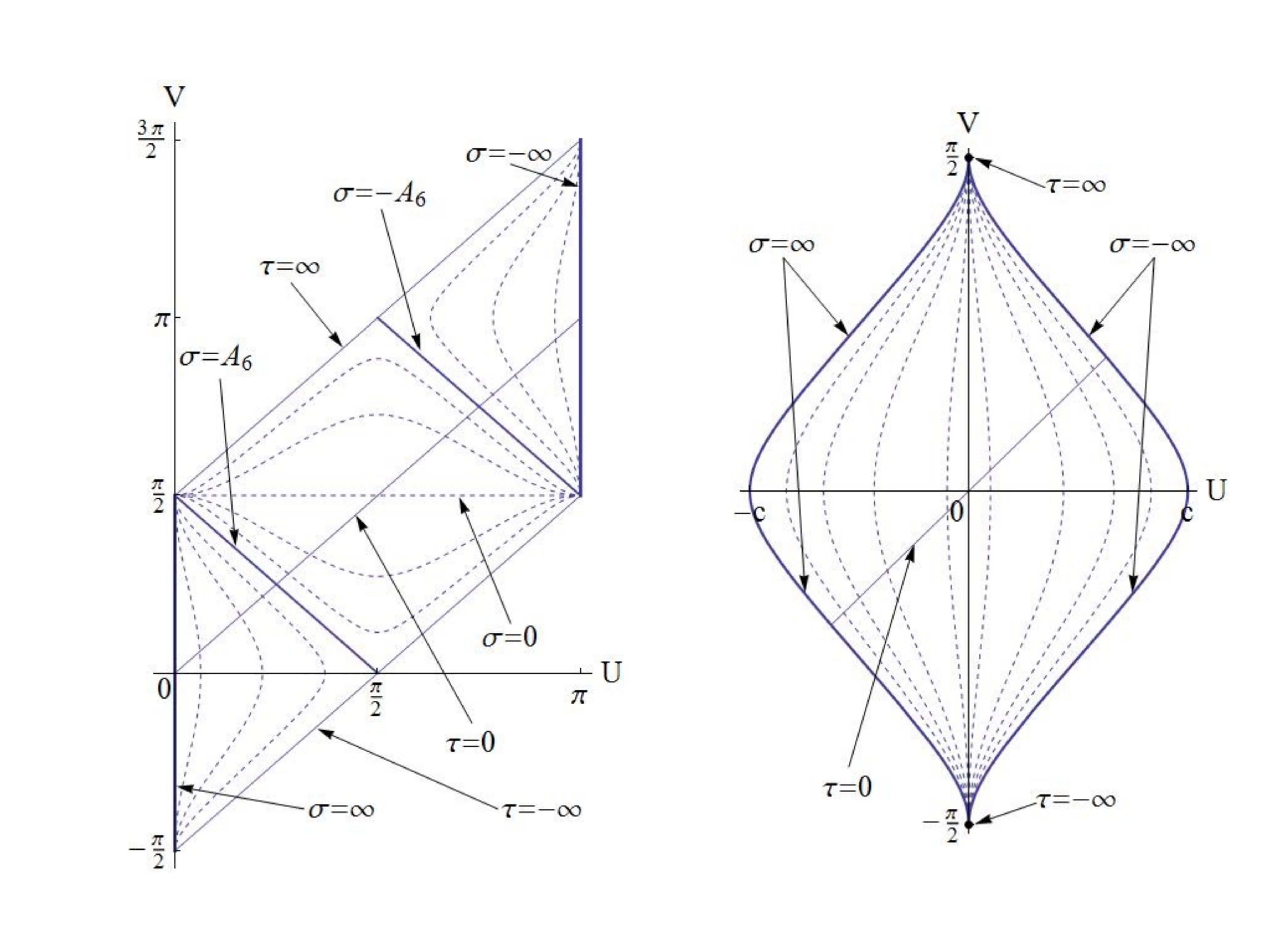}
\caption{\small The left diagram corresponds to $A_6^2>0$, with a subsequent causal structure which is divided in two regions by the $\sigma = A_6$ curve. The end points of the worldsheet, denoted by $\sigma = \pm \infty$ corresponds to two fundamental degrees of freedom with opposite charges. The right Penrose diagram corresponds to $A_6^2<0$, and the resulting causal structure does not possess any event horizon. Dashed lines on both diagrams correspond to curves with constant $\sigma$. This diagram is taken from \cite{Chernicoff:2010yv}, with the Authors' consent.}
\label{fig1}
\end{figure}

Given the above action in (\ref{ngact}), one can solve the classical equations of motion. Instead of choosing a gauge and solving for the equations of motion, let us review the general solution of \cite{Mikhailov:2003er} in the embedding space formalism. We begin with a description of the AdS$_5$ geometry as an hyperbolic submanifold in ${\mathbb R}^{2,4}$, described by
\begin{eqnarray}
Y^M Y_M = \eta_{MN} Y^M Y^N = - R^2 \ , \quad \eta_{MN} = {\rm diag} \left(-1,-1, 1, 1, 1, 1 \right) \ .  \label{embedAdS}
\end{eqnarray}
To describe a two-dimensional worldsheet, let us define a light-like vector $\ell^M(\tau)$ which obeys:
\begin{eqnarray}
\eta_{MN} \ell^M \ell^N = 0 \ , \quad \eta_{MN} \partial_\tau \ell^M \partial_\tau \ell^N = -R^2 \ .  \label{stringembed}
\end{eqnarray}
Using the relations above, it is easy to show that the general solution of (\ref{embedAdS}) is given by
\begin{eqnarray}
Y^M \left( \tau, \sigma \right) = \pm \partial_\tau \ell^M(\tau ) + \sigma \ell^M(\tau) \ . \label{mikhailov}
\end{eqnarray}
As demonstrated in \cite{Mikhailov:2003er}, (\ref{mikhailov}) is an extremal surface inside AdS$_5$. The overall sign in front of the first term in (\ref{mikhailov}) corresponds to a purely ingoing and outgoing non-linear wave solutions, with respect to the location of the source. The induced worldsheet metric is subsequently given by
\begin{eqnarray}
ds_{\rm ws}^2 = \left[\eta_{MN} \left( \partial_\tau^2 \ell^M \right)  \left( \partial_\tau^2 \ell^N  \right) + \sigma^2  \right] d\tau^2 - 2 d\tau d\sigma \ . 
\end{eqnarray}
Following the discussion in \cite{Chernicoff:2010yv}, we introduce the following:
\begin{eqnarray}
A_6^2 = - \eta_{MN} \left( \partial_\tau^2 \ell^M \right) \left( \partial_\tau^2 \ell^N  \right) \ , \quad d\tau = d\tilde{\tau} - \frac{d\sigma}{A_6^2 - \sigma^2} \ , 
\end{eqnarray}
where, $A_6$ represents the proper acceleration, defined on the embedding space ${\mathbb R}^{2,4}$. The resulting induced metric is given by
\begin{eqnarray}
ds_{\rm ws}^2 = - \left( \sigma^2 - A_6^2 \right) d\tilde{\tau}^2 + \frac{d\sigma^2}{\left( \sigma^2 - A_6^2 \right)} \ . \label{wsads2}
\end{eqnarray}
In general, $A_6$ is a $\tau$-dependent function. However, for cases when $A_6$ is a constants, the worldsheet describes an AdS$_2$-black hole. The corresponding causal structure is best described in terms of the Penrose diagram, which is shown in figure \ref{fig1}. 
The relevant change of coordinates are given in \cite{Chernicoff:2010yv}, which we also review below:
\begin{eqnarray}
V & = & \frac{1}{2} \left[\tan^{-1} \left( \tau + \frac{1}{A_6} \log\left( \frac{\sigma + A_6}{\sigma - A_6} \right) \right) + \tan^{-1}\tau \right]  \ , \label{V1} \\
U & = & \frac{1}{2} \left[\tan^{-1} \left( \tau + \frac{1}{A_6} \log\left( \frac{\sigma + A_6}{\sigma - A_6} \right) \right) - \tan^{-1}\tau \right]  \ , \label{U1}
\end{eqnarray}
for $\sigma > A_6$. For $\sigma < - A_6$, the corresponding patches are obtained by adding a $\pi$-shift to (\ref{V1}) and (\ref{U1}). For the region ${\rm Abs}(\sigma) < A_6$, the relevant coordinate changes are given by
\begin{eqnarray}
V & = & \frac{1}{2} \left[ \pi - \tan^{-1} \left( \tau + \frac{2}{A_6} \tanh^{-1} \left( \frac{\sigma}{A_6} \right) \right) + \tan^{-1} \tau \right]  \ , \label{V2} \\
U & = & \frac{1}{2} \left[ \pi - \tan^{-1} \left( \tau + \frac{2}{A_6} \tanh^{-1} \left( \frac{\sigma}{A_6} \right) \right) - \tan^{-1} \tau \right]  \ . \label{U2}
\end{eqnarray}
Note that, the worldsheet metric, in the $\{V, U\}$ coordinate takes the form $ds^2 \propto \left( dU^2 - dV^2 \right) $, where the proportionality factor is a function of the $U$ and $V$ coordinates. For the case when $A_6^2 < 0$, the following coordinate change\cite{Chernicoff:2010yv}:
\begin{eqnarray}
2 V & = & \tan^{-1} \left( \tau - \frac{2}{\sqrt{-A_6^2}} \tan^{-1} \left( \frac{\sigma}{\sqrt{-A_6^2}}  \right) + \tan^{-1}\tau  \right)   \ , \label{V3} \\
2 U & = & \tan^{-1} \left( \tau - \frac{2}{\sqrt{-A_6^2}} \tan^{-1} \left( \frac{\sigma}{\sqrt{-A_6^2}}  \right) - \tan^{-1}\tau  \right)   \ , \label{U3}
\end{eqnarray}
again brings the worldsheet metric back to the desired form. A detailed analysis of various trajectories are discussed in \cite{Chernicoff:2010yv}, which results in the Penrose diagram shown in figure \ref{fig1}. The crucial point to note is the presence of a causal horizon in the case $A_6^2 > 0$ (the left Penrose diagram in figure \ref{fig1}), which is absent for $A_6^2 < 0$ (the right Penrose diagram in figure \ref{fig1}). We will now discuss a few simple and explicit examples.

The string with a single end point at the UV, which is what is described above, can be written in a particularly recognizable form\cite{Chernicoff:2013iga}:
\begin{eqnarray}
&& t \left( t_r, z \right) = t_r \pm \frac{z}{\sqrt{1- \vec{v}(t_r)^2}} \ , \label{is1} \\
&& \vec{x} \left( t_r , z \right) = \vec{x}(t_r) \pm \frac{\vec{v}(t_r) z}{\sqrt{1- \vec{v}(t_r)^2}} \ , \label{is2}
\end{eqnarray}
where the background is written in the standard Poincar\'{e} patch of (\ref{back}). The positive sign corresponds to a retarded solution, in which energy flux propagates from the end point of the string to the Poincar\'{e} horizon and the negative sign corresponds to an advanced solution with a reverse energy flow. The time argument $t_r$ denotes retarded or advanced time, correspondingly. The above solution corresponds to an infinitely massive fundamental matter, for a non-vanishing but finite mass, the corresponding solutions are discussed in \cite{Chernicoff:2008sa, Chernicoff:2009re, Chernicoff:2009ff}.

\subsection{Asymptotically Uniform Acceleration}

We can describe the string profile in terms of the embedding space, however, in this case it is straightforward to adopt a Poincar\'{e} patch description of the profile. This can simply be done by using {\it e.g.}~Poincar\'{e} slicing of the embedding space and subsequently choosing an appropriate vector $\ell^M$. Evidently, various slicing of the embedding space yields various AdS-metrics, and correspondingly the manifold on which the CFT is defined. For an explicit exposition of such slices, see {\it e.g.}~\cite{Penedones:2016voo}.

Here we will discuss the particular case, in which an infinitely massive quark ({\it i.e.}~the string end point) undergoes an uniform acceleration. In the standard Poincar\'{e}-patch, this worldsheet embedding is given by\cite{Xiao:2008nr}
\begin{eqnarray}
x \left( t, z \right)^2 = a^{-2} + t^2 - z^2 \ . \label{uaws}
\end{eqnarray}
Clearly, we have chosen a static gauge $\sigma = z$, $\tau = t$ to describe the above solution of the Nambu-Goto action in (\ref{ngact}). The embedding function $x(t,z)$ is simply one of the Minkowski-direction coordinates of the CFT. There is a clear sign choice involved in the function $x(t,z)$, this corresponds to the overall charge of the string end point. For simplicity, we can easily think of them as quark and anti-quark degrees of freedom.

The causal structure induced from the embedding in (\ref{uaws}) is similar to a black hole in AdS-background, in which the event horizon is located at $z_{\rm h} = a^{-1}$. Moreover, the embedding in (\ref{uaws}) can be constructed by patching two sections of (\ref{is1}) and (\ref{is2}). Physically, because of the uniform acceleration, a given retarded string profile terminates at the worldsheet event horizon: $z_{\rm h} = a^{-1}$, where the local speed of propagation exceeds the speed of light. Therefore the rest of the profile needs to be completed accordingly. This can be done by smoothly patching the retarded solution with an advanced solution. This class of embeddings, along with various generalizations, were discussed in details in \cite{Chernicoff:2013iga, Jensen:2013ora}.

\begin{figure}[h!]
\centering
{\includegraphics[width = 3.3in]{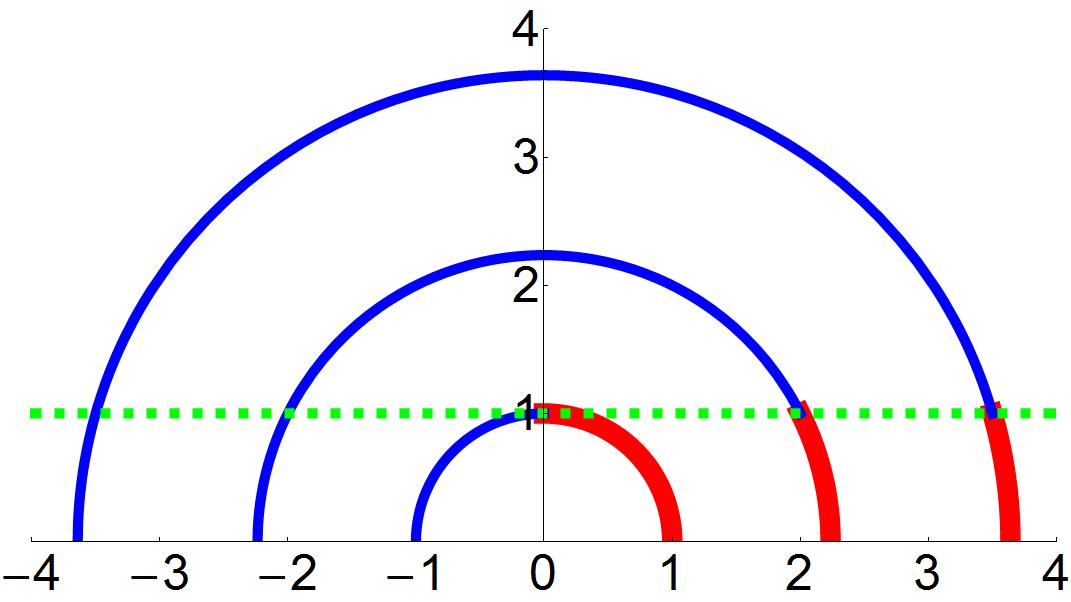}}
{\includegraphics[width = 3.3in]{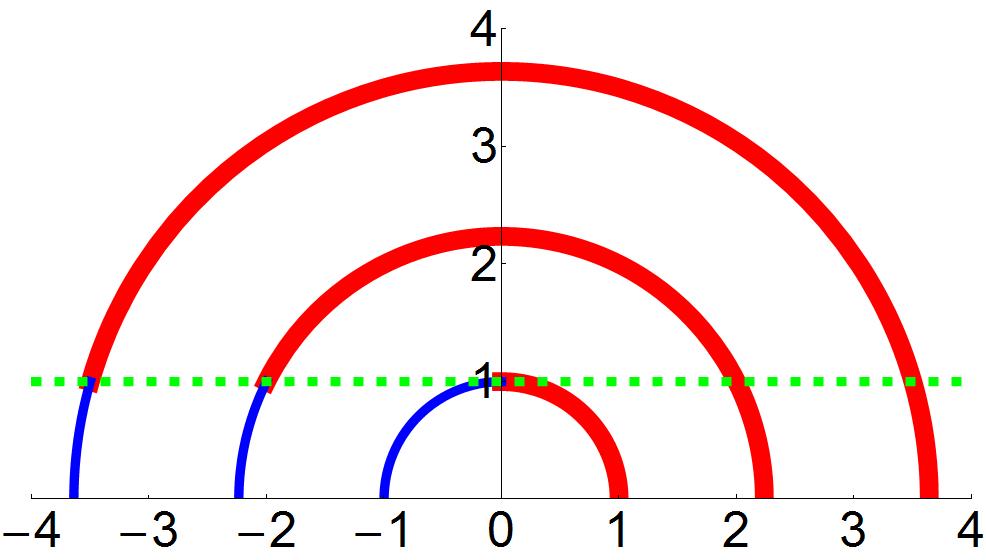}}
\caption{\small We have shown here various string profiles in $\{z, x\}$-plane. The vertical axis is the $z$-axis and the horizontal axis is the $x$-axis. The left diagram corresponds to $t<0$ and the semi-circular profiles shrink as time increases towards positive values. On the right, we have $t>0$ and the semi-circular profiles grow as time increases. In both, the advanced solution is denoted by the blue, thin curve and the retarded solution is represented by the red, thick curve. The green horizontal dashed line corresponds to the worldsheet event horizon, located at $z_{\rm h} = a^{-1}$. We have taken these diagrams from \cite{Chernicoff:2013iga}, with the Authors' consent.}
\label{uastring}
\end{figure}
Patching a retarded solution and an advanced solution has been discussed extensively in \cite{Chernicoff:2013iga}. The advanced and the retarded configurations cover non-symmetric regions of the configuration space. This construction is perhaps best represented in terms of the figures in \cite{Chernicoff:2013iga}, which we include in figure \ref{uastring}, with the Authors' consent.

\subsection{Uniform Circular Trajectory}

A simple but instructive example is to consider the string end point moving in a circle of constant radius. Evidently, the end point does undergo an acceleration. Let us review this based on \cite{Chernicoff:2010yv}. Let us begin with the Poincar\'{e} slicing of the embedding in (\ref{embedAdS}), given by
\begin{eqnarray}
Y^0 = \frac{R}{2z} \left( 1 + x^2 + z^2 \right) \ , \quad Y^A = \frac{R}{z} x^A \ , \quad Y^1 = \frac{R}{2z} \left( 1 - x^2 - z^2 \right) \ .
\end{eqnarray}
The induced AdS is given by the Poincar\'{e} disc and the dual field theory is defined on an ${\mathbb R}^{1,3}$. To describe a string embedding, further, we need to specify the vector $\ell^M$, which is given by
\begin{eqnarray}
&& \ell^{0} =\frac{1}{2} \left[ 1 - \left( x(\tau) \right)^2\right]   \ , \quad \ell^{1} = \frac{1}{2} \left[ 1 - \left( x(\tau) \right)^2\right] \ , \\
 && \ell^\mu = x^\mu(\tau) \ ,
\end{eqnarray}
where the $x^2$ is constructed by taking an inner product with a Minkowski metric $\eta_{\mu\nu}$ of the vector $x^\mu \in {\mathbb R}^{1,3}$. To satisfy (\ref{stringembed}), we further impose: $\left( \partial_\tau x^\mu \right)  \left( \partial_\tau x_\mu\right)  = 1$. The parameter $\tau$ is physically identified with the proper time associated to the string end point. Fixing $\sigma^{-1} = z$, the embedding in (\ref{mikhailov}) is given by
\begin{eqnarray}
x\left( \tau, z \right) = z \left( \partial_\tau x^\mu (\tau) \right) + x^\mu (\tau) \ .
\end{eqnarray}
Let us now describe the particular case, first discussed in \cite{Athanasiou:2010pv}. In this case, the string end point is moving with a constant angular velocity $\omega$, in a circle of radius $r_0$. The solution can be described by a collection of the functions $x^\mu(\tau)$. The non-trivial components are given by
\begin{eqnarray}
x^1 = r_0 \cos \left(\omega t_{\rm ret} (\tau)\right)  \ , \quad x^2 = r_0 \sin \left(\omega t_{\rm ret} (\tau)\right)  \ , 
\end{eqnarray}
while (\ref{stringembed}) yields:
\begin{eqnarray}
\partial_\tau t_{\rm ret} = \frac{1}{\sqrt{1-\omega^2 r_0^2}} = \gamma = \frac{1}{\sqrt{1-v^2}} \ , \quad \implies \quad t_{\rm ret} = \gamma \tau \ .
\end{eqnarray}
Thus, the full solution can be represented by
\begin{eqnarray}
&& x^0 = z\gamma + t_{\rm ret} \ , \\
&& x^1 = r_0 \cos \left(\omega t_{\rm ret} (\tau)\right)  \ , \quad x^2 = r_0 \sin \left(\omega t_{\rm ret} (\tau)\right)  \ , 
\end{eqnarray}
The induced geometry on the worldsheet inherits an event horizon at $z_{\rm h} = 1/ (\omega^2 \gamma v)$. Propagating modes across this radial scale become causally disconnected and therefore yield an effective temperature:
\begin{eqnarray}
T_{\rm ws} = \frac{v \gamma^2 \omega}{2\pi } \ .
\end{eqnarray}
On these simple examples, the general picture is already emerging and clear. We will not further explicitly discuss more possibilities, specially the ones that appear in the global patch of AdS. However, let us briefly summarize the physics.

In the global AdS patch, the boundary theory is kept at a finite volume and likewise develops a mass gap. Thus, a rotating string will not develop an event horizon on the world sheet for any value of the frequency, unlike the Poincar\'{e}-slice result. In this case, for sufficiently large angular frequency $\omega L_{\rm typical} > 1$, where $L_{\rm typical}$ is the typical length-scale associated with the finite volume, the worldsheet develops an event horizon and measures an effective thermal description on the worldsheet. For $\omega L_{\rm typical} < 1$, there is no event horizon and thus no effective thermal description holds. As pointed out in \cite{Chernicoff:2010yv}, this is similar to the behaviour of an Unruh-DeWitt detector\cite{Davies:1996ks}, undergoing a circular motion in a compact space. The corresponding Hawking radiation of the string worldsheet was analyzed in terms of a synchrotron radiation in the dual gauge theory in \cite{Athanasiou:2010pv}. There certainly are numerous interesting such string configurations which demonstrate very interesting physics, driven by the worldsheet event horizon. We will not elaborate more on examples, instead we will briefly discuss some recent advances in understanding the same physics from a slightly different perspective, in the next section.

\subsection{Chaos: A Recent Development}

A thermal effective description ensures that the string end point will undergo a stochastic motion, because of the thermal fluctuations. This results in a Brownian motion for the string end point. In \cite{deBoer:2008gu}, this effect was explored in details by considering fluctuations of the string around a classical saddle. The fluctuation degrees of freedom, effectively, propagate in a curved geometry with an event horizon and therefore exhibits random fluctuations due to the Hawking-Unruh radiation. In \cite{Son:2009vu} a Schwinger-Keldysh description of the stochastic was discussed, which is based on a Kruskal-type extension of the worldsheet geometry. Such Brownian motions are expected to be dissipated in the medium, since the system is thermalized.

Given a thermal system, how fast a small perturbation relaxes to the thermal value sets the thermalization time scale for such small excitations. This is closely related to the scrambling time for the system, which determines the speed at which a quantum system spreads a localized information. In \cite{Sekino:2008he}, it was conjectured that black holes are the fastest scramblers in Nature, for which the scrambling time scales as the logarithm of the number of degrees of freedom. In the context of Holography, therefore, a strongly coupled system in thermal equilibrium will also satisfy this bound.

The notion of ergodicity and thermalization are intimately related. In a semi-classical description, the physics of thermalization is further related to a particular notion of growth of $n$-point correlation functions, where $n \ge 4$. Also, the correlator must contain operators which are not time-ordered. Conventionally, these are known as the out-of-time order correlator, or OTOC in brief. The basic idea here is rather simple, which we quickly review below. 

\begin{figure}[h!]
\centering
\includegraphics[scale=0.5]{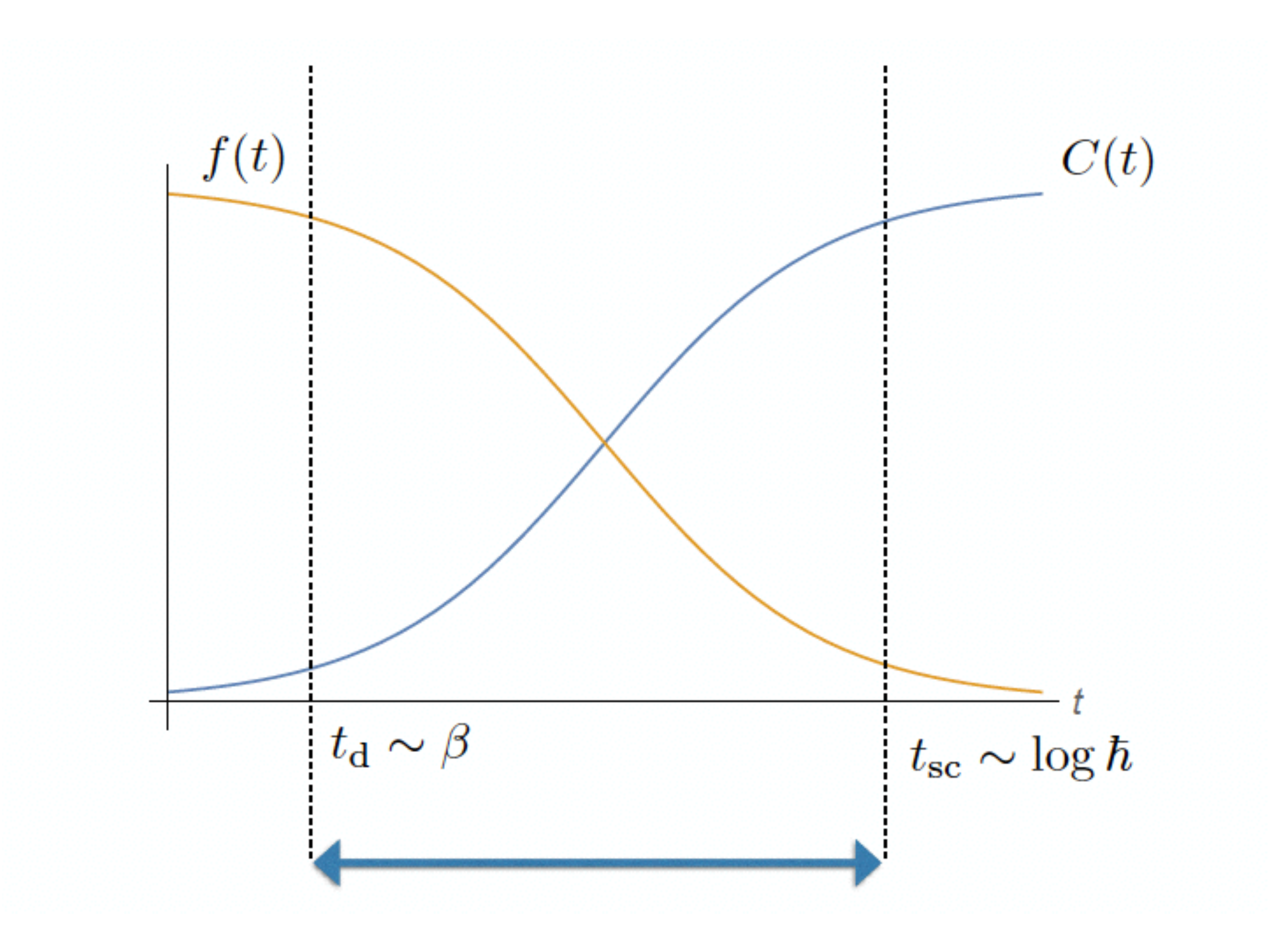}
\caption{\small A schematic representation of the chaos diagnostic function(s). Here, qualitatively, $f(t) = 1 - C(t)$, while both $f(t)$ and $C(t)$ individually carries the information of an exponential decay (for $f(t)$) or growth (for $C(t)$) of the corresponding correlator. The exponential behaviour occurs in a regime between a $\cO(1)$ time-scale, such as the dissipation time $t_d$, and a parametrically large time-scale, such as the scrambling time $t_{\rm sc}$. This hierarchy of $t_d$ and $t_{\rm sc}$ is ensured in the $\hbar \to 0$ limit.}
\label{lyasc}
\end{figure}
Given a classical system, a phase space description is provided in terms of canonical coordinates and momenta variables: $q(t)$ and $p(t)$, respectively. The notion of classical chaos is defined as a response of a classical trajectory at late times, in terms of the variation of it's initial value. Typically, a chaotic system is characterized by exponentially diverging trajectories at sufficiently late times:
\begin{eqnarray}
\left\{q(t), p \right\} \equiv \frac{\partial q(t)}{\partial q(0)} \sim {\rm exp} \left( \lambda_{\rm L} t \right) \ , 
\end{eqnarray}
where the real number $\lambda_{\rm L}$ is known as the Lyapunov exponent. The left most expression is the standard classical Poisson bracket. Within a semi-classical framework, the diagnostic of chaos, along with the notion of Lyapunov exponents, can be easily generalized for a quantum system, by replacing the Poisson bracket with a commutator, and finally, computing the square of this commutator to make sure that no accidental phase cancellation takes place.\footnote{Note that, it is expected that for higher powers of the commutator, one would observe a growing OTOC. For SYK-type models, this growth can be explicitly calculated in the form of a six-point OTOC, see {\it e.g.}~\cite{Bhattacharya:2018nrw}.} Thus, a chaos diagnostic can be defined in terms of the following correlator:
\begin{eqnarray}
C(t) = - \left \langle\left[ W(t), V(0)\right]^2  \right \rangle \ ,
\end{eqnarray}
where $W(t)$ and $V(0)$ are generic Hermitian operators. The function, $C(t)$ possess both time-ordered and out-of-time ordered correlator; but the exponential growth is visible in the OTOC sector only. In terms of this diagnostic function $C(t)$, the scrambling time is also defined as the time scale when $C(t) \sim \cO(1)$. The basic behaviour of the diagnostic function is pictorially demonstrated in figure \ref{lyasc}. The basic idea here is simple: Given a thermal state, one computes an OTOC based on a Schwinger-Keldysh construction. Now, in the limit $t \gg t_d$, where $t_d$ is the dissipation time-scale, one observes a growth in the correlator. In case this is an exponential growth, it is simple to extract the corresponding Lyapunov exponent. This particular notion of the growth of an operator is intrinsic to a semi-classical description.

It is not simple to calculate higher point OTOCs, in general and there are currently a handful of tractable examples exist. Nevertheless, in \cite{Maldacena:2015waa}, a bound for the Lyapunov exponent was derived $\lambda_{\rm L} \le 2\pi T$, for a system with temperature $T$, and in natural units. It was further conjectured that the maximal chaos limit is saturated for holographic theories. For a black hole in AdS, this saturation is simply guaranteed by the near horizon dynamics, where a local Rindler description holds. This intuition is based on recasting the four point OTOC in terms of a two-two high energy scattering amplitude in the thermofield double picture, see {\it e.g.}~\cite{Shenker:2013pqa, Shenker:2013yza, Shenker:2014cwa}.

Given the black hole like causal structure on the string worldsheet, it is therefore expected that such a saturation will hold on the worldsheet horizon as well. Indeed, it was explicitly shown in \cite{Murata:2017rbp, deBoer:2017xdk} that the maximal saturation occurs on the string worldsheet, with the effective temperature. Moreover, a corresponding soft sector effective action, which is a Schwarzian derivative action,  was explicitly obtained in \cite{Banerjee:2018twd, Banerjee:2018kwy}, and its coupling with various heavy modes were explicitly determined. The soft sector action can be simply obtained by embedding the string worldsheet in an AdS$_3$-background and using the Brown-Henneaux large diffeomorphisms, projected on the worldsheet. This yields:
\begin{eqnarray}
S_{\rm effective} = \frac{\epsilon_{\rm IR}}{\alpha'} \int dx \left\{ \varphi(x) , x\right\} \ , \quad \left\{ \varphi(x) , x\right\} \equiv \frac{3}{2} \frac{\varphi''^2}{\varphi'^2} - \frac{\varphi' \varphi'''}{\varphi'^2} \ , \label{schNG}
\end{eqnarray}
where $\alpha'$ is the inverse string tension, $\epsilon_{\rm IR}$ is a physical scale, typically given by the event horizon on the induced worldsheet, and $\varphi(x)$ is the dynamical degree of freedom. This effective description has a natural interpretation as an Euclidean theory, however, because of the high derivative coupling, the Lorentzian description has pathologies. 
\begin{figure}[h!]
\centering
\includegraphics[scale=0.45]{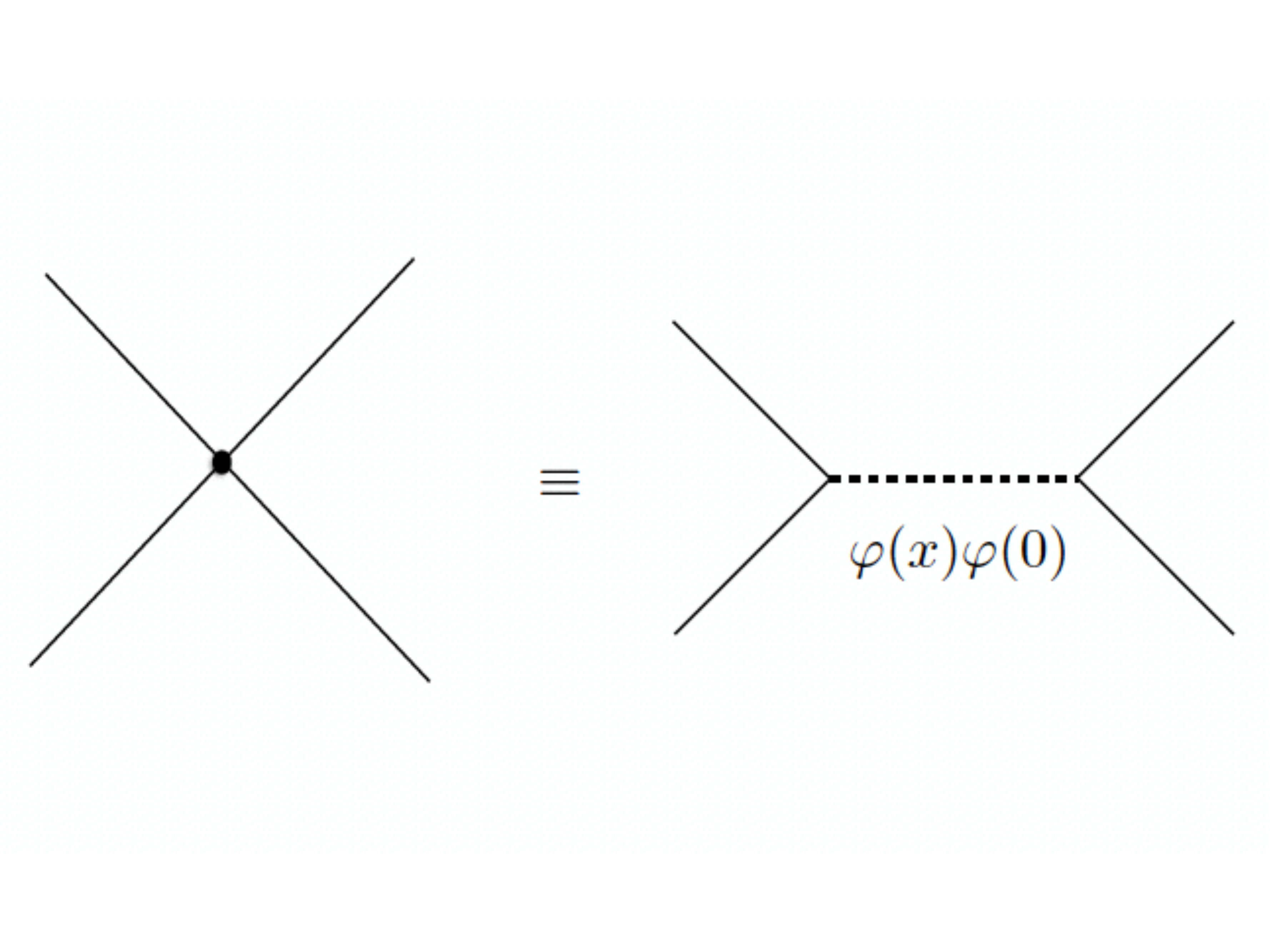}
\caption{\small The four point effective vertex contains two interaction vertices with the soft sector, along with a propagator of the same soft sector. The dashed horizontal line is the soft sector propagator, and the solid lines are external hard modes.}
\label{schint}
\end{figure}

The soft sector physics, described by (\ref{schNG}), is ultimately responsible for the maximal chaos on the worldsheet. Qualitatively, this is straightforward to understand. In the two-two elastic scattering process, the four point vertex of any semi-classical fluctuation of the string worldsheet can be resolved in terms of two interaction vertices of the fluctuation with the soft sector mode, and a propagator of the soft sector itself. This is, pictorially demonstrated in figure~\ref{schint}. This soft sector dynamics on the string worldsheet is in close resemblance with the soft sector dynamics of AdS$_2$ in the Jackiw-Teitelboim gravity\cite{Maldacena:2016upp}. Finally, by setting $C(t) \sim \cO(1)$, we can determine the scrambling time. Note, further, that a direct four point OTOC correlator yields a scrambling time $t_{\rm sc} \sim \beta \log \sqrt{\lambda}$, where $\lambda$ is the corresponding 't Hooft coupling\cite{deBoer:2017xdk}. The same result is obtained by analyzing the soft sector physics, and its' coupling with heavier modes in the string fluctuations\cite{Banerjee:2018twd, Banerjee:2018kwy}. It can now be anticipated that on a D-brane horizon, one would observe a similar physics, which is demonstrated explicitly in \cite{Banerjee:2018twd, Banerjee:2018kwy}, in which the corresponding scrambling time depends both on the rank of the gauge group, $N_c$, and the 't Hooft coupling.

We will end this brief section here, leaving untouched a remarkable amount of recent research in related topics. Chaotic properties of strongly coupled CFTs, with or without a holographic dual is a highly evolving field of research. Moreover, the notion of chaos in quantum mechanical systems is a rich field in itself and only some explicit calculations using a particular semi-classical prescription have been made. Thus, we will leave a detailed discussion for future, and for now, shift our attention to the effective thermal description of a D-brane.

\setcounter{equation}{0}
\section{The Brane Worldvolume Description: A Model} \label{n4sym}

Let us consider a stack of $N_c$ D3-branes sitting at the tip of a cone, with a Sasaki-Einstein $5$-manifold base, henceforth denoted by SE$_5$. When the SE$_5 \equiv S^5$, the $10$-dimensional geometry is given by AdS$_5$-Schwarzschild$\times S^5$ which is dual to the $\cN=4$ super Yang-Mills (SYM) theory with an $SU(N_c)$ gauge group. The corresponding gravity data are given by\footnote{We are using the notation used in \cite{Albash:2007bq}.}
\begin{eqnarray} \label{metric1}
&& ds^2 = \frac{1}{4 r^2 R^2 } \left( - \frac{f^2}{\tilde{f}} dt^2 + \tilde{f} d\vec{x}^2 \right) + \frac{R^2}{r^2} dr^2 + R^2 d\Omega_5^2 \ , \\
&& f = 4 r^4 - b^4 \ , \quad \tilde{f} = 4 r^4 + b^4 \ , \\
&& F_{(5)} = \left(1 + \star \right) {\rm vol}_{S^5} = d C_{(4)} +  d \tilde{C}_{(4)} \ ,
\end{eqnarray}
where $\{t, \vec{x}\}$ are the field theory space-time directions, $r \in [b, \infty)$	 is the AdS-radial direction, $R$ is the curvature of AdS and $d\Omega_5^2$ is the metric on a unit $5$-sphere. These branes source a self-dual $F_{(5)}$. Here $\star$ denotes the Hodge dual operator. The unit sphere metric can be written as
\begin{eqnarray}
&& d\Omega_5^2 = d\theta^2 + \cos^2 \theta d\Omega_3^2 + \sin^2\theta d\phi^2 \ , \label{s3collapse} \\
&& d\Omega_3^2 = d\psi^2 + \cos^2 \psi d\beta^2 + \sin^2\psi d\gamma^2 \ . \label{s3met}
\end{eqnarray}
The corresponding black hole temperature is given by
\begin{eqnarray}
T = \frac{b}{\pi R^2} \ .
\end{eqnarray}
The AdS curvature sets the 't Hooft coupling for the dual field theory {\it via} $R^4 = \alpha'^2 g_{\rm YM}^2 N_c$, where $\alpha'$ is the string tension and $g_{\rm YM}$ is the gauge theory coupling.

The matter content of $\cN=4$ SYM is: the gauge field $A_\mu$, four adjoint fermions $\lambda$ and three complex scalars $\Phi^a$ ($a=1,2,3$). This theory has an $SU(4) \sim SO(6)$ R-symmetry, which corresponds to the rotational symmetry of the $S^5$ in the dual gravitational description. To introduce fundamental matter, one introduces open string degrees of freedom which is equivalent to introducing additional probe branes, along the lines of \cite{Karch:2002sh}. As mentioned earlier, of particular interest is to add $N_f$ probe D$7$-branes, in the limit $N_f \ll N_c$ to to suppress backreaction. These probe branes are extended along $\{t, \vec{x}\}$ and wraps the $S^3 \subset S^5$. The codimension $2$ brane is described by $\{\theta(r), \phi(r) \}$. Isometry along the $\phi$-direction implies $\phi(r)=0$, without any loss of generality. Thus, the embedding is described by a single function $\theta(r)$.

In terms of the dual gauge theory side\footnote{In this part, we closely follow the discussions in \cite{Myers:2007we}.}, we introduce an $\cN=2$ hypermultiplets in the background of $\cN=4$ SYM. The hypermultiplets consists of two Weyl fermions, denoted by $\psi$ and $\bar{\psi}$ and two complex scalars, denoted by $q$ and $\bar{q}$. Here, $\{\psi, q\}$ transforms under the fundamental representation of SU$(N_c)$ and $\{\bar{\psi}, \bar{q}\}$ transforms under the anti-fundamental. The dual operators corresponding to the D$7$-brane profile functions $\theta$ and $\phi$ are given by
\begin{eqnarray}
&& \cos\theta \leftrightarrow \cO_m = i \bar{\psi} \psi + \bar{q} \left( m_q + \sqrt{2} \Phi^1\right) \bar{q}^{\dagger} + q^{\dagger} \left( m_q + \sqrt{2} \Phi^1\right) q + {\rm h.c.} \ , \\
&& \phi \leftrightarrow \cO_\phi = \bar{\psi} \psi + i \sqrt{2} \bar{q} \, \Phi^1 \bar{q}^{\dagger} + i \sqrt{2} q^{\dagger} \Phi^1 q + {\rm h.c.} \ ,
\end{eqnarray}
where $\Phi^1$ is a complex scalar field in the $\cN=4$ supermultiplet and $m_q$ is the mass of the fundamental quark.

The Lagrangian of the worldvolume theory can be written in the $\cN=1$ language, as follows:
\begin{eqnarray}
\cL & = & {\rm Im} \left[ \tau \int d^4\theta \left( {\rm tr} \left( \bar{\Phi}_I e^V \Phi_I e^{-V} \right) + Q_r^{\dagger} e^V Q^r  + \tilde{Q}_r^\dagger e^{-V} \tilde{Q}^r \right) \right. \nonumber\\  
 &+& \left. \tau \int d^2\theta \left( {\rm tr} \left(W^\alpha W_\alpha \right) + W \right) + {\rm c.c.} \right] \ ,
\end{eqnarray}
where
\begin{eqnarray}
W = {\rm tr} \left(\epsilon_{IJK} \Phi_i \Phi_j \Phi_K \right) + \tilde{Q}_r \left(m + \Phi_3 \right) Q^r \ .
\end{eqnarray}
Here $W_\alpha$ denotes the vector multiplet, $\Phi_I$, with $I=1,2,3$ denotes the chiral superfields. Both of these are obtained from the $\cN=4$ vector multiplet. On the other hand, $Q^r$, $\tilde{Q}_r$, where $r = 1, \ldots, N_f$ denotes the $\cN=2$ matter sector. Further details are summarized in table~\ref{table1}, where the $SO(4) \equiv SU(2)_\Phi \times SU(2)_R$ symmetry corresponds to the $S^3$ isometry, wrapped by the D$7$-brane. The transverse SO$(2)$ symmetry can be explicitly broken by giving a mass to the hypers: $m_q \not = 0$, which is proportional to the separation of the D$3$-branes and the D$7$-branes. 
\begin{table}[ht] 
\centering  
\begin{tabular}{| c | c | c | c | c | c | c |} 
\hline
fields & components & spin & $SU(2)_\Phi \times SU(2)_R$ & $U(1)_R$ & $\Delta$ & $U(N_f)$ \\
\hline
$\Phi_1, \Phi_2$ & $X^4, X^5, X^6, X^7$ & $0$ & $\left(\frac{1}{2}, \frac{1}{2}\right)$ &  $0$ & $1$ & $1$  \\
 & $\lambda_1$, $\lambda_2$ & $\frac{1}{2}$ & $\left(\frac{1}{2}, 0 \right)$ & $-1$ & $\frac{3}{2}$ & $1$ \\
\hline
$\Phi_3$, $W_\alpha$ & $X_V^A = \left(X^8, X^9\right)$ & $0$ & $\left(0, 0 \right)$ & $+2$ & $1$ & $1$ \\
                 & $\lambda_3$, $\lambda_4$ & $\frac{1}{2}$ & $\left(0, \frac{1}{2} \right)$ & $+1$ & $1$ & $1$ \\
                 &  $v_\mu$ & $1$ & $\left(0, 0 \right)$ & $0$ & $1$ & $1$ \\
\hline
$Q$, $\tilde{Q}$ & $q^m = \left(q, \bar{q}\right)$ & $0$ & $\left(0, \frac{1}{2} \right)$  & $0$ & $1$ & $N_f$ \\
                          & $\psi_i = \left(\psi, \psi^\dagger\right)$  & $\frac{1}{2}$ &  $\left(0, 0 \right)$ & $\mp 1$ & $\frac{3}{2}$ & $N_f$ \\
\hline	                                         
\end{tabular} \caption{Degrees of freedom in the dual field theory. }  \label{table1}
\end{table} 

The effective action for the probe D7-brane is given by the Dirac-Born-Infeld (DBI) Lagrangian with an Wess-Zumino term\footnote{Here we are using the Lorentzian signature.}
\begin{eqnarray}
S_{\rm D7} & = & - N_f T_{{\rm D}7} \int d^8 \xi \, \sqrt{ - {\rm det} \left(P[G_{ab} + B_{ab}]  + 2\pi\alpha' f_{ab}\right) } \nonumber\\
& + & \left(2 \pi \alpha' \right)^2 \frac{\mu_7}{2} \int f_{(2)} \wedge f_{(2)} \wedge \left( P\left[C_{(4)} \right] + P \left[\tilde{C}_{(4)}\right]  \right)\ . \label{dbi}
\end{eqnarray}
Here $P[G_{ab} + B_{ab}]$ denotes the pull-back of the NS-NS sector fields: $G$ denotes the closed-string background metric and $B$ denotes the NS-NS field; $f_{ab}$ is the worldvolume $U(1)$ gauge field on the probe. The four-form potentials $C_{(4)}$ and $\tilde{C}_{(4)}$ yield the self-dual five-form. Collectively, $\{\xi\}$ denotes D$7$-brane worldvolume coordinates, $T_{\rm D7} = \mu_7/g_s$ denotes the D$7$-brane tension. Here $g_s$ is the string coupling constant.\footnote{Note that, even though we take $N_f \not =1$, we work with an Abelian version of the effective action. This is clearly a truncation of the full non-Abelian action to an Abelian sector. We will assume such a truncation holds true. }

We intend to design a system in which energy is constantly pumped into the probe sector. This can be achieved by exciting the gauge field\footnote{We absorb the factor of $(2\pi\alpha')$ in the field strength.}
\begin{eqnarray} \label{gaugeansatz}
A = a_1 (r) dx^1  = \left( - E t + a(r)  \right) dx^1 \ ,
\end{eqnarray}
This ansatz (\ref{gaugeansatz}), which is consistent with the equations of motion, manifestly breaks the O$(3) \to$O$(2)$ in the $\vec{x}$-plane. The constant electric field $E$ is applied along the $x^1$-direction, to which only the probe sector couples, in the probe limit. The function $a(r)$ is governed by the equation of motion obtained from (\ref{dbi}), using the ansatz above.  The resulting effective action for the probe D$7$-brane is a functional of two functions $\theta(r)$ and $a(r)$. The physical meaning of these functions can be understood from the corresponding asymptotic behaviour:
\begin{eqnarray}
&& \lim_{r \to \infty} a (r) = 0 + \frac{J}{2 r^2} + \ldots \ , \\
&&  \lim_{r \to \infty} \theta (r) = \frac{m}{r} + \frac{c}{r^3} + \ldots \ .
\end{eqnarray}
In the dual gauge theory, the corresponding source and vevs are given by (in units $b=1$)
\begin{eqnarray}
&& \langle J^1 \rangle =  - (4 \pi^3 \alpha') N_f  V_{\mathbb R^3}  T_{\rm D7} J \ , \label{bcurrent} \\
&& m_q = \frac{m}{2\pi\alpha'} \ , \quad \langle \bar{\psi} \psi \rangle  = - 8 \pi^3 \alpha' V_{\mathbb R^3} N_f T_{\rm D7}  c \ . \label{bmc}
\end{eqnarray}
Here $ \langle J^1 \rangle$ is the expectation value of the current, $m_q$ is the mass of the hypermultiplet and $\langle \bar{\psi} \psi \rangle$. In the limit of vanishing electric field, thermal physics drives a first order phase transition in this system, see {\it e.g.}~\cite{Mateos:2006nu, Albash:2006ew}, as $(m_q/T)$ is tuned. This transition separates two phases: one containing bound mesonic degrees of freedom and a plasma phase, consisting of the $\cN =4$ adjoint and $\cN=2$ hypermultiplet degrees of freedom. A similar physics exists in the global patch of AdS as well\cite{Karch:2006bv}.

In the presence of $E$ but not background event horizon, {\it i.e.}~$b=0$, the boundary current in (\ref{bcurrent}) by demanding a reality condition of the on-shell effective action, much like \cite{Gubser:2006bz}. This exercise yields\cite{Karch:2007pd}:
\begin{eqnarray} \label{ohm1}
J^2 = \frac{\left( 4 r_*^4 -1 \right)^2 \left( 4 r_*^4 + 1 \right)^3 \cos^6\theta\left(r_*\right)}{64 r_*^6 \left( 4 r_*^4 + 1 \right)^2} \ , \quad r_*^2 = \frac{E+ \sqrt{E^2 + 1 }}{2} \ .
\end{eqnarray}
The gauge theory current expectation value vanishes when $E=0$, which is expected; it also vanishes when $\theta(r_*) =0$. The latter corresponds to a shrinking of the $S^3$ (which is wrapped by the D$7$-brane) before the brane can reach the radius $r_*$. For all our purposes, this radius $r_*$ acts as an event horizon on the D-brane worldvolume, thereby defining a causal structure similar to a black hole in AdS. We will make this connection more precise later.

\begin{figure}[h!]
\centering
{\includegraphics[width = 1.5in]{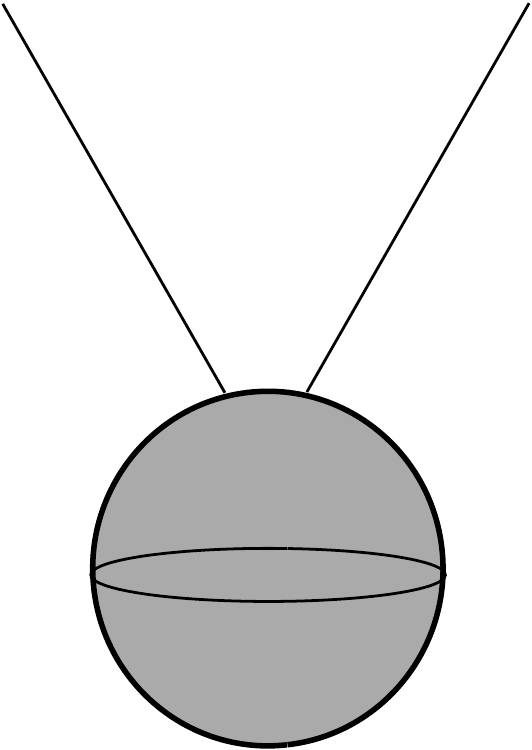}}
{\includegraphics[width = 1.5in]{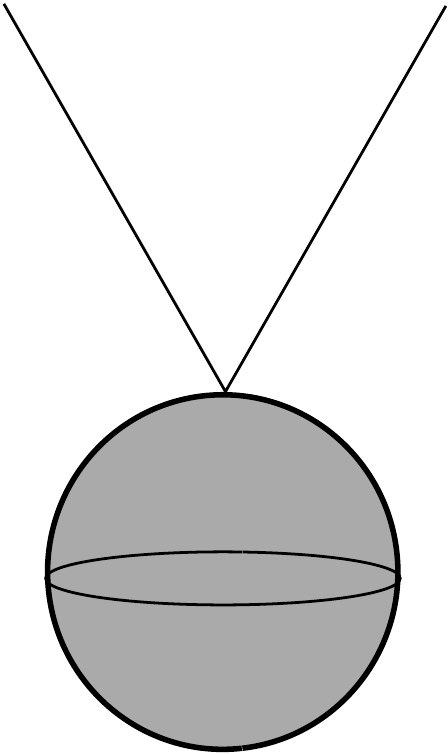}}
{\includegraphics[width = 1.5in]{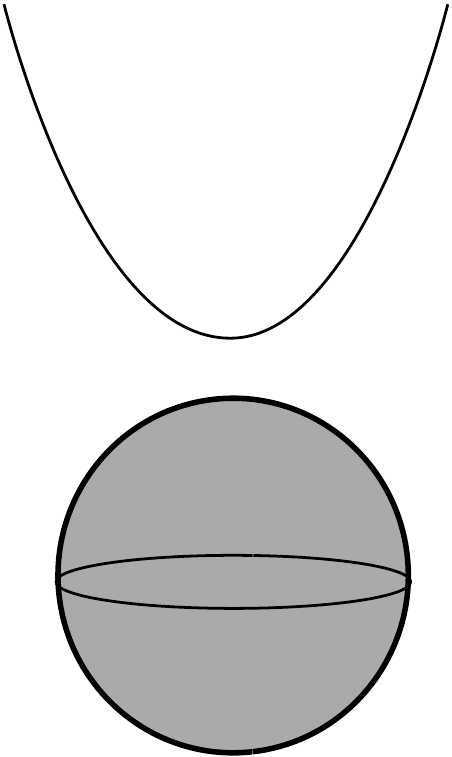}}
\caption{\small In the vanishing electric field limit, the D-brane (denoted by the curves above) can stay above the black hole (right-most), fall into the black hole (left-most). These two phases are separated by a critical embedding, corresponding to the middle one.}
\label{figbh0}
\end{figure}

\begin{figure}[h!]
\centering
{\includegraphics[width = 1.5in]{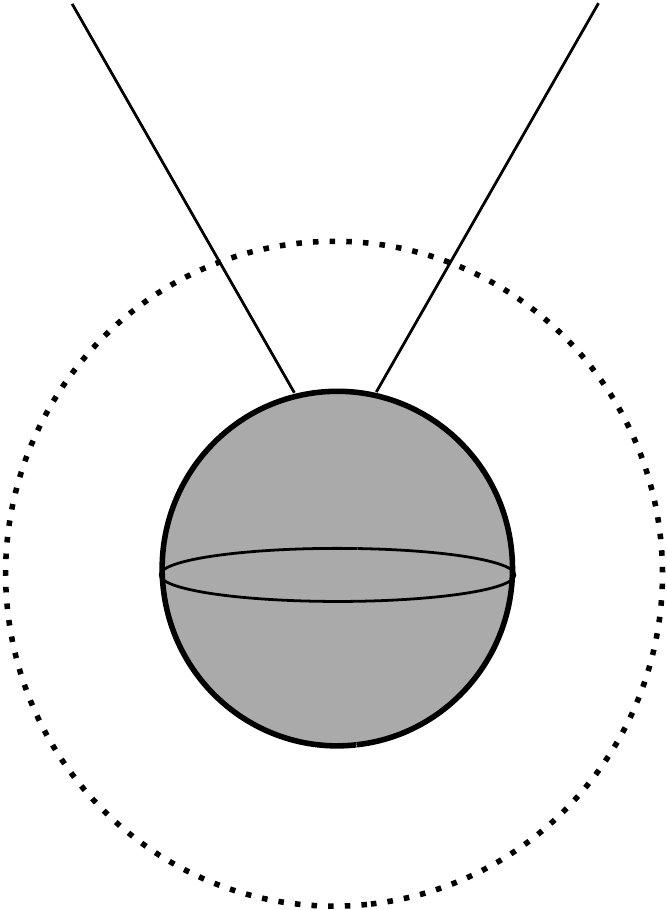}}
{\includegraphics[width = 1.5in]{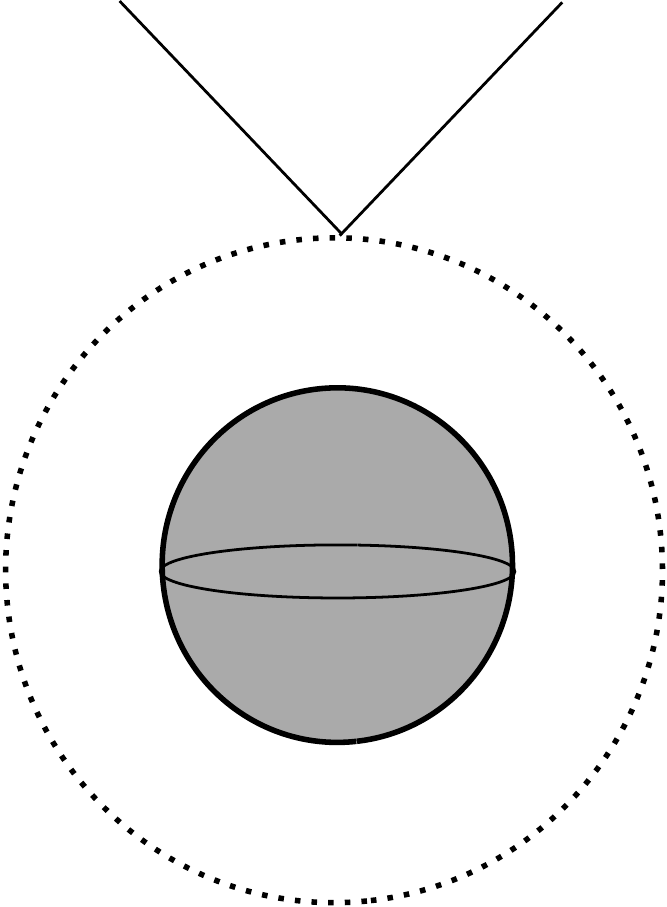}}
{\includegraphics[width = 1.5in]{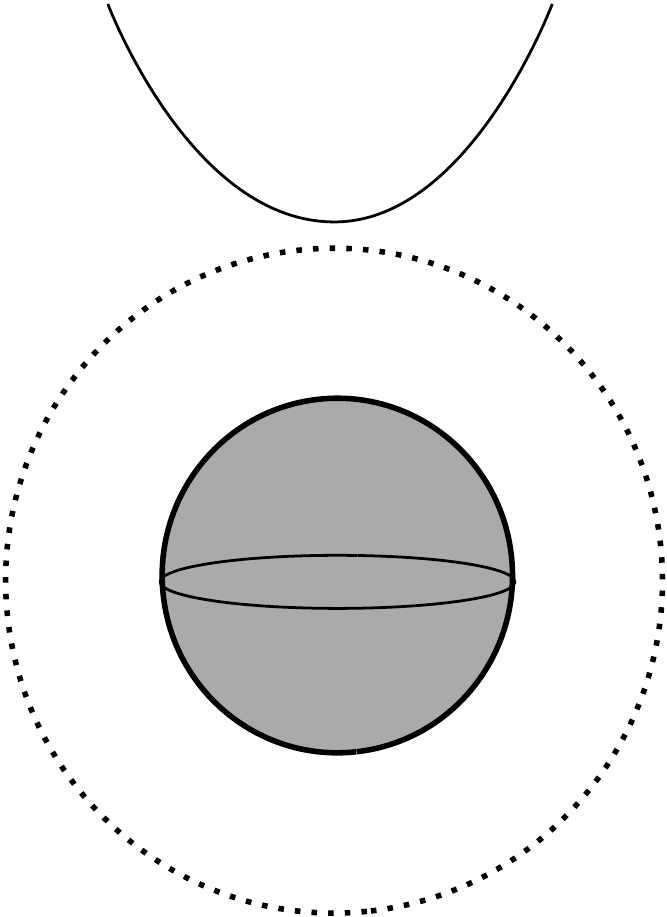}}
\caption{\small With a non-vanishing electric field, the new radial location $r_*$ emerges, which is shown as the dashed circular curve above. In this case, also, the embedding functions can be divided into two categories, separated by a critical one. The critical curve is subtle to define, since the conical singularity in the critical curve of figure \ref{figbh0} can occur either at the location of the background event horizon, or at $r_*$. Here, we will choose the latter, which will be further justified later.}
\label{figbh}
\end{figure}

Geometrically, therefore, there is a close parallel to the purely thermal physics. When there is a black hole present in the background, the probe brane can either fall inside the black hole, or stay above it. This depends on the asymptotic separation between the D$3$ branes and the D$7$-branes. This is how a first order phase transition separates the two phases, as the parameter $\left( m_q/T \right)$ is tuned. This is pictorially demonstrated in figure \ref{figbh0}. 
Now, when the background black hole is absent, but a non-vanishing electric field is excited on the worldvolume, one can identify two inequivalent class of probe brane profiles. These profiles are distinguished in terms of their boundary condition at the IR, namely:
\begin{eqnarray} \label{bcmin}
&& \left. \theta \right|_{r_{\rm min}} = \pi/2 \ , \quad \left. \theta' \right|_{r_{\rm min}} = - \infty \ , \quad {\rm bound \, \, state} \ , \\
&& \left. \theta \right|_{r_{*}} = \theta_0 \ , \quad \left. \theta' \right|_{r_*} = - \frac{1}{\hat{r}_*} \tan\frac{\theta_0}{2} \ , \quad {\rm plasma \, \, state} \ .
\end{eqnarray}
These boundary conditions are obtained either by regularity of the embedding function, or directly from the equations of motion. Pictorially, this is demonstrated in figure \ref{figbh}.
This transition is discussed in details in \cite{Albash:2007bk, Erdmenger:2007bn}, and figures \ref{figbh0} and \ref{figbh} are taken from \cite{Albash:2007bk}. Having discussed the classical physics, we now turn to analyze the physics of fluctuations of the probe brane sector.

\subsection{Fluctuations: Bosonic Sector}

The D$7$-brane fluctuations correspond to the meson operators in the dual $\cN=2$ field theory. For example, the scalar meson operators are given by\footnote{In reviewing this section, we follow the discussions in \cite{Kirsch:2006he}.}
\begin{eqnarray}
\cO_{\rm scalar}^{A \ell} = \bar{\psi}_i \sigma_{ij}^A \, X^\ell \psi_j + \bar{q}^m X_V^A X^\ell q^m \ , \quad i , m = 1,2 \ .
\end{eqnarray}
These operators have a conformal dimensions $\Delta = 3 + \ell$. Here, $\sigma^A = \left(\sigma_1, \sigma_2\right)$ is the Pauli matrices doublet, while $X_V^A$, $q^m$, $\psi_i$ are defined in table \ref{table1}. Also, $X^\ell$ denotes the symmetric traceless operator $X^{\{ i_1 \ldots i_\ell \}}$ of $\ell$ adjoint scalars $X^i$, for $i = 4,5,6,7$. The mesonic operators, $\cO_{\rm scalar}^{A\ell}$ transform in the $\left(\frac{\ell}{2}, \frac{\ell}{2}\right)$ of the SO$(4)$ and have $+2$ charge under the SO$(2) \equiv {\rm U}(1)_R$.

To obtain the fluctuation action, we can expand (\ref{dbi}) around the classical profile discussed above:
\begin{eqnarray}
&& \theta = \theta^{(0)} + \delta\theta \ , \quad \phi = \phi^{(0)} + \delta\phi \ , \\
&& f_{ab} = f_{ab}^{(0)} + \delta f_{ab} \ ,
\end{eqnarray}
where all the fluctuations $\delta \theta$, $\delta \phi$ and $\delta f_{ab}$ depend on all the worldvolume coordinates of the D$7$-brane. In general, the fluctuations can be coupled and therefore quite complicated to analyze. Substantial simplification occurs at $\theta_0(r) = 0$, which corresponds to the massless case. Also, we can consistently truncate the fluctuations oscillating only along the Minkowski-directions.

With these, the effective scalar and vector fluctuation actions are given by
\begin{eqnarray}
S_{\rm scalar} & = & - T_{\rm D7} \int d^8\xi \, \sqrt{- {\rm det} E^{(0)}} \, \frac{1}{2} \cS^{ab} \left[ G_{\theta\theta} \left(\partial_a \delta\theta\right) \left(\partial_b \delta\theta\right)  \right] \ , \label{scasim1} \\
S_{\rm vector} & = & - T_{\rm D7} \int d^8\xi \, \sqrt{- {\rm det} E^{(0)}} \, \frac{1}{4} \cS^{aa'} \cS^{bb'}  \left(\delta f_{a'b}\right) \left(\delta f_{b'a} \right) \ . \label{vecsim1}
\end{eqnarray}
Here the emergent metric $\cS$ is given by
\begin{eqnarray}
\cS = \cS_{tx^1 u} \otimes \cS_{x^2x^3} \otimes \cS_{S^3} \ , 
\end{eqnarray}
where $\cS_{x^2x^3}$ is identical to the metric components in that plane, $\cS_{S^3}$ is given by metric components along the $S^3$. The components of $\cS$ in the $\{t, x^1\equiv x, u\}$-plane are given by
\begin{eqnarray}
&& \cS_{tt} = G_{tt} + \frac{E^2}{G_{xx}} \ , \quad \cS_{tu} = \frac{E a'}{G_{xx}} = \cS_{ut} \ ,  \label{sym1} \\
&&  \cS_{xx} = \frac{E^2}{G_{tt}} + G_{xx} + \frac{a'^2}{g_{uu}} \ , \quad \cS_{uu} = g_{uu} + \frac{a'^2}{G_{xx}} \ , \quad g_{uu} = G_{uu} + \theta'^2 G_{\theta\theta}  \ . \label{sym2}
\end{eqnarray}
This metric $\cS$ is known as the open string metric\cite{Seiberg:1999vs}, which we will elaborate on subsequently. Evidently, there is no reason that the background geometry, denoted by $G$, and the emergent metric $\cS$ has the same causal structure. We will discuss some generic features of this emergent metric, however, for now, let us calculate the fermionic part of the action.

\subsection{Fluctuations: Fermionic Sector}

The fermionic fluctuations of the D$7$-brane correspond to the supersymmetric partners of the mesonic operators, discussed in the previous section. These operators are of two types\footnote{Here also we follow the discussions in \cite{Kirsch:2006he}.}
\begin{eqnarray}
&& \cF_\alpha^\ell \sim  \bar{q} X^\ell \tilde{\psi}_\alpha^\dagger + \tilde{\psi}_\alpha X^\ell  q \ , \\
&& \cG_\alpha^\ell \sim \bar{\psi}_i \sigma_{ij}^A \lambda_{\alpha B} X^\ell \psi_j + \bar{q}^m X_V^A \lambda_{\alpha B} X^\ell q^m \ , \quad  A, B = 1, 2 \ ,
\end{eqnarray}
with conformal dimensions $\Delta = \frac{5}{2} + \ell$ and $\Delta = \frac{9}{2} + \ell$. The quadratic order fluctuation action is somewhat involved in explicit form, and therefore we will simply write down the final massaged from, obtained in \cite{Kundu:2015qda}, which is given by
\begin{eqnarray} \label{ferm4}
&& S_{\rm spinor}  =  \frac{T_{\rm D7}}{2} \int d^8\xi \sqrt{-{\rm det} E^{(0)}}  \left[ \bar{\Psi} \gamma^a D_a \Psi - \cM \bar{\Psi}  \Psi  \right]  \ , \\
&& \left\{\gamma^a , \gamma^b\right\} = 2 \cS^{ab} \ ,
\end{eqnarray}
where $\cM$ is a mass matrix which we do not specify here. The corresponding equation of motion: 
\begin{eqnarray}
\gamma^a D_a \Psi - \cM \Psi = 0 \ .
\end{eqnarray}
It is clear that the fluctuation equation for the fermionic sector also couples to the effective metric $\cS$, which couples to the scalar and the vector fluctuations in the bosonic sector. These equations can be subsequently solved for the spectrum. It can already be said that the spectrum will have a quasinormal mode spectrum, analogous to a black hole geometry, when $\cS$ contains an event horizon. It turns out that, with a non-vanishing electric field on the classical probe profile, the metric $\cS$ indeed inherits an event horizon at $r_*$, which governs an effective thermal spectrum.

\subsection{Open String Metric: Some Features}

It is clear that the fluctuations in the D-brane, {\it i.e.}~the degrees of freedom living on the probe, couple to the same emergent metric. This hints towards an {\it open string equivalence principle} of the probe sector. Let us now discuss some generic features of the emergent open string metric, based on the examples above. The non-trivial structure exists in the $\{t, u, x\}$-sub-space. Here the line-element is given by
\begin{eqnarray}
ds_{\rm eff}^2 & = & \cS_{tt} dt^2 + 2 \cS_{ut} du dt + \cS_{uu} du^2 + \cS_{xx} dx^2 \nonumber\\
& = & \cS_{tt} d\tau^2 + \left(\cS_{uu} - \frac{\cS_{ut}^2}{\cS_{tt}}\right) du^2 + \cS_{xx} dx^2 \ ,
\end{eqnarray}
where we have redefined the time coordinate
\begin{eqnarray}
t = \tau + h(u) \ , \quad {\rm where} \quad h'(u) = \frac{\cS_{ut}}{\cS_{tt}} \ .
\end{eqnarray}
The full metric, hence, is given by
\begin{eqnarray} \label{osm1}
&& ds_{\rm osm}^2  =  - \frac{u^2}{R^2} \left( f - \frac{E^2R^4}{u^4} \right) d\tau^2 + \gamma_{uu} du^2 + \gamma_{xx} dx^2 + \frac{u^2}{R^2} dx_\perp^2  + R^2 \cos^2\theta d\Omega_3^2 \ , \\
&& f(u) = 1 - \frac{u_H^4}{u^4} \ , \quad dx_\perp^2 = dx_2^2 + dx_3^2 \ . 
\end{eqnarray}
For brevity, we do not explicitly write down the expressions for $\gamma_{uu}$, $\gamma_{xx}$. The background in (\ref{osm1}) is asymptotically AdS, and in the IR, it has en event horizon. This is similar to AdS-BH geometry. However, (\ref{osm1}) has a $u$-dependent curvature scalar and a singularity at $u=0$. This singularity is present event in the $u_H=0$ limit, and is solely supported by the worldvolume gauge field $E$.

With $\theta_0 =0$ and $u_H=0$, (\ref{osm1}) takes a simpler form: 
\begin{eqnarray}\label{osmred}
ds_5^2 = - \frac{u^2}{R^2} \left( 1 - \frac{E^2R^4}{u^4} \right) d\tau^2 + \frac{R^2 u^4}{u^6 - J^2 R^6} du^2 + \frac{u^4}{R^2} \frac{E^2R^4-u^4}{J^2R^6-u^6} dx^2 + \frac{u^2}{R^2} dx_\perp^2 \ .
\end{eqnarray}
Here we have suppressed the $S^3$-directions. Given this geometry, we can easily check some global features, such as the status of energy conditions. For this, let us choose the null vector: $n_\mu = (n_\tau, 1, 0, 0, 0)$ with $n_\tau^2 = \left(- g^{xx}/g^{\tau\tau} \right)$. It can be checked that ${\cal E}_{\mu\nu} n^\mu n^\nu <0$, where ${\cal E}_{\mu\nu} $ is the Einstein tensor corresponding to the metric (\ref{osmred}). On physical grounds, this is not unexpected and in keeping with the idea that such an emergent geometry can not be obtained from any low energy closed string sector. Thus, this is an intrinsic open string description. 

{\it relation between $\tau$ and $t$ are needed.}

An effective temperature can be defined from the metric in (\ref{osmred}): by Euclidean continuation of $\tau \to i \tau_E$, periodically compactifying the $\tau_E$ direction and demanding Euclidean regularity yields:
\begin{eqnarray} \label{temp1}
&& T_{\rm eff} = \sqrt{\frac{3}{2}} \, \frac{\sqrt{E}}{\pi R} \ , \quad {\rm with} \quad u_H = 0 \ , \label{teff0} \\
 && T_{\rm eff} = \frac{1}{2\pi R} \frac{\left(6 E^2 + 4 \left(\pi T R\right)^4\right)^{1/2}}{\left(E^2 + \left(\pi T R\right)^4 \right)^{1/4}} \ , \quad T = \frac{u_H}{\pi R^2} \ , \label{teff}
\end{eqnarray}
It is clear from the above formulae, that $T_{\rm eff} > T$. While the closed string low energy sector measures a temperature $T$, the corresponding open string sector measures $T_{\rm eff}$. The system is, thus, inherently non-equilibrium. However, in the $N_f \ll N_c$ limit, any heat exchange is suppressed and an equilibrium-like description holds. Note that (\ref{teff0}) and (\ref{teff}) corresponds to the massless case: $\theta_0 = 0$. When $m_q \not = 0$, the general formula is obtained in {\it e.g.}~\cite{Kim:2011qh}.

\begin{figure}[h!]
\centering
\includegraphics[scale=0.30]{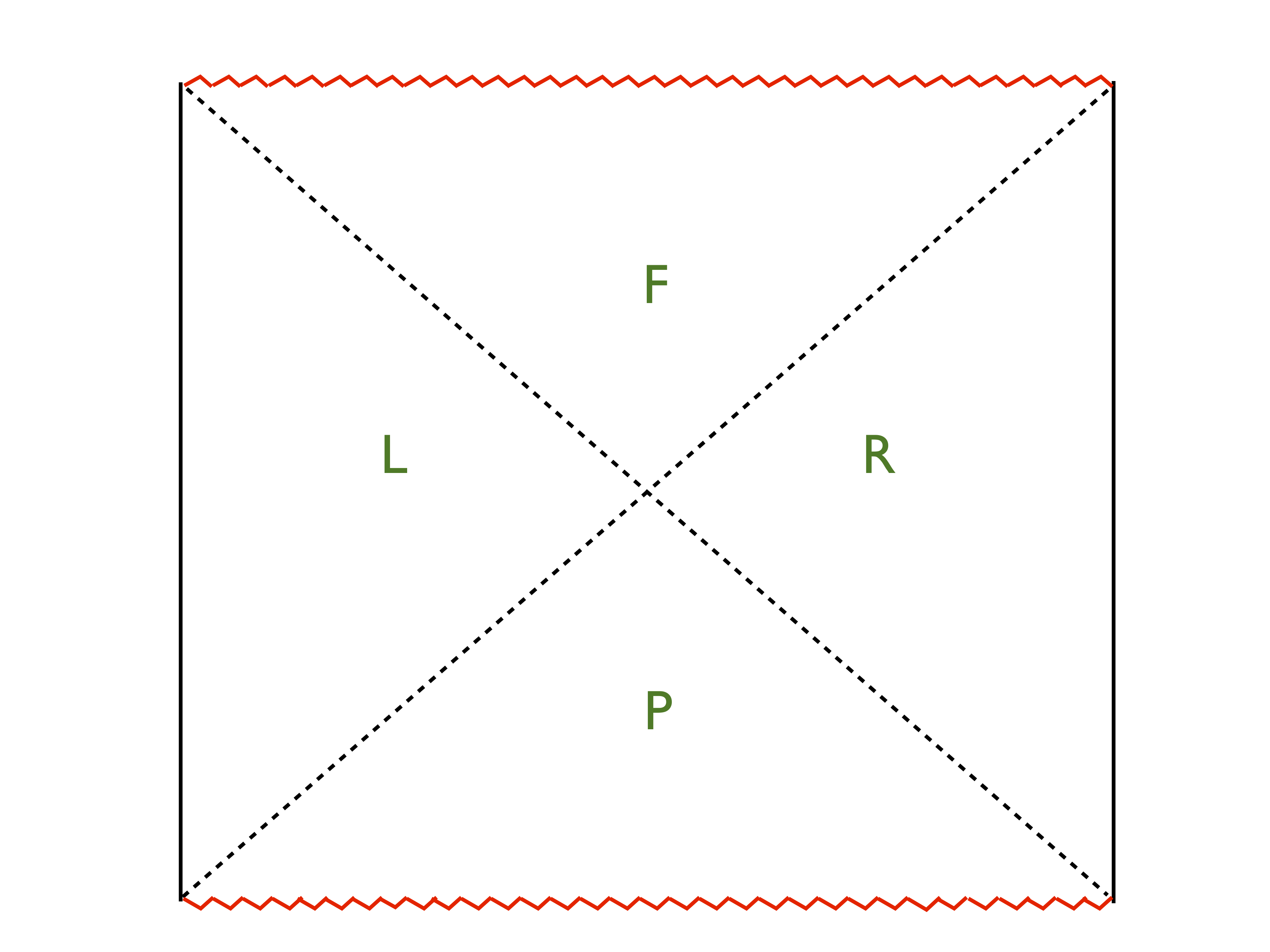}
\caption{\small A qualitative picture of the Kruskal extension of the open string metric. We have drawn the Penrose diagram in a ``square" form, which is not strictly correct. We will elaborate more on the Penrose diagram in later sections. This diagram is taken from \cite{Kundu:2015qda}.}
\label{pen1}
\end{figure}

Let us go back to (\ref{osmred}). The similarity of the causal structure with a black hole can be established by going through a set of coordinate transformations that finally describes (\ref{osmred}) is a Kruskal-type patch. Focussing on the near-horizon region, this can be achieved by:
\begin{eqnarray}
&& \alpha_* = \alpha + 2M \log \left| \frac{\alpha - 2M}{2M}\right| \ , \\
&& v = \tau' + \alpha_* \ , \quad u = \tau' - \alpha_* \ , \quad U = - e^{-u/4M} \ , \quad V = e^{v/4M} \ .
\end{eqnarray}
The metric in (\ref{osmred}), in the $\{\tau, u\}$-plane can be written as:
\begin{eqnarray}
ds_{\rm osm}^2 = - \frac{32 M^3}{\alpha} e^{-\alpha/2M} dU dV + ds_{\perp}^2 \ .
\end{eqnarray}
For our purposes, $ds_\perp^2$ will play no role. Correspondingly Kruskal-Szekres time and radial coordinates can be defined as: $t_K =  V + U$, $x_K = V - U$. Thus, a Penrose diagram can be drawn which, schematically takes the form in figure \ref{pen1}.
Finally, consider the $\tau'={\rm const}$ hypersurfaces in the Kruskal patch. In the following coordinate system:
\begin{eqnarray}
ds_{\rm osm}^2 = - \left(\frac{1- \frac{M}{2\beta}}{1 + \frac{M}{2\beta}}\right)^2 d\tau'^2 + \left(1 + \frac{M}{2\beta}\right)^4 d\beta^2 + ds_{\perp}^2 \ , \quad \alpha = \left(1 + \frac{M}{2\beta}\right)^2 \beta \ ,
\end{eqnarray}
the $\tau'={\rm const}$ hypersurfaces are simply ${\mathbb R}_\beta \otimes {\mathbb R}_x \otimes {\mathbb R}^2$.  Nevertheless, for a given $\alpha$, there are two roots of $\beta$, which are related by the symmetry: $\beta \to M^2 / 4 \beta$. The corresponding two regions, parametrized by the two roots of $\beta$, are connected by a constant-size Einstein-Rosen bridge. All these features are very similar to a standard black hole geometry.

\setcounter{equation}{0}
\section{D-Brane Description: A Different Model}

It is possible to further construct explicit examples of such probe brane configurations, in a background geometry. Since such constructions can be varied in richness, we will not attempt to provide a classification, rather we will present another explicit and instructive example. In this case, we are motivated by the analytical control the model provides us with. We begin with the standard AdS$_5 \times \cM_5$, where $\cM_5$ is some Einstein manifold (Sasaki-Einstein, if we want to preserve, at least, $\cN=1$ supersymmetry). Let us choose $\cM_5 \equiv S^5$. The probe degrees of freedom are now introduced by adding an $N_f$ D$5$-branes, as shown in table \ref{table2}.
\begin{table}[ht] 
\centering  
\begin{tabular}{| c | c | c | c | c | c | c | c | c | c | c | } 
\hline
Brane & 0 & 1 & 2 & 3 & 4 & 5 & 6 & 7 & 8 & 9  \\
\hline	                         
($N_c$) D3 & -- & -- & -- & -- & X & X & X & X & X & X \\
\hline                
($N_f$) D5 & -- & -- & -- & X & -- & -- & -- & X & X & X \\
\hline
\end{tabular} \caption{Fundamental sector introduced in the background of $N_c$ D$3$-branes.}  \label{table2}
\end{table} 
In table \ref{table2}, the notation $-$ stands for a direction along the worldvolume of the brane and $X$ represents the directions transverse to it. The directions $\{0,1,2,3\}$ represents the Minkowski directions. The rest of the directions are an ${\mathbb R}^6$ transverse to the stack of D$3$-branes. The D$3$-branes are extended along the Minkowski directions, and let us split ${\mathbb R}^6 \equiv {\mathbb R}^3 \times {\mathbb R}^3$. Now, let us use $\{\rho, \Omega_2\}$ and $\{\ell, \tilde{\Omega}_2\}$ to represent the two ${\mathbb R}^3$'s. Thus, $\{4,5,6\}$ corresponds to $\{\rho, \Omega_2\}$ and $\{7,8,9\}$ represents $\{\ell, \tilde{\Omega}_2\}$. This configuration, which preserves eight real supercharges, was studied initially in \cite{DeWolfe:2001pq, Filev:2009ai}, from the perspective of analyzing a defect CFT system. The degrees of freedom are: an adjoint vector multiplet and a hypermultiplet coming from the D$3$-branes in $(3+1)$-dimensions. The probe D$5$ sector yields a $(2+1)$-dimensional hypermultiplet in the fundamental representation.

The geometry is given by
\begin{eqnarray}
&& ds^2 = ds_{{\rm AdS}_5}^2  + ds_{S^5}^2 \ , \\
&& ds_{S^5}^2 = R^2 \left( d\psi^2 + \cos^2\psi d\Omega_2^2 + \sin^2 \psi d\tilde{\Omega}_2^2 \right) \ , \\
&& \rho = u \cos\psi \ , \quad \ell = u \sin\psi \ .
\end{eqnarray}
The D$5$-brane embedding function can be parametrized by a single function: $\psi(u)$. We assume that $P[\tilde{\Omega}_2] = 0$, where $\tilde{\Omega}_2$ denotes the metric components which yields the line element $d\tilde{\Omega}_2^2$. As in the previous case, there are distinct families of embedding which corresponds to different physics in the probe sector. Also, the analysis becomes simple when the fundamental sector is massless. This corresponds to setting $\psi = 0$. The induced worldvolume metric is simply an AdS$_4 \times S^2$.

To study a non-trivial dynamics of the D$5$-brane, we can excite a similar U$(1)$-flux on the worldvolume. The corresponding fluctuation analysis, by virtue of the open string equivalence principle\footnote{One can explicitly check this to be true. See for example \cite{Sonner:2012if, Banerjee:2018kwy} for a brief discussion on this, and \cite{Banerjee:2018kwy} for an explicit calculation of the same.}, exhibits a similar coupling with the open string metric. The metric is given by
\begin{eqnarray}
&& ds_{\rm eff}^2 = - \frac{u^2}{R^2} \left(f - \frac{E^2R^4}{u^4}\right) d\tau^2 + \gamma_{xx} dx^2 + \gamma_{uu} du^2 + \frac{u^2}{R^2} dy^2 + \cos^2\psi d\Omega_2^2 \ , \label{osm2} \\
&& f = 1 - \frac{u_H^4}{u^4} \ .
\end{eqnarray}
Again, we do not explicitly write down the expressions of $\gamma_{xx}$ and $\gamma_{uu}$. On the embedding $\psi(u) =0$, the open string metric in (\ref{osm2}) yields AdS$_4 \times S^2$ when both $u_H=0$ and $E=0$, or asymptotically. The IR-behaviour is different from an AdS$_4$-BH, since the functional form of $f(u)$ is different here. Also, demanding positivity of $\gamma_{xx}$ readily yields the expectation value of the current in the dual gauge theory.\footnote{In this case one gets, $j = E R^2$, independent of the background temperature $T = u_H/ (\pi R^2)$.} The corresponding effective temperature is now given by
\begin{eqnarray}
T_{\rm eff} = \left(T^4 + \frac{E^2}{\pi^2 R^4}\right)^{1/4} \ , \quad T = \frac{u_H}{\pi R^2} \ .
\end{eqnarray}
The near-horizon geometry has a similar structure as a black hole and a corresponding Penrose diagram can be drawn. Further note that, due to a substantial analytical control of this system, in  \cite{Sonner:2012if}, current-current correlation function, in the probe sector, was explicitly calculated. This correlation function turns out to possess a thermal behaviour with a temperature $T_{\rm eff}$. In particular, this effective temperature appears in the fluctuation-dissipation relation. For other (scalar or fermionic) sectors, even though analytical calculations may not be feasible, because of the open string equivalence principle, a similar behaviour is expected.

Further note that, this thermal imprint continues to hold beyond standard $2$-point functions. In \cite{Banerjee:2018twd, Banerjee:2018kwy}, a four-point out-of-time order (OTOC) correlator is explicitly calculated in the same vector sector discussed in \cite{Sonner:2012if}. This OTOC exhibits an exponential growth with real-time and a Lyapunov exponent which satisfies the maximal chaos bound: $\lambda_{\rm L} = 2 \pi T_{\rm eff}$.  This bound is shown to hold for a generic system, and is expected to saturate for systems with large number of degrees of freedom with a gravity dual. From a gravitational perspective, this saturation is universally described by the near-horizon dynamics of a black hole. In our case, a similar statement holds true for a geometry with a causal structure which is similar to a black hole.

Continuing on this theme, we can explicitly construct other examples where the adjoint matter sector is not described by a CFT, unlike the $\cN=4$ SYM. Such gauge theories can be constructed by considering the near-horizon dynamics of the D$p$-branes with $p \not = 3$. We will, however, not explicitly discuss such examples, but point the reader to {\it e.g.}~Sakai-Sugimoto model considered in \cite{Bergman:2008sg, Johnson:2008vna}.

\setcounter{equation}{0}
\section{Branes in Spacetime: A General Description}

In this section we will discuss the essential physics, without referring to any explicit example. The basic idea is as follows: Given a low energy closed string data, in string frame, $\{G, B, \phi\}$, where $G$ is the metric, $B$ is the NS-NS $2$-form and $\phi$ is the dilaton, along with the RR sector fluxes, the geometry is essentially described by a solution of type II supergravity. In this ten-dimensional background, we imagine describing a defect D-brane, as an embedded hypersurface. This hypersurface spontaneously breaks translational symmetry of the background and the corresponding low energy sector is described by the so-called DBI action, along with Wess-Zumino terms. The latter is topological in nature, and will not affect directly our discourse and therefore let us ignore such terms. The background RR-fluxes couple only to this topological sector and therefore we can ignore them for our discussions.

The essential dynamics is given by the following action, in the notation of \cite{Kundu:2013eba}:
\begin{eqnarray}
S_{\rm DBI} = T_{{\rm D}q} \int d^{q+1} \xi e^{- \phi} \sqrt{ - {\rm det} \left[ P[G+B] + \left( 2\pi \alpha' F \right) + i \bar{\psi} \gamma \nabla \psi \right) } + \ldots \ , \label{dbieff}
\end{eqnarray}
where $T_{{\rm D}q}$ encodes the tension of the brane, $F$ is the U$(1)$-flux on the worldvolume of the brane. The action also contains fermionic degrees of freedom, denoted by $\psi$, which can be completely fixed by demanding supersymmetry of the bosonic DBI action. Here, $\gamma$ denotes a gamma-matrix and $\nabla$ is an appropriately constructed covariant derivative. For our purpose, this schematic description suffices.

Around a classical saddle of (\ref{dbieff}), there are three kinds of fluctuations modes: (i) scalar, (ii) vector and (iii) spinor. Schematically, these fluctuations can be written as:
\begin{eqnarray}
&& X^\mu = X_{(0)}^\mu + \varphi^i \delta_\mu^i \ , \\
&& F_{ab} = F_{ab}^{(0)} + {\cal F}_{ab} \ , \quad \psi_a \ ,
\end{eqnarray}
where the classical saddle is described by the data $\{X_{(0)}^\mu, F_{ab}^{(0)}\}$. Here $\mu, \nu$ run over the entire spacetime directions, $a,b$ run over the worldvolume directions and $i,j $ run over the directions transverse to the D-brane. In the scalar sector we consider only the transverse fluctuations since longitudinal fluctuations can be gauged away by using worldvolume diffeomorphism. The Volkov-Akulov type fermionic terms is defined with gamma matrices that the following algebra:
\begin{eqnarray}
\left\{ \gamma_a, \gamma_b \right \} = P[G]_{ab} \ . 
\end{eqnarray}
To calculate fluctuations around the saddle $\{X_{(0)}^\mu, F_{ab}^{(0)}\}$, one needs to invert the matrix $M = P[G+B] + \left( 2\pi \alpha' F \right)$. This inversion yields\cite{Seiberg:1999vs}:
\begin{eqnarray}
M^{ab}  = \left( \left( P[G+B] + \left( 2\pi \alpha' F \right) \right)^{-1} \right)^{ab} = \cS^{ab} + \cA^{ab} \ , 
\end{eqnarray}
where
\begin{eqnarray}
&& \cS^{ab} = \left( \frac{1}{P[G+B] + \left( 2\pi \alpha' F^{(0)} \right) } \cdot P[G] \cdot \frac{1}{P[G - B] - \left( 2\pi \alpha' F^{(0)} \right) } \right) ^{ab} \ , \\
&& \cS_{ab} = P[G]_{ab} - \left( F^{(0)} \cdot P[G]^{-1} \cdot F^{(0)} \right)_{ab} \ , \\
&& \cA^{ab} = - \left( \frac{1}{P[G+B] + \left( 2\pi \alpha' F^{(0)} \right) } \cdot F^{(0)} \cdot \frac{1}{P[G - B] - \left( 2\pi \alpha' F^{(0)} \right) } \right) ^{ab} \ , 
\end{eqnarray}
Finally, the quadratic fluctuation Lagrangians are obtained as:
\begin{eqnarray}
&& \cL_{\rm scalar} =  - \frac{\kappa}{2} e^{-\phi} \sqrt{-{\rm det} M } \, \cS^{ab} \partial_a \varphi^i \partial_b \varphi^i + \ldots \ , \\
&& \cL_{\rm vector} = - \frac{\kappa}{4} e^{-\phi} \sqrt{-{\rm det} M } \, \cS^{ab} \cS^{cd} {\cal F}_{ac} {\cal F}_{bd} + \ldots \ , \\
&& \cL_{\rm spinor} = i  \kappa  e^{-\phi} \sqrt{-{\rm det} M } \, \bar{\psi}  \cS^{ab} \Gamma_a \nabla_b \psi + \ldots \ ,
\end{eqnarray}
where 
\begin{eqnarray}
\Gamma^{a} = \left( \cS^{ab} + \cA^{ab} \right) \gamma_b  \quad  \implies \quad \left\{ \Gamma^a, \Gamma^b \right \} =  2 \cS^{ab} \ . 
\end{eqnarray}
The indies above are raised and lowered with the metric $G$, as usual. Therefore, corresponding to two different metric $G$ and $\cS$, we have two inequivalent causal structures. Note that, such a possibility has already been discussed in the literature in \cite{Gibbons:2001gy, Gibbons:2000xe}, irrespective of holography. We will now discuss some specific aspects of this open string metric, in more details.

\subsection{Event Horizons: Some Comments}

In \cite{Kundu:2013eba}, a general formula for $T_{\rm eff}$ was obtained. This is the effective temperature that the probe sector measures. In general, one always observes $T_{\rm eff} > T$, however there are known exceptions, such as \cite{Nakamura:2013yqa}. Let us now consider a general case, exciting other fluxes on the worldvolume. Depending on the flux, this can correspond to turning on a chemical potential in the dual gauge theory (in the probe sector), and/or introducing a constant magnetic field.\footnote{See {\it e.g.}~\cite{Albash:2007bk, Erdmenger:2007bn, Mateos:2007vc, Evans:2011mu} for a detailed discussion of such physics in the D3-D7 system, and \cite{Bergman:2008sg, Johnson:2008vna, Johnson:2009ev, Bergman:2008qv, Thompson:2008qw} for similar physics in the Sakai-Sugimoto model.}

In general, we can arrange the electric and the magnetic field to be (i) parallel, or (ii) perpendicular. Assume that we have an ${\mathbb R}^{1,3}$-submanifold as the conformal boundary, where the dual gauge theory is defined. Consider, now, the case when (i) $E \parallel B$. The corresponding ansatz is:
\begin{eqnarray}
f = - E dt \wedge dx^1 - a_x' dx^1 \wedge du - B dx^2 \wedge dx^3 + a_t' dt \wedge du \ ,
\end{eqnarray}
where $a_x(u)$ contains the information about the gauge theory current and $a_t(u)$ contains the information about the chemical potential. The open string metric can be obtained as 
\begin{eqnarray}
 ds_{\rm eff}^2 & = & \left(\cS_{tt} - \frac{\cS_{tx}^2}{\cS_{xx}}\right) d\tau^2 + \left(\frac{\cS_{tx}}{\sqrt{\cS_{xx}}} d\tau + dX\right)^2 + \cS_{\perp} dx_{\perp}^2 , \nonumber\\
& + & \left(\frac{\cS_{tx}^2 \cS_{uu} - 2 \cS_{tu} \cS_{tx} \cS_{ux} + \cS_{tt} \cS_{ux}^2 - \cS_{tt} \cS_{uu} \cS_{xx}}{\cS_{tx}^2 - \cS_{tt} \cS_{xx}}\right) du^2  \ , \label{metabs2} \\
 dt & = & d\tau  - \frac{\cS_{tx} \cS_{ux} - \cS_{tu} \cS_{xx}}{\cS_{tx}^2 - \cS_{tt} \cS_{xx}} du \ , \\
dx^1 & = &  \frac{dX}{\sqrt{\cS_{xx}}} - \frac{\cS_{tu}\cS_{tx} - \cS_{tt} \cS_{ux}}{\cS_{tx}^2 - \cS_{tt} \cS_{xx}} du \ , 
\end{eqnarray}
with
\begin{eqnarray}
&& \cS_{\perp} = G_{xx} + \frac{B^2}{G_{xx}}   \ , \quad dx_{\perp}^2 = \left(dx^2\right)^2 + \left(dx^3\right)^2 \ ,
\end{eqnarray}
and 
\begin{eqnarray}
&& \cS_{tt} = G_{tt} + \frac{E^2}{G_{xx}} + \frac{a_t'^2}{G_{uu}} \ , \quad \cS_{tx} = \frac{a_t' a_x'}{G_{uu}} \ , \quad \cS_{tu} = \frac{E a_x'}{G_{xx}} \ , \\
&& \cS_{xx} = G_{xx} + \frac{E^2}{G_{tt}} + \frac{a_x'^2}{G_{uu}} \ , \quad  \cS_{uu} = G_{uu} + \frac{a_t'^2}{G_{tt}} + \frac{a_x'^2}{G_{uu}} \ .
\end{eqnarray}
The location of the open string metric event-horizon is given by
\begin{eqnarray}
 \cS_{tx}^2 - \cS_{tt} \cS_{xx} = 0  \quad & \implies& \quad \left. \tilde{d}^2 e^{2\phi} G_{tt} + G_{xx} \left( B^2 G_{tt} + G_{tt} G_{xx}^2 + e^{2\phi} j^2 \right) \right|_{u_*} = 0 \nonumber\\
& \implies & \quad  \left.G_{tt}G_{xx} + E^2 \right|_{u_*} = 0 \ ,
\end{eqnarray}
with the following definitions:
\begin{eqnarray}
j = \frac{\partial \cL_{\rm DBI}}{\partial a_x'} \ , \quad \tilde{d} = \frac{\partial \cL_{\rm DBI}}{\partial a_t'} \ ,
\end{eqnarray}
From the boundary gauge theory perspective, the fundamental sector current is proportional to $j$ and the charge density is, up to an overall constant, given by $\tilde{d}$. As before, the expectation value of the current can be determined by imposing regularity on the OSM metric. Note that, a corresponding membrane-paradigm description, for computing transport properties of the dual gauge theory, was developed and explored in \cite{Kim:2011qh}. Along similar lines, one can explore the case when (ii) $E \perp B$. In this case, an additional Hall current flows, which was analyzed in {\it e.g.}~\cite{O'Bannon:2007in}.

\subsection{Ergoplane: A Special Feature}

The open string metric can also contain an ``ergoplane" and therefore similar related black hole physics. This is explicitly demonstrated with the D$3$-D$7$ system in \cite{Kim:2011qh}. To see this, we excite a chemical potential. By setting $B=0$ in (\ref{metabs2}), one obtains:
\begin{eqnarray}
ds_{\rm eff}^2 & = & \left(G_{tt}G_{xx} +E^2\right) \left[ \frac{dT^2}{G_{xx}} + \frac{dX^2}{G_{xx}} \right] + \frac{1}{G_{uu}} \left( a_t' dT + a_x' dX \right)^2 + \sum_{i=2}^m G_{xx} dx^i dx^i \nonumber\\
&+& \left(G_{uu} + \frac{a_t'^2 G_{tt} + a_x'^2 G_{xx}}{G_{tt} G_{xx} + E^2}\right) du^2 \ , \\
dT & = & dt + \frac{E a_x'}{G_{tt} G_{xx} + E^2} du \ , \quad dX= dx - \frac{E a_t'}{G_{tt} G_{xx} + E^2} du \ .
\end{eqnarray}
The event horizon is located at $(G_{tt} G_{xx} + E^2) = 0$, and the ergoplane is located at:
\begin{eqnarray}
 \frac{G_{tt}G_{xx} + E^2}{G_{xx}} + \frac{a_t'^2}{G_{uu}} = 0  \quad & {\implies}& \quad \left(G_{tt}G_{xx} + E^2\right) = 0  \ , \nonumber\\
& {\rm or} & \quad \left(G_{tt} G_{xx}^m + e^{2\phi} j^2 \right) = 0 \ .
\end{eqnarray}
This directly yields a root $u_{\rm erg} \not = u_*$. As before, $j^2$ is determined from a simple algebraic equation: $\left. \tilde{d}^2 e^{2\phi} G_{tt} + \left(G_{tt} G_{xx}^m + e^{2\phi} j^2 \right) G_{xx} \right|_{u_*}= 0$.

Let us take a simple example, in which the background if given by an AdS$_{d+2}$, with a vanishing dilaton. One then obtains: 
\begin{eqnarray}
u_{\rm erg}^2 = E R^2 \left( 1+ \frac{\tilde{d}^2}{E^d}\right)^{1/d} > u_*^2 = E R^2 \ , 
\end{eqnarray}
The event horizon and the ergoplane merge in the limit $(\tilde{d}^2/E^d) \to 0$ and/or $(1/d) \to 0$.\footnote{The same statement holds for an AdS$_{d+2}$-BH background geometry as well.}

\subsection{Exactly Solvable Toy Models: More Examples}

In addition to the explicit D-brane constructions that have been discussed before, here we briefly mention some examples in which explicit analytical calculations can be performed, to arrive at the same conclusion. These are the so-called Lifshitz background of the following form:
\begin{eqnarray}
ds^2 = - \frac{dt^2}{v^{2z}} + \frac{1}{v^2} d\vec{x}^2 + \frac{dv^2}{v^2} \ , \label{lifback}
\end{eqnarray}
where $\vec{x}$ is a $2$-dimensional vector and $v$ is the radial coordinate ($v \to 0$ corresponds to boundary and $v\to\infty$ corresponds to the deep-IR). As before, we can consider a similar DBI-dynamics in the (\ref{lifback}), with a worldvolume flux: $a_x = - E t + h(v)$. One readily obtains an equation of motion for the function $h(v)$, which can be solved in terms of a first integral of motion. As before, this first integral of motion can be subsequently determined in terms of $e = \left( 2\pi \alpha' \right) E$.

Around this classical saddle, we consider fluctuations which yields:
\begin{eqnarray}
S_{\rm gauge} = - \frac{1}{4} \int dt d^dx \sqrt{- {\rm det}(G+f)} \cS^{aa'} \cS^{bb'} \delta f_{ab} \delta f_{a'b'} \ ,
\end{eqnarray}
where $\delta f$ denotes gauge field fluctuations. This action results in the following equation of motion
\begin{eqnarray}
\partial_a \left[\sqrt{-{\rm det} (G+f)} \cS^{aa'} \cS^{bb'} \delta f_{a'b'} \right] = 0 \ .
\end{eqnarray}
In the $\{t,v\}$-plane the corresponding OSM is given by
\begin{eqnarray}
&& ds^2 = - \frac{g(v)}{v^{2z}} d\tau^2 + \frac{1}{g(v)} \frac{dv^2}{v^2} \ , \quad g(v) = 1 - e^2 v^{2z+2} \ , \\
&& \tau = t + s(v) \ , \quad \frac{ds}{dv} = -\frac{e^2 v^{1+ 3z}} {1- e^2 v^{2z+2}}
\end{eqnarray}

With the following ansatz:
\begin{eqnarray}
\delta a_\tau = \delta a_\tau (v) e^{-i\omega \tau} \ , \quad \delta a_i = \delta a_i (v) e^{- i \omega \tau} \ ,
\end{eqnarray}
some explicit solutions, with purely ingoing boundary condition, can be obtained as follows:
\begin{eqnarray}
 z & = &  \frac{1}{2} \ , \nonumber\\
 \delta a_x(v) & = &  c_x \, \left( \frac{1 - \sqrt{v}}{1+ \sqrt{v}} \right)^{- \frac{i \omega}{3}} \left( \frac{1 - \sqrt{v} + v }{1+ \sqrt{v} + v } \right)^{- \frac{i \omega}{6}} \, {\rm exp} \left( \frac{i\omega}{\sqrt{3}} \arctan\left[  \frac{\sqrt{3} v}{1- v}\right] \right) \ ,  \label{z12} \\
 z & = &  1 \ ,  \nonumber\\
\delta a_x(v) &  = & c_x \,  \left( \frac{1 - v}{1+v} \right)^{- \frac{i \omega}{4}}   \, {\rm exp} \left( \frac{i\omega}{2} \arctan (v) \right) \ . \label{z1} \\
z & = & 2 \ , \nonumber \\
\delta a_x(v) & = & c_x \, \left( 1- v^2 \right)^{\frac{ - i \omega}{6}} \left( 1 + v^2 + v^ 4  \right)^{\frac{i \omega}{12}}  \, {\rm exp} \left( \frac{i\omega}{2\sqrt{3}}  \arctan \left( - \frac{\sqrt{3}}{1 + v^2} \right) \right) \ . \label{z2}
\end{eqnarray}
Here $c_x$ is a constant. Using these explicit solutions, we calculate {\it e.g.}~two point correlator, similar to \cite{Sonner:2012if}. As a result, we get a fluctuation-dissipation relation, with an effective temperature: $T_{\rm eff} = \frac{z+1}{\pi} e^{z/ (z+1)}$. While this is not surprising, it adds more explicit evidence to the theme.

\subsection{The Probe Limit: Validity}

Our entire discussion is based on the probe limit. Since this is a crucial ingredient in our construction, let us briefly review what this entails. At the level of the equations of motion, the Einstein tensor of a given solution must be parametrically large compared to the stress-tensor of the probe sector. Consider the decoupling limit of $N_c$ coincident D$p$-branes. These, in the string frame, are given by\cite{Itzhaki:1998dd}
\begin{eqnarray}
ds^2 & = & \left(\frac{u}{L}\right)^{\frac{7-p}{2}} \left ( - dt^2 + d\vec{x}_p^2\right ) + \left(\frac{L}{u}\right)^{\frac{7-p}{2}} \left( du^2 +  u^2 d\Omega_{8-p}^2 \right) \ , \\
e^\phi & = & \left(\frac{u}{L}\right)^{\frac{(p-3)(7-p)}{4}} \ , \\
F_{8-p} & = & \left( 7 - p\right) L^{7-p} \omega_{8-p} \ .
\end{eqnarray}
Here $d\Omega_{8-p}$ denotes the line element of an $(8-p)$-sphere and $\omega_{8-p}$ denotes the corresponding volume form. For these geometries, Einstein tensor behaves as:
\begin{eqnarray}
E_{tt} \sim \frac{u^{5-p}}{L^{7-p}} \sim E_{xx} \ , \quad E_{uu} \sim \frac{1}{u^2} \ , \quad E_{\alpha\beta} \sim u^{0} \, \eta_{\alpha\beta}  \ . \label{testbem}
\end{eqnarray}
Here $\alpha, \beta$ run over the gauge theory directions.

Consider $N_f$-number of probe D$(p+4)$-branes. These span $\{t, \vec{x}_p\}$-directions and wraps three-cycle $\cX_3 \subset S^{8-p}$. As before, we excite a U$(1)$-flux: $A_{x^1} = - E t + a_1 (u)$.  The probe energy-momentum tensor is given by
\begin{eqnarray}
T_{tt} \to \frac{u^2}{L^3} \sim T_{xx} \sim T_{x^1 x^1} \ , \quad T_{uu} \to \frac{1}{u^{5-p}} \ , \quad T_{\alpha\beta} \to \frac{1}{u^{3-p}} \, \eta_{\alpha\beta} \quad {\rm as} \quad u\to \infty \ , \label{testpemUV}
\end{eqnarray}
and
\begin{eqnarray}
T_{tt} & \to & \left( E J \right) u^{(p-9)/2} \ , \quad T_{x^1x^1} \to \left( \frac{J}{E} \right) u^{(5-p)/2} \ , \quad T_{xx} \to \left( \frac{E}{J} \right) u^{(p+3)/2} \ , \\
T_{uu} & \to & \left( E J \right) u^{(3p-23)/2} \ , \quad T_{\alpha\beta } \to \left( \frac{E}{J} \right) u^{(3p-7)/2} \quad {\rm as} \quad u \to 0 \ .  \label{testpemIR}
\end{eqnarray}
A comparison between (\ref{testbem}) and (\ref{testpemUV}) establishes validity of the probe limit, except the case with $p=4$. Similarly, comparing (\ref{testbem}) with (\ref{testpemIR}), we conclude that the IR will be heavily modified. In fact, this modification can only be controlled by placing an event horizon in the bulk. Further note that, $T_{tt} \sim (E \cdot J)$ has an Ohmic dissipation nature, in terms of the dual gauge theory. Similar conclusions were reached in {\it e.g.}~\cite{Hartnoll:2009ns, Bigazzi:2013jqa}.

\subsection{Effective Thermodynamics: A Summary}

The basic statement of gauge-gravity duality is an equivalence between the bulk gravitational path integral and the dual field theory path integral:
\begin{eqnarray}
&& \cZ_{\rm bulk} = \cZ_{\rm QFT} \ , \quad \cZ_{\rm bulk} = e^{i S_{\rm bulk}} \ , \quad S_{\rm bulk} = S_{\rm grav} + S_{\rm matter} \ , \nonumber \\
&& {\rm subsequently} \quad \cZ_{\rm QFT} = e^{i S_{\rm QFT}} \ , \quad  S_{\rm QFT} = S_{\rm gauge} + S_{\rm matter} \ .
\end{eqnarray}
In writing the above, we have explicitly assumed that the dual field theory is a gauge theory, with some matter content. When the matter sector consists of probe degrees of freedom, the corresponding path integral factorizes into two parts:  $ \cZ_{\rm bulk} = \cZ_{\rm sugra} \cZ_{\rm sugra}$. In the $N_c \to \infty$ limit, a semi-classical description is viable by considering small perturbations around a classical saddle:
\begin{eqnarray}
\cZ_{\rm sugra} = \int D\left[ \delta \phi \right] D\left[ \delta G \right] e^{- N_c^2 \left( \cS_{\rm sugra}^{(0)} + \cS_{\rm sugra}^{(2)} \left[ \delta \phi, \delta G \right] + \ldots \right) } \ , 
\end{eqnarray}
where $\cS_{\rm sugra}^{(2)}$ is the quadratic fluctuation term. Similarly, the probe sector also has a semi-classical description. Thus, at the semi-classical level, the entire path integral factorizes into a classical piece and a quadratic fluctuation piece. Schematically, these take the following form\cite{Banerjee:2016qeu}:
\begin{eqnarray}
\cZ_{\rm classical} = e^{- N_c^2  \cS_{\rm sugra}^{(0)} - g_s N_c N_f \cS_{\rm DBI}^{(0)} - g_s N_f^2 \cS_{\rm back-reac}^{(1)} + \ldots } \ , 
\end{eqnarray}
where $g_s$ is the string coupling and $\cS_{\rm back-reac}^{(1)}$ denotes the backreaction of the probe sector on the classical saddle. Naively, by tuning $N_f \ll N_c$, we can safely ignore the backreaction. Note, however, that this conclusion is subtle to make, as we have already demonstrated in the previous section.

The quadratic fluctuation part can be schematically represented as:
\begin{eqnarray}
\cZ_{\rm fluc} = \int D\left[ \varphi_{\rm grav}\right] e^{- N_c^2  \cS_{\rm sugra}^{(2)} + \ldots  }  \int D \left[ \varphi_{\rm brane}\right] e^{- g_s N_c N_f  \cS_{\rm DBI}^{(2)} + \ldots  }  \ , 
\end{eqnarray}
where $\varphi_{\rm grav}$ and $\varphi_{\rm brane}$ represent gravity and brane fluctuation modes, respectively. From this, we can already compare the relative scaling of two-point functions in the gravity and in the brane sectors, which are given by $\langle \varphi_{\rm grav} \varphi_{\rm grav}\rangle \sim 1/ N_c^2$ and $\langle \varphi_{\rm brane} \varphi_{\rm brane} \rangle \sim 1/ (N_c N_f)$.

Let us focus on the D-brane sector. In this sector, let us rewrite the generic form of the scalar and vector fluctuations:
\begin{eqnarray}
 \S_{\rm scalar} & = & - \frac{\kappa}{2} \int d\xi^8\left(\frac{{\rm det} G}{{\rm det} \S} \right)^{1/4} \sqrt{-{\rm det} \S} \,  \S^{ab} \, \partial_a \varphi^i \,  \partial_b \varphi^i  + \ldots  , \label{fscalar} \\
 \S_{\rm vector} & = & - \frac{\kappa}{4}  \int d\xi^8 \left(\frac{{\rm det} G}{{\rm det} \S} \right)^{1/4} \sqrt{-{\rm det} \S} \, \S^{ab}\S^{cd} \, \F_{ac} \F_{bd} + \ldots \ . \label{fvector} 
\end{eqnarray}
Here, $\kappa$ denotes an overall constant and $\ldots$ represent various interaction terms. The fields $\varphi^i$, $\F$ represent the scalar and vector fluctuation modes, respectively. In (\ref{fscalar})-(\ref{fvector}), $\S$ is the open string metric, which we have defined before. The kinetic term in (\ref{fscalar}) and (\ref{fvector}) can be written in a more canonical form:
\begin{eqnarray}
 \sqrt{- {\rm det} \tilde{\S}} \ \tilde{\S}^{ab} \left(\partial_a \varphi \right)  \left(\partial_b \varphi \right) \ , \quad {\rm and} \quad  \sqrt{- {\rm det} \tilde{\S}} \ \tilde{\S}^{ab} \tilde{\S}^{cd} \F_{ac} \F_{bd} \ , \label{canokin} 
\end{eqnarray}
The conformal factor $\Omega$ needs to be determined for each case, separately.

Before discussing the Euclidean path integral ({\it i.e.}~the partition function), let us briefly comment on the stress-tensor of the dual field theory. The corresponding data can be represented primarily in terms of the open string metric data, in the following form\cite{Banerjee:2015cvy}: 
\begin{eqnarray}
\langle T_\nu^\mu \rangle \propto \int du e^{\gamma \phi} \left( \frac{{\rm det} G}{{\rm det} \cS} \right)^{1/4} \sqrt{- {\rm det} \cS} \, \cS_\nu ^\mu \ ,
\end{eqnarray}
where $\gamma$ is a constant and $\phi$ is the dilaton field and $\mu. \nu$ are the directions along the dual field theory. In \cite{Banerjee:2015cvy}, several examples were discussed with explicit form of the energy-momentum tensor in the dual field theory.

In the Euclidean patch, the path integral yields a thermodynamic description in terms of a partition function. It is now expected that the probe sector thermodynamics will be simply determined in terms of the effective temperature $T_{\rm eff}$. However, the entropy, which can be obtained from the partition function itself, is not given in terms of the area of the OSM event horizon. It was also argued in \cite{Karch:2008uy}, that, in the probe limit, only free energy can be reliably calculated by computing on-shell action in the probe sector. For thermal entropy and such, the contribution coming from backreaction of the defect degrees of freedom mixes with the probe sector contribution.  {\it check the exact statement here.}

For the case in consideration, a proposed thermodynamic free energy was discussed in \cite{Alam:2012fw}, and subsequently generalized in \cite{Kundu:2013eba}. The Helmholtz free energy is given by
\begin{eqnarray}
\cF_{\rm H} & = & \left. T_{\rm eff} \, S_{\rm DBI}^{(\rm E)} \right |_{\rm on-shell} \nonumber \\
& \sim & \int_0^{u_*} du d^p\xi \left( \cL_{\rm on-shell} - j a_x' \right) \ .
\end{eqnarray}
Here $u_*$ is the OSM event-horizon, $S_{\rm DBI}^{(\rm E)}$ is the Euclidean DBI action. This is the only extensive quantity that we can define. The free energy, in the special case of AdS$_3$, contains a universal term of $\left(T_{\rm eff}^2 - T^2 \right)$, where $T_{\rm eff}$ and $T$ are the probe and the background sector temperatures. Such a term, intuitively, captures the heat exchange across the two systems in a two dimesional CFT.

Furthermore, the presence of the OSM event horizon, and the open string data: {\it specify this}
\begin{eqnarray}
&& \cS = G - \left( B \cdot G^{-1} \cdot B \right) \\
&& \cG_s = \left( \frac{{\rm det} \left( G + B \right)}{\rm det G} \right)^{1/2} \ ,
\end{eqnarray}
where $G$ denotes the worldvolume metric, $B$ denotes the anti-symmetric two-form (with factors of $\alpha'$ absorbed). Here $\cG_s$ is the open string coupling\cite{Seiberg:1999vs}. There is a natural geometric area which defines {\it an entropic quantity}, given by
\begin{eqnarray}
s \sim \left. \frac{1}{\cG_s} {\rm Area} \left( \cS \right) \right |_{\tau= {\rm const}, \, u = u_*} \ .
\end{eqnarray}
However, the physical meaning of this is unclear and it is certainly not the thermal entropy. An intriguing possibility was suggested in \cite{Sonner:2013mba, Jensen:2013ora, Jensen:2014lua}, in which one identifies this entropy with the entanglement entropy of the pair produced in the presence of a strong electric field. Before leaving this section, let us note that, a slightly different non-equilibrium thermodynamics has been proposed and explored in \cite{Nakamura:2012ae, Hoshino:2014nfa}, in which the prescription is provided in the Lorentzian section of the geometry, unlike in the Euclidean section which we have discussed here.

\subsection{Some Causal Features}

\begin{figure}[h!]
\centering
{\includegraphics[width = 3.3in]{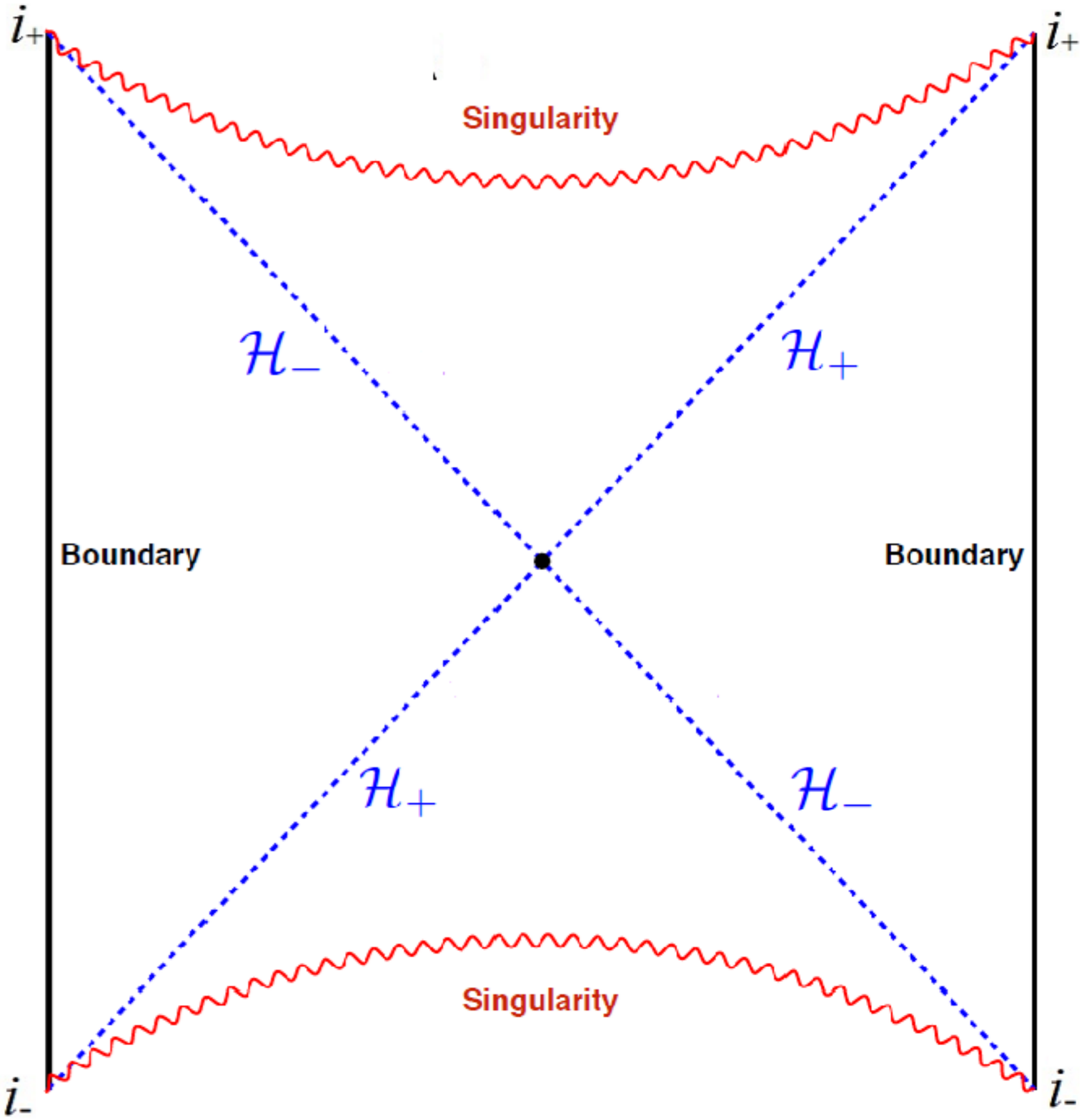}}
{\includegraphics[width = 3.3in]{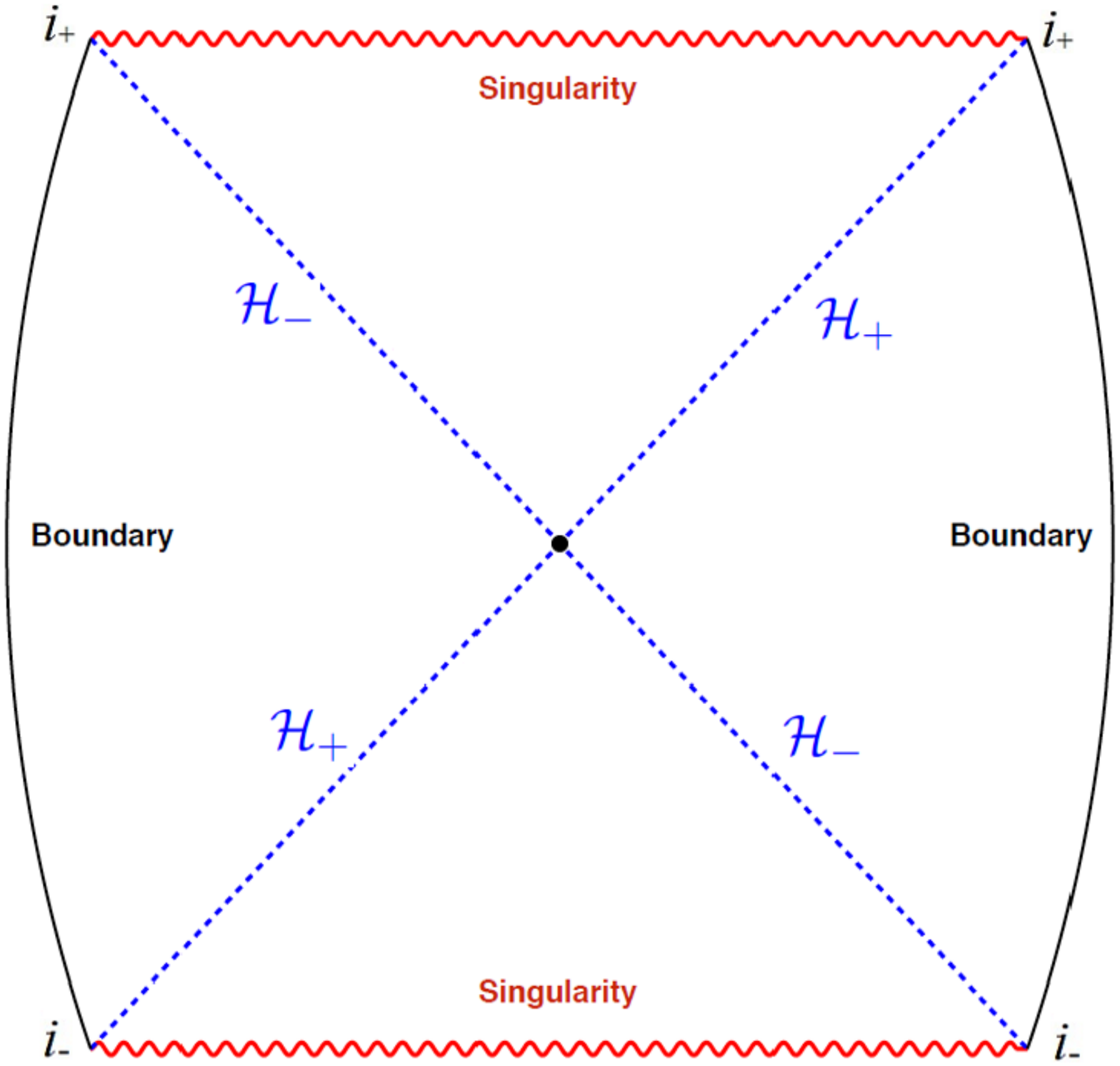}}
\caption{\small Penrose diagram corresponding to an OSM embedded in a purely AdS$_3$ (left) and a purely AdS$_4$ (right) background. The black dot at the centre corresponds to the bifurcation surface. In both these cases, there is a singularity denoted by the red curvy line.}
\label{pen34}
\end{figure}

A detailed analysis of the causal structure of such OSM geometry was discussed in \cite{Banerjee:2016qeu}. We will briefly review them here. As simple examples, let us discuss cases when the background is AdS$_3$ and AdS$_4$. We imagine a space-filling D-brane (and thus a corresponding DBI-action), with the U$(1)$ flux turned on. The explicit form of the metric are given in \cite{Banerjee:2016qeu}, and we will only discuss the qualitative physics here. One usually begins with a Poincar\'{e} patch description, and by going to a Kruskal extension, eventually ends up with a Penrose diagram. The resulting Penrose diagram for a purely AdS$_3$ and a purely AdS$_4$ background is shown in figure \ref{pen34}. 
The Penrose diagrams are qualitatively similar to what one obtains by solving Einstein gravity with an AdS asymptotics. Note, however, that in asymptotically AdS$_3$, the standard BTZ-type black holes do not have any singularity. For the OSM embedded in AdS$_3$, this is not the case and a curvature singularity exists. This singularity is visible by computing the Ricci-scalar itself.

Let us discuss some aspects of ``energy conditions" for the OSM geometry. A simple way to define such conditions is to demand that the OSM geometry can be obtained by solving a Einstein equations, with a suitable matter sector and translating the energy conditions on the corresponding matter sector. Thus, imagine:
\begin{eqnarray}
\G_{\mu\nu} + \Lambda g_{\mu\nu} = \T_{\mu\nu} \ , \label{einpretend}
\end{eqnarray}
where $\Lambda = - d(d-1) / 2$ is the cosmological constant in asymptotically AdS$_{d+1}$-background, $\G_{\mu\nu}$ is the Einstein tensor of the OSM and $\T_{\mu\nu}$ is the putative matter field.

Defined this way, an AdS$_3$-OSM yields 
\begin{eqnarray}
\T_{\mu\nu} t^\mu t^\nu \ge 0 \ , 
\end{eqnarray}
where $t^\mu$ is an arbitrary timelike vector; and hence Weak Energy Condition (WEC) is satisfied. Similarly, Strong Energy Condition (SEC) is also satisfied. The Null Energy Condition (NEC), on the other hand, yields: $\T_{\mu\nu} n^\mu n^\nu < 0$, for arbitrary null vectors $n^\mu$, and is violated. For the AdS$_4$-osm, with a similar choice for the timelike and the null vector, WEC yileds: $\T_{\mu\nu} t^\mu t^\nu  < 0$ and is thus violated. On the other hand, the NEC evaluates to $\T_{\mu\nu} n^\mu n^\nu = 0$.

In view of (\ref{canokin}), one may explore similar questions for the conformal metric $\tilde{\S}$. For the conformal OSM, in asymptotically AdS$_3$, one obtains:
\begin{eqnarray}
&& \tilde{\T}_{\mu\nu} t^\mu t^\nu < 0 \quad \implies \quad {\rm WEC \, \, violated} \ , \\
&&  \tilde{\T}_{\mu\nu} n^\mu n^\nu > 0 \quad \implies \quad {\rm NEC \, \, satisfied} \ ,
\end{eqnarray}
For asymptotically AdS$_4$, the conformal factor is identity. We can also check that a conformal AdS$_5$-OSM violates WEC, while both NEC and SEC are satisfied. In brief, the the conformal OSM always violates the WEC. Therefore there cannot be any area-increasing theorem for such event horizons.

\subsection{A Dynamical Example}

In a special example, discussed in \cite{Karch:2010kt}, a dynamical OSM geometry can also be constructed. Let us begin with the following background:
\begin{eqnarray} \label{EFmet}
ds^2 = - u^2 f(u) dV^2 + 2 dV du + u^2 dx_d^2 \ ,
\end{eqnarray}
where we have used Eddington-Finkelstein ingoing coordinate
\begin{eqnarray}
dt = dV + h(u) du \ , \quad {\rm with} \quad h(u) = - \frac{1}{u^2 f(u)} \ .
\end{eqnarray}
The metric takes the usual black hole metric form in the $\{t, u\}$-patch. The corresponding DBI action takes the following schematic form:
\begin{eqnarray}
S_{\rm DBI} = - \tau \int d^{d+2} \xi e^{-\phi} G_{xx}^{d/2} \, X \ , \quad X = \left[ 1+ G_{xx}^{-1} \left( 2 f_{xu} f_{xV} - f_{xu}^2 G_{tt}\right)\right]^{1/2} \ ,
\end{eqnarray}
where the U$(1)$-flux on the worldvolume is given by $f = f_{xu}(u) dx \wedge du + f_{xV} (u) dx \wedge dV$

A simple solution of the resulting equations of motion can be obtained for AdS$_4$\cite{Karch:2010kt}, characterized by one undetermined function: $f_{xV} = E(V)$. The corresponding OSM now takes the form:
\begin{eqnarray}
ds_{\rm osm}^2 = \left(G_{tt} + \frac{E(V)^2}{G_{xx}}\right) dV^2 + 2 dV du + u^2 dx_2^2 \ , \label{osmdyn}
\end{eqnarray}
which has an AdS-Vaidya form, with a dynamically evolving apparent and event horizon. Choosing an appropriate function $E(V)$, an event horizon formation on the brane can be easily described. In the dual gauge theory, one subsequently obtains a time-dependent current $\langle j(t) \rangle \sim E(t)$.

\setcounter{equation}{0}
\section{Conclusions}

In this review, we have discussed various examples in which the non-linear dynamics can give rise to an effective causal structure. When this causal structure is similar to that of a black hole, many similarities to classical as well as quantum properties of black holes become manifest. This bears substantial resemblance to other non-linear systems that describe gravity-like phenomena, known as {\it Analogue Gravity}, see {\it e.g.}~\cite{Barcelo:2005fc} for a detailed review on this. For the cases considered here, this causal structure emerges from an open string equivalence principle and this is what forms the analogue to gravity.

However, our attempts have been to briefly review the essential ideas, some explicit and simple examples on which these ideas are rather manifest. We have certainly not made any attempt to review the vast literature on the physics of probes traveling through a thermal medium in a strongly coupled gauge theory. The basic ingredients used to study this are open strings in a supergravity background. In the probe limit, the corresponding string profile can develop a worldsheet event horizon, which plays a crucial role in determining energy loss, drag force, stochastic Brownian motion of the probe. For a more extensive review on these, we will refer the reader to \cite{Hubeny:2010ry}. Beyond the probe limit, the open string backreaction estimates { \it e.g.}~also the radiation produced by the quark-like degree of freedom. See {\it e.g.}~\cite{Gubser:2007ga, Gubser:2007zr} for some of the early studies including the string backreaction. In a similar time-dependent context, far less explicit examples are known with backreaction from D-branes, since, technically, this a much harder problem to tackle. See {\it e.g.}~\cite{Sahoo:2010sp} for a perturbative analysis of the backreaction.

Staying in the probe limit, while there are similarities with a black hole, there are unsettled issues as well. One of the main issues is about the physical meaning of the area of the event horizon in all of the cases we have discussed above. For a black hole, this corresponds to the thermal entropy, moreover, one can prove area increasing theorems (with certain energy conditions for the matter field) in coherence with the second law of thermodynamics. This, however, is in stark contrast with what we have reviewed here. First, we can identify the Euclidean path integral as the corresponding thermal free energy and, therefore, derive the thermal entropy from this free energy. It is rather easy to see that this does not equal the area of the event horizon on the string worldsheet, or of the open string metric geometry. Secondly, as we have explicitly discussed {\it energy conditions} for such cases, there is no area increase theorem for such horizons.

One intriguing proposal of \cite{Sonner:2013mba} is to view this area as a measure of {\it entanglement entropy}, since, the open string event horizon is created due to a Schwinger pair creation on the D-brane and is therefore intimately related to entanglement. This is explicitly demonstrated by establishing that the Lorentzian section of the string worldsheet in \cite{Xiao:2008nr} is the analytic continuation of worldsheet instantons that describe Schwinger pair production of fundamental degrees of freedom. A similar statement on D-branes is desirable, however, to the best of our knowledge no such explicit connection has been made yet.

Complexity is a well-defined notion in quantum systems, which measures the minimum number of unitary operations required to reach a certain state starting from a base-state. Based on the eternal black holes in AdS, recently is has been suggested in \cite{Susskind:2014rva} that computational complexity in the dual CFT is measured, geometrically, by the Einstein-Rosen bridge in the eternal black hole spacetime. Since, at present, it remains unclear how to precisely define complexity in a QFT framework, we will not delve deeper into this issue. We will, however, simply note that from a geometric point of view, the string worldsheet or the D-brane open string metric that we have reviewed here, naturally allows us to define such a quantity. Indeed, in \cite{deBoer:2017xdk}, it has already been explicitly discussed. It would be very interesting to explore this issue further, since the two-dimensional worldsheet allows for an excellent analytical control over such issues.

Finally, we end by briefly commenting on one of the most interesting questions in black hole physics: the information paradox. Since the causal structure in the probe sector is closely similar to the causal structure of a black hole, in the eternal patch ({\it i.e.}~the thermofield double), one can address the nature of this paradox, following the proposal in \cite{Maldacena:2001kr}. In the standard thermofield double scenario, the mismatch between the gravity correlator and the unitary CFT correlation occurs at an order $e^{- c_1 S}$, where $S$ is the entropy of the system and $c_1$ is an irrelavant constant. For a purely gravitational system, $S \sim 1/ G_N$ and therefore the mismatch in correlators occurs at a non-perturbative order in $1/G_N$. This regime requires a highly interacting string theory, since $G_N \sim g_s^2$.

On a string worldsheet, a na\'{i}ve analysis following \cite{Maldacena:2001kr} would imply that the mismatch between the geometric correlator and the CFT correlator occurs at a non-perturbative order in $e^{- c_2 S}$, where $c_2$ is, as before, an unimportant constant. In this case, $S \sim 1 /\ell_s^2$, where $\ell_s$ is the string length and it does not depend on the string coupling. The Nambu-Goto theory receives no correction in $\ell_s$ and therefore, it may be possible to make quantitative progress in estimating such non-perturbative effects.

Finally, we end this review with the mention of the Lieb-Robinson bound, which provides an emergent upper bound on how fast information can propagate in a generic non-relativistic quantum mechanical system, with local interactions. While this limiting velocity is system specific, it rather intriguingly suggests a {\it relativity-like} structure for quantum non-relativistic systems, at least in the kinematic sense. It will be interesting to understand the physics of this better and perhaps explore a possible connection to what we have discussed in this review.

\section{Acknowledgements}

We thank Sumit Das, Jacques Distler, Roberto Emparan, Willy Fischler, Vadim Kaplunovsky, Sandipan Kundu, Hong Liu, Gautam Mandal, David Mateos, Andy O'Bannon, Bala Sathiapalan, Julian Sonner and Javier Tarr\'{i}o for numerous discussions and conversations on related topics. We specially thank Tameem Albash, Avik Banerjee, Veselin Filev, Clifford Johnson, Sandipan Kundu, Rohan Poojary on various collaborations that helped form the bulk of this review article. We also thank Mariano Chernicoff, Alberto Guijosa and Juan Pedraza for giving the permission to use several figures from their papers. This work is supported by the Department of Atomic Energy, Govt.~of India.




\begin{thebibliography}{99}


\bibitem{Feynman:1963fq}
R.~Feynman and J.~Vernon, F.L., 
``The Theory of a general quantum system interacting with a linear dissipative system,"
  Annals Phys. {\bf 24}  (1963) 118--173.
  


\bibitem{Schwinger:1960qe}
J.~S. Schwinger, 
``Brownian motion of a quantum oscillator,"  
J.Math.Phys. {\bf 2} (1961) 407--432.
  

\bibitem{Keldysh:1964ud}
L.~Keldysh, 
``Diagram technique for nonequilibrium processes,"  
  Zh.Eksp.Teor.Fiz. {\bf 47} (1964) 1515--1527.
  
  
\bibitem{Chou:1984es}
K.-c. Chou, Z.-b. Su, B.-l. Hao, and L.~Yu, 
``Equilibrium and Nonequilibrium Formalisms Made Unified,"
  Phys.Rept. {\bf 118} (1985) 1.
  

\bibitem{Landsman:1986uw}
N.~P. Landsman and C.~G. van Weert, 
``Real and Imaginary Time Field Theory at Finite Temperature and Density,"  
Phys. Rept. {\bf 145} (1987) 141.



\bibitem{Kamenev:2009jj}
A.~Kamenev and A.~Levchenko, 
``Keldysh technique and nonlinear sigma-model: Basic principles and applications,"
[arXiv:0901.3586 [hep-th]].  
  
  
  
\bibitem{Haehl:2016pec} 
  F.~M.~Haehl, R.~Loganayagam and M.~Rangamani,
  ``Schwinger-Keldysh formalism. Part I: BRST symmetries and superspace,''
  JHEP {\bf 1706}, 069 (2017)
  doi:10.1007/JHEP06(2017)069
  [arXiv:1610.01940 [hep-th]].
  
  
\bibitem{Haehl:2016uah} 
  F.~M.~Haehl, R.~Loganayagam and M.~Rangamani,
  ``Schwinger-Keldysh formalism. Part II: thermal equivariant cohomology,''
  JHEP {\bf 1706}, 070 (2017)
  doi:10.1007/JHEP06(2017)070
  [arXiv:1610.01941 [hep-th]].
  
  
\bibitem{Haehl:2017qfl} 
  F.~M.~Haehl, R.~Loganayagam, P.~Narayan and M.~Rangamani,
  ``Classification of out-of-time-order correlators,''
  arXiv:1701.02820 [hep-th].
  

\bibitem{1975tu}  
Y. Takahashi and H. Umezawa, Collect. Phenom. 2, 55 (1975).  


\bibitem{Israel:1976ur} 
  W.~Israel,
  ``Thermo field dynamics of black holes,''
  Phys.\ Lett.\ A {\bf 57}, 107 (1976).
  doi:10.1016/0375-9601(76)90178-X
  
  
\bibitem{Maldacena:2001kr} 
  J.~M.~Maldacena,
  ``Eternal black holes in anti-de Sitter,''
  JHEP {\bf 0304}, 021 (2003)
  doi:10.1088/1126-6708/2003/04/021
  [hep-th/0106112].
  
  
\bibitem{Maldacena:2013xja} 
  J.~Maldacena and L.~Susskind,
  ``Cool horizons for entangled black holes,''
  Fortsch.\ Phys.\  {\bf 61}, 781 (2013)
  doi:10.1002/prop.201300020
  [arXiv:1306.0533 [hep-th]].
  
  
\bibitem{Aamodt:2010pa} 
  K.~Aamodt {\it et al.}  [ALICE Collaboration],
  ``Elliptic flow of charged particles in Pb-Pb collisions at 2.76 TeV,''
  Phys.\ Rev.\ Lett.\  {\bf 105}, 252302 (2010)
  [arXiv:1011.3914 [nucl-ex]].
  

\bibitem{O'Hara:2002} 
  K.~M.~O'Hara {\it et al.},
  ``Observation of a strongly interacting degenerate Fermi gas of atoms,''
  Science\  {\bf 298}, 2179 (2002)
  [arXiv:0212463 [cond-mat]].
  
  
  
\bibitem{Maldacena:1997re} 
  J.~M.~Maldacena,
  ``The Large N limit of superconformal field theories and supergravity,''
  Int.\ J.\ Theor.\ Phys.\  {\bf 38}, 1113 (1999)
  [Adv.\ Theor.\ Math.\ Phys.\  {\bf 2}, 231 (1998)]
  doi:10.1023/A:1026654312961, 10.4310/ATMP.1998.v2.n2.a1
  [hep-th/9711200].
  
  
\bibitem{Gubser:1998bc} 
  S.~S.~Gubser, I.~R.~Klebanov and A.~M.~Polyakov,
  ``Gauge theory correlators from noncritical string theory,''
  Phys.\ Lett.\ B {\bf 428}, 105 (1998)
  doi:10.1016/S0370-2693(98)00377-3
  [hep-th/9802109].
  
  
\bibitem{Witten:1998qj} 
  E.~Witten,
  ``Anti-de Sitter space and holography,''
  Adv.\ Theor.\ Math.\ Phys.\  {\bf 2}, 253 (1998)
  doi:10.4310/ATMP.1998.v2.n2.a2
  [hep-th/9802150].
  
  
\bibitem{Aharony:1999ti} 
  O.~Aharony, S.~S.~Gubser, J.~M.~Maldacena, H.~Ooguri and Y.~Oz,
  ``Large N field theories, string theory and gravity,''
  Phys.\ Rept.\  {\bf 323}, 183 (2000)
  doi:10.1016/S0370-1573(99)00083-6
  [hep-th/9905111].
  

\bibitem{Son:2002sd} 
  D.~T.~Son and A.~O.~Starinets,
  ``Minkowski space correlators in AdS / CFT correspondence: Recipe and applications,''
  JHEP {\bf 0209}, 042 (2002)
  doi:10.1088/1126-6708/2002/09/042
  [hep-th/0205051].
  
\bibitem{Herzog:2002pc} 
  C.~P.~Herzog and D.~T.~Son,
  ``Schwinger-Keldysh propagators from AdS/CFT correspondence,''
  JHEP {\bf 0303}, 046 (2003)
  doi:10.1088/1126-6708/2003/03/046
  [hep-th/0212072].
  
  
\bibitem{Iqbal:2009fd} 
  N.~Iqbal and H.~Liu,
  ``Real-time response in AdS/CFT with application to spinors,''
  Fortsch.\ Phys.\  {\bf 57}, 367 (2009)
  doi:10.1002/prop.200900057
  [arXiv:0903.2596 [hep-th]].
  
  
\bibitem{Skenderis:2008dh} 
  K.~Skenderis and B.~C.~van Rees,
  ``Real-time gauge/gravity duality,''
  Phys.\ Rev.\ Lett.\  {\bf 101}, 081601 (2008)
  doi:10.1103/PhysRevLett.101.081601
  [arXiv:0805.0150 [hep-th]].
  
  
\bibitem{Skenderis:2008dg} 
  K.~Skenderis and B.~C.~van Rees,
  ``Real-time gauge/gravity duality: Prescription, Renormalization and Examples,''
  JHEP {\bf 0905}, 085 (2009)
  doi:10.1088/1126-6708/2009/05/085
  [arXiv:0812.2909 [hep-th]].
  
  
\bibitem{CasalderreySolana:2011us} 
  J.~Casalderrey-Solana, H.~Liu, D.~Mateos, K.~Rajagopal and U.~A.~Wiedemann,
  ``Gauge/String Duality, Hot QCD and Heavy Ion Collisions,''
  book:Gauge/String Duality, Hot QCD and Heavy Ion Collisions. Cambridge, UK: Cambridge University Press, 2014
  doi:10.1017/CBO9781139136747
  [arXiv:1101.0618 [hep-th]].
  

\bibitem{DeWolfe:2013cua} 
  O.~DeWolfe, S.~S.~Gubser, C.~Rosen and D.~Teaney,
  ``Heavy ions and string theory,''
  Prog.\ Part.\ Nucl.\ Phys.\  {\bf 75}, 86 (2014)
  doi:10.1016/j.ppnp.2013.11.001
  [arXiv:1304.7794 [hep-th]].
  

\bibitem{Gubser:2009md} 
  S.~S.~Gubser and A.~Karch,
  ``From gauge-string duality to strong interactions: A Pedestrian's Guide,''
  Ann.\ Rev.\ Nucl.\ Part.\ Sci.\  {\bf 59}, 145 (2009)
  doi:10.1146/annurev.nucl.010909.083602
  [arXiv:0901.0935 [hep-th]].
  
  
\bibitem{Hartnoll:2009sz} 
  S.~A.~Hartnoll,
  ``Lectures on holographic methods for condensed matter physics,''
  Class.\ Quant.\ Grav.\  {\bf 26}, 224002 (2009)
  doi:10.1088/0264-9381/26/22/224002
  [arXiv:0903.3246 [hep-th]].
  
  
\bibitem{Hartnoll:2016apf} 
  S.~A.~Hartnoll, A.~Lucas and S.~Sachdev,
  ``Holographic quantum matter,''
  arXiv:1612.07324 [hep-th].
  
  
  
\bibitem{Das:2016eao} 
  S.~R.~Das,
  ``Old and New Scaling Laws in Quantum Quench,''
  PTEP {\bf 2016}, no. 12, 12C107 (2016)
  doi:10.1093/ptep/ptw146
  [arXiv:1608.04407 [hep-th]].
  
  
\bibitem{Hubeny:2010ry} 
  V.~E.~Hubeny and M.~Rangamani,
  ``A Holographic view on physics out of equilibrium,''
  Adv.\ High Energy Phys.\  {\bf 2010}, 297916 (2010)
  doi:10.1155/2010/297916
  [arXiv:1006.3675 [hep-th]].
  
  
\bibitem{PhysRevLett.95.267001}
A.~G. Green and S.~L. Sondhi, Phys. Rev. Lett. {\bf 95},  267001  (2005).

\bibitem{PhysRevLett.97.227003}
A.~G. Green, J.~E. Moore, S.~L. Sondhi, and A. Vishwanath, Phys. Rev. Lett.
  {\bf 97},  227003  (2006).
  
\bibitem{Karch:2010kt} 
  A.~Karch and S.~L.~Sondhi,
  ``Non-linear, Finite Frequency Quantum Critical Transport from AdS/CFT,''
  JHEP {\bf 1101}, 149 (2011)
  
  
\bibitem{PhysRevLett.103.206401}
S.~Kirchner and Q.~Si, Phys. Rev. Lett. {\bf 103},  206401  (2009).

\bibitem{Cugliandolo.97}
L.~F.~Cugliandolo, J.~Kurchan, L.~Peliti, 
Phys. Rev. E {\bf 55}, 3898 (1997).


\bibitem{Babington:2003vm} 
  J.~Babington, J.~Erdmenger, N.~J.~Evans, Z.~Guralnik and I.~Kirsch,
  ``Chiral symmetry breaking and pions in nonsupersymmetric gauge / gravity duals,''
  Phys.\ Rev.\ D {\bf 69}, 066007 (2004)
  doi:10.1103/PhysRevD.69.066007
  [hep-th/0306018].


\bibitem{Mateos:2006nu} 
  D.~Mateos, R.~C.~Myers and R.~M.~Thomson,
  ``Holographic phase transitions with fundamental matter,''
  Phys.\ Rev.\ Lett.\  {\bf 97}, 091601 (2006)
  doi:10.1103/PhysRevLett.97.091601
  [hep-th/0605046].
  

\bibitem{Albash:2006ew} 
  T.~Albash, V.~G.~Filev, C.~V.~Johnson and A.~Kundu,
  ``A Topology-changing phase transition and the dynamics of flavour,''
  Phys.\ Rev.\ D {\bf 77}, 066004 (2008)
  doi:10.1103/PhysRevD.77.066004
  [hep-th/0605088].
  
  
\bibitem{Karch:2006bv} 
  A.~Karch and A.~O'Bannon,
  ``Chiral transition of N=4 super Yang-Mills with flavor on a 3-sphere,''
  Phys.\ Rev.\ D {\bf 74}, 085033 (2006)
  doi:10.1103/PhysRevD.74.085033
  [hep-th/0605120].
  

\bibitem{Gubser:2006bz} 
  S.~S.~Gubser,
  ``Drag force in AdS/CFT,''
  Phys.\ Rev.\ D {\bf 74}, 126005 (2006)
  doi:10.1103/PhysRevD.74.126005
  [hep-th/0605182].


\bibitem{Herzog:2006gh} 
  C.~P.~Herzog, A.~Karch, P.~Kovtun, C.~Kozcaz and L.~G.~Yaffe,
  ``Energy loss of a heavy quark moving through N=4 supersymmetric Yang-Mills plasma,''
  JHEP {\bf 0607}, 013 (2006)
  doi:10.1088/1126-6708/2006/07/013
  [hep-th/0605158].
  
  
\bibitem{Peeters:2007ti} 
  K.~Peeters and M.~Zamaklar,
  ``Dissociation by acceleration,''
  JHEP {\bf 0801}, 038 (2008)
  doi:10.1088/1126-6708/2008/01/038
  [arXiv:0711.3446 [hep-th]].
  

\bibitem{Hubeny:2014kma} 
  V.~E.~Hubeny and G.~W.~Semenoff,
  ``String worldsheet for accelerating quark,''
  JHEP {\bf 1510}, 071 (2015)
  doi:10.1007/JHEP10(2015)071
  [arXiv:1410.1171 [hep-th]].
  
  
\bibitem{deBoer:2008gu} 
  J.~de Boer, V.~E.~Hubeny, M.~Rangamani and M.~Shigemori,
  ``Brownian motion in AdS/CFT,''
  JHEP {\bf 0907}, 094 (2009)
  doi:10.1088/1126-6708/2009/07/094
  [arXiv:0812.5112 [hep-th]].
  
  
\bibitem{Son:2009vu} 
  D.~T.~Son and D.~Teaney,
  ``Thermal Noise and Stochastic Strings in AdS/CFT,''
  JHEP {\bf 0907}, 021 (2009)
  doi:10.1088/1126-6708/2009/07/021
  [arXiv:0901.2338 [hep-th]].
  
  
\bibitem{Giecold:2009cg} 
  G.~C.~Giecold, E.~Iancu and A.~H.~Mueller,
  ``Stochastic trailing string and Langevin dynamics from AdS/CFT,''
  JHEP {\bf 0907}, 033 (2009)
  doi:10.1088/1126-6708/2009/07/033
  [arXiv:0903.1840 [hep-th]].
  

\bibitem{CasalderreySolana:2009rm} 
  J.~Casalderrey-Solana, K.~Y.~Kim and D.~Teaney,
  ``Stochastic String Motion Above and Below the World Sheet Horizon,''
  JHEP {\bf 0912}, 066 (2009)
  doi:10.1088/1126-6708/2009/12/066
  [arXiv:0908.1470 [hep-th]].
   
  
  
\bibitem{Chernicoff:2011xv} 
  M.~Chernicoff, J.~A.~Garcia, A.~Guijosa and J.~F.~Pedraza,
  ``Holographic Lessons for Quark Dynamics,''
  J.\ Phys.\ G {\bf 39}, 054002 (2012)
  doi:10.1088/0954-3899/39/5/054002
  [arXiv:1111.0872 [hep-th]].
  
\bibitem{Jensen:2013ora} 
  K.~Jensen and A.~Karch,
  ``Holographic Dual of an Einstein-Podolsky-Rosen Pair has a Wormhole,''
  Phys.\ Rev.\ Lett.\  {\bf 111}, no. 21, 211602 (2013)
  doi:10.1103/PhysRevLett.111.211602
  [arXiv:1307.1132 [hep-th]].
  
  
\bibitem{Sonner:2013mba} 
  J.~Sonner,
  ``Holographic Schwinger Effect and the Geometry of Entanglement,''
  Phys.\ Rev.\ Lett.\  {\bf 111}, no. 21, 211603 (2013)
  doi:10.1103/PhysRevLett.111.211603
  [arXiv:1307.6850 [hep-th]].
  
  
\bibitem{Jensen:2014lua} 
  K.~Jensen and J.~Sonner,
  ``Wormholes and entanglement in holography,''
  Int.\ J.\ Mod.\ Phys.\ D {\bf 23}, no. 12, 1442003 (2014)
  doi:10.1142/S0218271814420036
  [arXiv:1405.4817 [hep-th]].
  
  
\bibitem{Chernicoff:2013iga} 
  M.~Chernicoff, A.~Güijosa and J.~F.~Pedraza,
  ``Holographic EPR Pairs, Wormholes and Radiation,''
  JHEP {\bf 1310}, 211 (2013)
  doi:10.1007/JHEP10(2013)211
  [arXiv:1308.3695 [hep-th]].
  
  
\bibitem{Nakamura:2012ae} 
  S.~Nakamura,
  ``Nonequilibrium Phase Transitions and Nonequilibrium Critical Point from AdS/CFT,''
  Phys.\ Rev.\ Lett.\  {\bf 109}, 120602 (2012)
  doi:10.1103/PhysRevLett.109.120602
  [arXiv:1204.1971 [hep-th]].
  

\bibitem{Nakamura:2013yqa} 
  S.~Nakamura and H.~Ooguri,
  ``Out of Equilibrium Temperature from Holography,''
  Phys.\ Rev.\ D {\bf 88}, no. 12, 126003 (2013)
  doi:10.1103/PhysRevD.88.126003
  [arXiv:1309.4089 [hep-th]].
  
  
\bibitem{Hoshino:2014nfa} 
  H.~Hoshino and S.~Nakamura,
  ``Effective temperature of nonequilibrium dense matter in holography,''
  Phys.\ Rev.\ D {\bf 91}, no. 2, 026009 (2015)
  doi:10.1103/PhysRevD.91.026009
  [arXiv:1412.1319 [hep-th]].
  
  
\bibitem{Matsumoto:2018ukk} 
  M.~Matsumoto and S.~Nakamura,
  ``Critical Exponents of Nonequilibrium Phase Transitions in AdS/CFT Correspondence,''
  arXiv:1804.10124 [hep-th].
  
  
\bibitem{Hoshino:2018vne} 
  H.~Hoshino and S.~Nakamura,
  ``Lorentz transformation of temperature and effective temperature,''
  arXiv:1807.10132 [hep-th].
  
\bibitem{Das:2010yw} 
  S.~R.~Das, T.~Nishioka and T.~Takayanagi,
  ``Probe Branes, Time-dependent Couplings and Thermalization in AdS/CFT,''
  JHEP {\bf 1007}, 071 (2010)
  doi:10.1007/JHEP07(2010)071
  [arXiv:1005.3348 [hep-th]].
  
  
\bibitem{AliAkbari:2012hb} 
  M.~Ali-Akbari and H.~Ebrahim,
  ``Meson Thermalization in Various Dimensions,''
  JHEP {\bf 1204}, 145 (2012)
  doi:10.1007/JHEP04(2012)145
  [arXiv:1203.3425 [hep-th]].
  
\bibitem{AliAkbari:2012vt} 
  M.~Ali-Akbari and H.~Ebrahim,
  ``Thermalization in External Magnetic Field,''
  JHEP {\bf 1303}, 045 (2013)
  doi:10.1007/JHEP03(2013)045
  [arXiv:1211.1637 [hep-th]].

\bibitem{tempreview}
https://www.physik.uni-augsburg.de/theo1/hanggi/Casas.pdf


\bibitem{2011JPhA...44V3001C}
L.~F.~Cugliandolo,
``The effective temperature,"
Journal of Physics A Mathematical General, {\bf 44}, 483001 (2011)
[arXiv:1104.4901 [cond-mat.stat-mech]]


\bibitem{CaronHuot:2011dr} 
  S.~Caron-Huot, P.~M.~Chesler and D.~Teaney,
  ``Fluctuation, dissipation, and thermalization in non-equilibrium AdS$_5$ black hole geometries,''
  Phys.\ Rev.\ D {\bf 84}, 026012 (2011)
  doi:10.1103/PhysRevD.84.026012
  [arXiv:1102.1073 [hep-th]].


\bibitem{Castorina:2007eb} 
  P.~Castorina, D.~Kharzeev and H.~Satz,
  ``Thermal Hadronization and Hawking-Unruh Radiation in QCD,''
  Eur.\ Phys.\ J.\ C {\bf 52}, 187 (2007)
  doi:10.1140/epjc/s10052-007-0368-6
  [arXiv:0704.1426 [hep-ph]].
  
  
\bibitem{Novello:1999pg} 
  M.~Novello, V.~A.~De Lorenci, J.~M.~Salim and R.~Klippert,
  ``Geometrical aspects of light propagation in nonlinear electrodynamics,''
  Phys.\ Rev.\ D {\bf 61}, 045001 (2000)
  doi:10.1103/PhysRevD.61.045001
  [gr-qc/9911085].
   
  
\bibitem{Kharzeev:2005iz} 
  D.~Kharzeev and K.~Tuchin,
  ``From color glass condensate to quark gluon plasma through the event horizon,''
  Nucl.\ Phys.\ A {\bf 753}, 316 (2005)
  doi:10.1016/j.nuclphysa.2005.03.001
  [hep-ph/0501234].
  

\bibitem{Kharzeev:2006zm} 
  D.~Kharzeev, E.~Levin and K.~Tuchin,
  ``Multi-particle production and thermalization in high-energy QCD,''
  Phys.\ Rev.\ C {\bf 75}, 044903 (2007)
  doi:10.1103/PhysRevC.75.044903
  [hep-ph/0602063].
  

\bibitem{BraunMunzinger:2003zd} 
  P.~Braun-Munzinger, K.~Redlich and J.~Stachel,
  ``Particle production in heavy ion collisions,''
  In *Hwa, R.C. (ed.) et al.: Quark gluon plasma* 491-599
  [nucl-th/0304013].
  
  
\bibitem{Cleymans:1992zc} 
  J.~Cleymans and H.~Satz,
  ``Thermal hadron production in high-energy heavy ion collisions,''
  Z.\ Phys.\ C {\bf 57}, 135 (1993)
  doi:10.1007/BF01555746
  [hep-ph/9207204].
  
  
\bibitem{Becattini:1997rv} 
  F.~Becattini and U.~W.~Heinz,
  ``Thermal hadron production in p p and p anti-p collisions,''
  Z.\ Phys.\ C {\bf 76}, 269 (1997)
  Erratum: [Z.\ Phys.\ C {\bf 76}, 578 (1997)]
  doi:10.1007/s002880050551
  [hep-ph/9702274].
  
  

\bibitem{tHooft:1973alw} 
  G.~'t Hooft,
  ``A Planar Diagram Theory for Strong Interactions,''
  Nucl.\ Phys.\ B {\bf 72}, 461 (1974).
  doi:10.1016/0550-3213(74)90154-0
  
  
\bibitem{tHooft:1993dmi} 
  G.~'t Hooft,
  ``Dimensional reduction in quantum gravity,''
  Conf.\ Proc.\ C {\bf 930308}, 284 (1993)
  [gr-qc/9310026].
  
  
\bibitem{Susskind:1994vu} 
  L.~Susskind,
  ``The World as a hologram,''
  J.\ Math.\ Phys.\  {\bf 36}, 6377 (1995)
  doi:10.1063/1.531249
  [hep-th/9409089].

  
\bibitem{Itzhaki:1998dd} 
  N.~Itzhaki, J.~M.~Maldacena, J.~Sonnenschein and S.~Yankielowicz,
  ``Supergravity and the large N limit of theories with sixteen supercharges,''
  Phys.\ Rev.\ D {\bf 58}, 046004 (1998)
  doi:10.1103/PhysRevD.58.046004
  [hep-th/9802042].
  
  
\bibitem{Karch:2002sh} 
  A.~Karch and E.~Katz,
  ``Adding flavor to AdS / CFT,''
  JHEP {\bf 0206}, 043 (2002)
  [hep-th/0205236].
  
  
\bibitem{Caceres:2010rm} 
  E.~Caceres, M.~Chernicoff, A.~Guijosa and J.~F.~Pedraza,
  ``Quantum Fluctuations and the Unruh Effect in Strongly-Coupled Conformal Field Theories,''
  JHEP {\bf 1006}, 078 (2010)
  doi:10.1007/JHEP06(2010)078
  [arXiv:1003.5332 [hep-th]].
  

\bibitem{Chernicoff:2010yv} 
  M.~Chernicoff and A.~Paredes,
  ``Accelerated detectors and worldsheet horizons in AdS/CFT,''
  JHEP {\bf 1103}, 063 (2011)
  doi:10.1007/JHEP03(2011)063
  [arXiv:1011.4206 [hep-th]].
  

\bibitem{FG}
C. Fefferman, C. R. Graham, Conformal invariants, in \'{E}lie Cartan et les Math\'{e}matiques d'Aujourd'hui, (Ast\'{e}risque, 1985), 95.


\bibitem{deHaro:2000vlm} 
  S.~de Haro, S.~N.~Solodukhin and K.~Skenderis,
  ``Holographic reconstruction of space-time and renormalization in the AdS / CFT correspondence,''
  Commun.\ Math.\ Phys.\  {\bf 217}, 595 (2001)
  doi:10.1007/s002200100381
  [hep-th/0002230].
  

\bibitem{Skenderis:2000in} 
  K.~Skenderis,
  ``Asymptotically Anti-de Sitter space-times and their stress energy tensor,''
  Int.\ J.\ Mod.\ Phys.\ A {\bf 16}, 740 (2001)
  doi:10.1142/S0217751X0100386X
  [hep-th/0010138].
  
  
\bibitem{Skenderis:2002wp} 
  K.~Skenderis,
  ``Lecture notes on holographic renormalization,''
  Class.\ Quant.\ Grav.\  {\bf 19}, 5849 (2002)
  doi:10.1088/0264-9381/19/22/306
  [hep-th/0209067].
  
  
\bibitem{Mikhailov:2003er} 
  A.~Mikhailov,
  ``Nonlinear waves in AdS / CFT correspondence,''
  hep-th/0305196.
  

\bibitem{Chernicoff:2008sa} 
  M.~Chernicoff and A.~Guijosa,
  ``Acceleration, Energy Loss and Screening in Strongly-Coupled Gauge Theories,''
  JHEP {\bf 0806}, 005 (2008)
  doi:10.1088/1126-6708/2008/06/005
  [arXiv:0803.3070 [hep-th]].
  
  
\bibitem{Chernicoff:2009re} 
  M.~Chernicoff, J.~A.~Garcia and A.~Guijosa,
  ``Generalized Lorentz-Dirac Equation for a Strongly-Coupled Gauge Theory,''
  Phys.\ Rev.\ Lett.\  {\bf 102}, 241601 (2009)
  doi:10.1103/PhysRevLett.102.241601
  [arXiv:0903.2047 [hep-th]].
  
  
\bibitem{Chernicoff:2009ff} 
  M.~Chernicoff, J.~A.~Garcia and A.~Guijosa,
  ``A Tail of a Quark in N=4 SYM,''
  JHEP {\bf 0909}, 080 (2009)
  doi:10.1088/1126-6708/2009/09/080
  [arXiv:0906.1592 [hep-th]].
  
  
\bibitem{Penedones:2016voo} 
  J.~Penedones,
  ``TASI lectures on AdS/CFT,''
  arXiv:1608.04948 [hep-th].
  
 
\bibitem{Xiao:2008nr} 
  B.~W.~Xiao,
  ``On the exact solution of the accelerating string in AdS(5) space,''
  Phys.\ Lett.\ B {\bf 665}, 173 (2008)
  doi:10.1016/j.physletb.2008.06.017
  [arXiv:0804.1343 [hep-th]].
  

\bibitem{Davies:1996ks} 
  P.~C.~W.~Davies, T.~Dray and C.~A.~Manogue,
  ``The Rotating quantum vacuum,''
  Phys.\ Rev.\ D {\bf 53}, 4382 (1996)
  doi:10.1103/PhysRevD.53.4382
  [gr-qc/9601034].
  
  
\bibitem{Athanasiou:2010pv} 
  C.~Athanasiou, P.~M.~Chesler, H.~Liu, D.~Nickel and K.~Rajagopal,
  ``Synchrotron radiation in strongly coupled conformal field theories,''
  Phys.\ Rev.\ D {\bf 81}, 126001 (2010)
  Erratum: [Phys.\ Rev.\ D {\bf 84}, 069901 (2011)]
  doi:10.1103/PhysRevD.81.126001, 10.1103/PhysRevD.84.069901
  [arXiv:1001.3880 [hep-th]].
  
  
\bibitem{Sekino:2008he} 
  Y.~Sekino and L.~Susskind,
  ``Fast Scramblers,''
  JHEP {\bf 0810}, 065 (2008)
  doi:10.1088/1126-6708/2008/10/065
  [arXiv:0808.2096 [hep-th]].
  
  
\bibitem{Bhattacharya:2018nrw} 
  R.~Bhattacharya, D.~P.~Jatkar and A.~Kundu,
  ``Chaotic Correlation Functions with Complex Fermions,''
  arXiv:1810.13217 [hep-th].
  
  
\bibitem{Maldacena:2015waa} 
  J.~Maldacena, S.~H.~Shenker and D.~Stanford,
  ``A bound on chaos,''
  JHEP {\bf 1608}, 106 (2016)
  doi:10.1007/JHEP08(2016)106
  [arXiv:1503.01409 [hep-th]].
  

\bibitem{Shenker:2013pqa} 
  S.~H.~Shenker and D.~Stanford,
  ``Black holes and the butterfly effect,''
  JHEP {\bf 1403}, 067 (2014)
  doi:10.1007/JHEP03(2014)067
  [arXiv:1306.0622 [hep-th]].
  

\bibitem{Shenker:2013yza} 
  S.~H.~Shenker and D.~Stanford,
  ``Multiple Shocks,''
  JHEP {\bf 1412}, 046 (2014)
  doi:10.1007/JHEP12(2014)046
  [arXiv:1312.3296 [hep-th]].
  

\bibitem{Shenker:2014cwa} 
  S.~H.~Shenker and D.~Stanford,
  ``Stringy effects in scrambling,''
  JHEP {\bf 1505}, 132 (2015)
  doi:10.1007/JHEP05(2015)132
  [arXiv:1412.6087 [hep-th]].
  

\bibitem{Murata:2017rbp} 
  K.~Murata,
  ``Fast scrambling in holographic Einstein-Podolsky-Rosen pair,''
  JHEP {\bf 1711}, 049 (2017)
  doi:10.1007/JHEP11(2017)049
  [arXiv:1708.09493 [hep-th]].
  
  
\bibitem{deBoer:2017xdk} 
  J.~de Boer, E.~Llabrés, J.~F.~Pedraza and D.~Vegh,
  ``Chaotic strings in AdS/CFT,''
  Phys.\ Rev.\ Lett.\  {\bf 120}, no. 20, 201604 (2018)
  doi:10.1103/PhysRevLett.120.201604
  [arXiv:1709.01052 [hep-th]].
  
  
\bibitem{Banerjee:2018twd} 
  A.~Banerjee, A.~Kundu and R.~R.~Poojary,
  ``Strings, Branes, Schwarzian Action and Maximal Chaos,''
  arXiv:1809.02090 [hep-th].
  
  
\bibitem{Banerjee:2018kwy} 
  A.~Banerjee, A.~Kundu and R.~Poojary,
  ``Maximal Chaos from Strings, Branes and Schwarzian Action,''
  arXiv:1811.04977 [hep-th].
    
  
\bibitem{Maldacena:2016upp} 
  J.~Maldacena, D.~Stanford and Z.~Yang,
  ``Conformal symmetry and its breaking in two dimensional Nearly Anti-de-Sitter space,''
  PTEP {\bf 2016}, no. 12, 12C104 (2016)
  doi:10.1093/ptep/ptw124
  [arXiv:1606.01857 [hep-th]].
  
  
\bibitem{Albash:2007bq} 
  T.~Albash, V.~G.~Filev, C.~V.~Johnson and A.~Kundu,
  ``Quarks in an external electric field in finite temperature large N gauge theory,''
  JHEP {\bf 0808}, 092 (2008)
  [arXiv:0709.1554 [hep-th]].
  

\bibitem{Karch:2007pd} 
  A.~Karch and A.~O'Bannon,
  ``Metallic AdS/CFT,''
  JHEP {\bf 0709}, 024 (2007)
  [arXiv:0705.3870 [hep-th]].
  

\bibitem{Myers:2007we} 
  R.~C.~Myers, A.~O.~Starinets and R.~M.~Thomson,
  ``Holographic spectral functions and diffusion constants for fundamental matter,''
  JHEP {\bf 0711}, 091 (2007)
  [arXiv:0706.0162 [hep-th]].
  
  
\bibitem{Albash:2007bk} 
  T.~Albash, V.~G.~Filev, C.~V.~Johnson and A.~Kundu,
  ``Finite temperature large N gauge theory with quarks in an external magnetic field,''
  JHEP {\bf 0807}, 080 (2008)
  [arXiv:0709.1547 [hep-th]].
  
  
\bibitem{Erdmenger:2007bn} 
  J.~Erdmenger, R.~Meyer and J.~P.~Shock,
  ``AdS/CFT with flavour in electric and magnetic Kalb-Ramond fields,''
  JHEP {\bf 0712}, 091 (2007)
  [arXiv:0709.1551 [hep-th]].
  

\bibitem{Kirsch:2006he} 
  I.~Kirsch,
  ``Spectroscopy of fermionic operators in AdS/CFT,''
  JHEP {\bf 0609}, 052 (2006)
  [hep-th/0607205].
  
  
\bibitem{Seiberg:1999vs} 
  N.~Seiberg and E.~Witten,
  ``String theory and noncommutative geometry,''
  JHEP {\bf 9909}, 032 (1999)
  [hep-th/9908142].
  
  
\bibitem{Kim:2011qh} 
  K.~-Y.~Kim, J.~P.~Shock and J.~Tarrio,
  ``The open string membrane paradigm with external electromagnetic fields,''
  JHEP {\bf 1106}, 017 (2011)
  [arXiv:1103.4581 [hep-th]].
  
  
\bibitem{Kundu:2015qda} 
  A.~Kundu,
  ``Effective Temperature in Steady-state Dynamics from Holography,''
  JHEP {\bf 1509}, 042 (2015)
  doi:10.1007/JHEP09(2015)042
  [arXiv:1507.00818 [hep-th]].
  
  
\bibitem{DeWolfe:2001pq} 
  O.~DeWolfe, D.~Z.~Freedman and H.~Ooguri,
  ``Holography and defect conformal field theories,''
  Phys.\ Rev.\ D {\bf 66}, 025009 (2002)
  [hep-th/0111135].
  
  
\bibitem{Filev:2009ai} 
  V.~G.~Filev,
  ``Hot Defect Superconformal Field Theory in an External Magnetic Field,''
  JHEP {\bf 0911}, 123 (2009)
  [arXiv:0910.0554 [hep-th]].
  
\bibitem{Sonner:2012if} 
  J.~Sonner and A.~G.~Green,
  ``Hawking Radiation and Non-equilibrium Quantum Critical Current Noise,''
  Phys.\ Rev.\ Lett.\  {\bf 109}, 091601 (2012)
  [arXiv:1203.4908 [cond-mat.str-el]].

  
\bibitem{Bergman:2008sg} 
  O.~Bergman, G.~Lifschytz and M.~Lippert,
  ``Response of Holographic QCD to Electric and Magnetic Fields,''
  JHEP {\bf 0805}, 007 (2008)
  [arXiv:0802.3720 [hep-th]].
  
  
\bibitem{Johnson:2008vna} 
  C.~V.~Johnson and A.~Kundu,
  ``External Fields and Chiral Symmetry Breaking in the Sakai-Sugimoto Model,''
  JHEP {\bf 0812}, 053 (2008)
  [arXiv:0803.0038 [hep-th]].
  
\bibitem{Kundu:2013eba} 
  A.~Kundu and S.~Kundu,
  ``Steady-state Physics, Effective Temperature Dynamics in Holography,''
  Phys.\ Rev.\ D {\bf 91}, no. 4, 046004 (2015)
  [arXiv:1307.6607 [hep-th]].
  
  
\bibitem{Gibbons:2001gy} 
  G.~W.~Gibbons,
  ``Aspects of Born-Infeld theory and string / M theory,''
  Rev.\ Mex.\ Fis.\  {\bf 49S1}, 19 (2003)
  [AIP Conf.\ Proc.\  {\bf 589}, 324 (2001)]
  doi:10.1063/1.1419338
  [hep-th/0106059].
  
  
\bibitem{Gibbons:2000xe} 
  G.~W.~Gibbons and C.~A.~R.~Herdeiro,
  ``Born-Infeld theory and stringy causality,''
  Phys.\ Rev.\ D {\bf 63}, 064006 (2001)
  doi:10.1103/PhysRevD.63.064006
  [hep-th/0008052].
  

\bibitem{Mateos:2007vc} 
  D.~Mateos, S.~Matsuura, R.~C.~Myers and R.~M.~Thomson,
  ``Holographic phase transitions at finite chemical potential,''
  JHEP {\bf 0711}, 085 (2007)
  [arXiv:0709.1225 [hep-th]].
  
\bibitem{Evans:2011mu} 
  N.~Evans, A.~Gebauer and K.~-Y.~Kim,
  ``E, B, mu, T Phase Structure of the D3/D7 Holographic Dual,''
  JHEP {\bf 1105}, 067 (2011)
  [arXiv:1103.5627 [hep-th]].
  
  
\bibitem{Johnson:2009ev} 
  C.~V.~Johnson and A.~Kundu,
  ``Meson Spectra and Magnetic Fields in the Sakai-Sugimoto Model,''
  JHEP {\bf 0907}, 103 (2009)
  [arXiv:0904.4320 [hep-th]].
  
  
\bibitem{Bergman:2008qv} 
  O.~Bergman, G.~Lifschytz and M.~Lippert,
  ``Magnetic properties of dense holographic QCD,''
  Phys.\ Rev.\ D {\bf 79}, 105024 (2009)
  [arXiv:0806.0366 [hep-th]].
  
  
\bibitem{Thompson:2008qw} 
  E.~G.~Thompson and D.~T.~Son,
  ``Magnetized baryonic matter in holographic QCD,''
  Phys.\ Rev.\ D {\bf 78}, 066007 (2008)
  [arXiv:0806.0367 [hep-th]].
  
  
\bibitem{O'Bannon:2007in} 
  A.~O'Bannon,
  ``Hall Conductivity of Flavor Fields from AdS/CFT,''
  Phys.\ Rev.\ D {\bf 76}, 086007 (2007)
  [arXiv:0708.1994 [hep-th]].
  

\bibitem{Hartnoll:2009ns} 
  S.~A.~Hartnoll, J.~Polchinski, E.~Silverstein and D.~Tong,
  ``Towards strange metallic holography,''
  JHEP {\bf 1004}, 120 (2010)
  [arXiv:0912.1061 [hep-th]].
  
  
\bibitem{Bigazzi:2013jqa} 
  F.~Bigazzi, A.~L.~Cotrone and J.~Tarrio,
  ``Charged D3-D7 plasmas: novel solutions, extremality and stability issues,''
  JHEP {\bf 1307}, 074 (2013)
  [arXiv:1304.4802 [hep-th]].
  
  
\bibitem{Banerjee:2016qeu} 
  A.~Banerjee, A.~Kundu and S.~Kundu,
  ``Emergent Horizons and Causal Structures in Holography,''
  JHEP {\bf 1609}, 166 (2016)
  doi:10.1007/JHEP09(2016)166
  [arXiv:1605.07368 [hep-th]].
  

\bibitem{Banerjee:2015cvy} 
  A.~Banerjee, A.~Kundu and S.~Kundu,
  ``Flavour Fields in Steady State: Stress Tensor and Free Energy,''
  JHEP {\bf 1602}, 102 (2016)
  doi:10.1007/JHEP02(2016)102
  [arXiv:1512.05472 [hep-th]].
  
  
\bibitem{Karch:2008uy} 
  A.~Karch, A.~O'Bannon and E.~Thompson,
  ``The Stress-Energy Tensor of Flavor Fields from AdS/CFT,''
  JHEP {\bf 0904}, 021 (2009)
  [arXiv:0812.3629 [hep-th]].
  
  
\bibitem{Alam:2012fw} 
  M.~S.~Alam, V.~S.~Kaplunovsky and A.~Kundu,
  ``Chiral Symmetry Breaking and External Fields in the Kuperstein-Sonnenschein Model,''
  JHEP {\bf 1204}, 111 (2012)
  [arXiv:1202.3488 [hep-th]].
  
\bibitem{Barcelo:2005fc} 
  C.~Barcelo, S.~Liberati and M.~Visser,
  ``Analogue gravity,''
  Living Rev.\ Rel.\  {\bf 8}, 12 (2005)
  [Living Rev.\ Rel.\  {\bf 14}, 3 (2011)]
  doi:10.12942/lrr-2005-12
  [gr-qc/0505065].
  
  
\bibitem{Gubser:2007ga} 
  S.~S.~Gubser, S.~S.~Pufu and A.~Yarom,
  ``Sonic booms and diffusion wakes generated by a heavy quark in thermal AdS/CFT,''
  Phys.\ Rev.\ Lett.\  {\bf 100}, 012301 (2008)
  doi:10.1103/PhysRevLett.100.012301
  [arXiv:0706.4307 [hep-th]].
  
  
\bibitem{Gubser:2007zr} 
  S.~S.~Gubser, S.~S.~Pufu and A.~Yarom,
  ``Shock waves from heavy-quark mesons in AdS/CFT,''
  JHEP {\bf 0807}, 108 (2008)
  doi:10.1088/1126-6708/2008/07/108
  [arXiv:0711.1415 [hep-th]].
   
  
\bibitem{Sahoo:2010sp} 
  B.~Sahoo and H.~U.~Yee,
  ``Electrified plasma in AdS/CFT correspondence,''
  JHEP {\bf 1011}, 095 (2010)
  doi:10.1007/JHEP11(2010)095
  [arXiv:1004.3541 [hep-th]].
  
\bibitem{Susskind:2014rva} 
  L.~Susskind,
  ``Computational Complexity and Black Hole Horizons,''
  [Fortsch.\ Phys.\  {\bf 64}, 24 (2016)]
  Addendum: Fortsch.\ Phys.\  {\bf 64}, 44 (2016)
  doi:10.1002/prop.201500093, 10.1002/prop.201500092
  [arXiv:1403.5695 [hep-th], arXiv:1402.5674 [hep-th]].


\end{thebibliography}
\end{document}